\documentclass[12pt]{article}
\usepackage{color}
\usepackage[dvips]{graphicx}
\usepackage{dcolumn}
\usepackage{amsmath}
\usepackage{url}

\begin{document}
\Large
\begin{center}
  Diagrammatic Multiplet-Sum Method (MSM) Density-Functional Theory (DFT):
  Investigation of the Transferability of Integrals in ``Simple'' 
  DFT-Based Approaches to Multi-Determinantal Problems
\end{center}
\normalsize

\vspace{0.5cm}

\noindent
Abraham PONRA\\
{\em Department of Physics, Faculty of Science, 
University of Maroua, P.O.\ Box 814, Maroua, CAMEROON\\
e-mail: abraponra@yahoo.com}

\vspace{0.5cm}

\noindent
Carolyne BAKASA\\
{\em Technical University of Kenya, P.O.\ Box 52428-00200,
Haile Selassie Avenue, Nairobe, KENYA\\
e-mail: carolyne.bakasa@gmail.com}

\vspace{0.5cm}

\noindent
Anne Justine ETINDELE\\
{\em Higher Teachers Training College, University of Yaounde I,
P.O.\ Box 47, Yaounde, CAMEROON\\
e-mail: anne.etindele@univ-yaounde1.cm}

\vspace{0.5cm}

\noindent
Mark E.\ CASIDA\\
{\em Laboratoire de Spectrom\'etrie, Interactions et Chimie th\'eorique 
(SITh),
D\'epartement de Chimie Mol\'eculaire (DCM, UMR CNRS/UGA 5250),
Institut de Chimie Mol\'eculaire de Grenoble (ICMG, FR2607), 
Universit\'e Grenoble Alpes (UGA)
301 rue de la Chimie, BP 53, F-38041 Grenoble Cedex 9, FRANCE\\
e-mail: mark.casida@univ-grenoble-alpes.fr} 

\vspace{0.5cm}

\begin{center}
{\bf Abstract}
\end{center}

\begin{quote}
\noindent
Static correlation is a difficult problem for density-functional
theory (DFT) as it arises in cases of degenerate or quasi-degenerate
states where a multideterminantal wave function provides the 
simplest reasonable first approximation to the true interacting wave function.
This is also where Kohn-Sham DFT may also fail to be noninteracting
$v$-representible (NVR).  In contrast, Kohn-Sham DFT typically works
well for describing the missing dynamic correlation when a 
single-determinantal reference wave function provides a good first
approximation to the true interacting wave function.  Multiplet sum method (MSM)
DFT [{\em Theor.\ Chim.\ Acta} {\bf 4}, 877 (1977)] provides one of the
earliest and simplest ways to include static correlation in DFT.  MSM-DFT
assumes that DFT provides a good description of single-determant energies 
and uses symmetry and simple ansatzes to include the effects of 
static correlation.
This is equivalent to determining the off-diagonal matrix elements in a small 
configuration interaction (CI) eigenvalue problem. We have developed a 
diagrammatic approach to MS-DFT facilitates comparison with wave function
CI and so allows educated guesses of off-diagonal CI matrix elements even
in the absence of symmetry.  In every case, an additional exchange-only
ansatz (EXAN) allows the MSM-DFT formulae to be transformed into wave function
formulae.  This EXAN also works for transforming time-dependent DFT
into time-dependent Hartree-Fock.  Although not enough to uniquely guess
DFT formulae from wave function formulae, the diagrammatic approach and
the EXAN provide important constraints on any guesses that might be 
used.  Some alternative guesses are tried out for problems concerning the 
ground and excited states of H$_2$, LiH, and O$_2$ in order to assess how 
much difference might be involved for different DFT guesses for off-diagonal 
matrix elements.
\end{quote}

\section{Introduction}
\label{sec:intro}

\marginpar{\color{blue} DFT}
John P.\ Perdew is a giant figure in the field of density-functional 
theory (DFT).  Speaking in the context
of the present article, John Perdew embodies, for us, the need 
to comprehend
\marginpar{\color{blue} xc}
the behavior of the {\em exact} exchange-correlation (xc) functional and the
creation of {\em ab initio} xc functionals which satisfy as many of the
conditions of the exact xc functional as possible.  He, as we do, believes 
that this is the best way to ensure that the realm of applicability of a 
density-functional approximation extends all the way from atoms to solids, 
via molecules.  The present, modest article, attempts in its own way 
imitate this admirable aspect of John's research in the sense of seeking 
structure and patterns which should be obeyed when building new theoretical 
models.

We wish now to talk about the unspeakable in Kohn-Sham DFT \cite{KS65}, or 
at least of an issue that is badly understood, and so rarely mentioned 
outside of specialized DFT circles.  
It is a problem of ground-state Kohn-Sham theory which resurfaces 
frequently in photochemical applications involving electronic excited states.
This is the inability of even exact DFT to describe static correlation due
to quasidegenerate states, made worse when using approximate xc functionals.
Photochemical funnels (i.e., avoided crossings and conical intersections)
are key for understanding many photophenomena and yet are exactly where
static correlation is strongest.  When these funnels are connectedhttps://www.overleaf.com/project/64dfc992f85a039220b96748 to the
ground state, then they also have a large impact on the ground-state
\marginpar{\color{blue} PES}
potential energy surface (PES).  The two most common DFT ways nowadays used
to treat excited states are the time-dependent (TD) DFT and multiplet sum
method (MSM) DFT.  Each of these methods has its advantages and disadvantages.
Here we will focus on MSM-DFT whose major disadvantage is its dependence on
symmetry arguments.  In fact, the principle motivation and {\em raison 
\marginpar{\color{blue} WFT}
d'\^etre} of the present article is to show how wave function theory (WFT)
like diagrams may be developed for MSM-DFT, with the hope that this may
then lead to the extension of MSM-DFT beyond high-symmetry problems.
We show by example for H$_2$, LiH, and O$_2$ that this provides a justification
for substituting some double excitation terms with terms associated with
single excitations (or, more properly, double replacement terms with single
replacement terms).  

The present reality is that most users of DFT are little educated in the 
fundamental theorems of DFT and may even be apathetic about this topic
as long as their calculations allow them to solve their pet problem.  
{\em Yet formal theory matters!}  The fundamental DFT theorems of 
Hohenberg-Sham \cite{HK64} and the reformulation of DFT in terms of fictious
auxiliary orbitals by Kohn and Sham \cite{KS65} present a well-defined
exact theory for which xc-functionals are to be developed.  These functionals
are to lead to a theory that behaves, not like WFT, but like formal DFT, 
obeying the exact conditions of exact DFT.  The downside is that formal
DFT can and does fail when trying to describe the ground states of certain
types of systems.  Hohenberg-Sham DFT assumed both $N$- and 
$v$-representability, which were later questioned.  Put simply, given 
a positive reasonable-looking function which integrates to $N$, can we
prove that it is the density of an $N$-electron wave function (the 
$N$-representability problem) and can we prove the existance of an external
potential $v$ whose ground state wave function leads to that positive 
reasonable-looking function which integrates to $N$ (the $v$-representability
problem)?  The answer to the $N$-representability problem is ``yes''
\cite{H81} but the answer to the $v$-representability problem is ``no'' 
\cite{EE83}.  It is now relatively well-known that the need for
$v$-representability may be circumvented by the Levy-Lieb constrained 
search reformulation of DFT \cite{L79,L82}.  However Kohn-Sham DFT 
still requires noninteracting
\marginpar{\color{blue} NVR}
$v$-representability (NVR).  As Levy had shown that there is always
a noninteracting system with the same density as the interacting 
system, but it is not necessarily a ground state density \cite{L82}. 
Put more clearly, failure of NVR means that the lowest energy solution 
of the Kohn-Sham problem
\marginpar{\color{blue} HOMO}
\marginpar{\color{blue} LUMO}
will have a lowest unoccupied molecular orbital (LUMO) which is lower
in energy than the highest occupied molecular orbital (HOMO) or, in the
jargon of the field, has ``a hole below the fermi level.''  
Failure of NVR has been shown in some cases by taking the best possible
from WFT and showing the impossibility of calculating the corresponding 
exact Kohn-Sham potential.  It is also manifest,
since it is easy to calculate, that the ground state of H$_2$ is in 
principle NVR. Nevertheless, a hole below the fermi level is often seen
when using approximate xc-functionals and which we call effective
failure of NVR \cite{CH12}.  This is exactly where Casida's formulation
of linear response TD-DFT can and does fail and which has an interesting 
connection with spin-wave instabilities.  An ensemble formulation of DFT cannot
fix this as it may lead to ensemble average geometries which are far from
what we want if we are calculating realistic PESs.  Alternatively, solving
the NVR problem may seem to require an orbital dependence of the xc-functional 
which is also problematic and counter to the principles of Kohn-Sham DFT.  
These problems arise from quasidegenerate states and seem built into 
Kohn-Sham DFT, causing problems for ground-state DFT, but also particularly 
severe problems for DFT methods that treat excited states.  
(See Refs.~\cite{TTR+08,CH12} for a more complete discussion of effective 
failure of NVR.)  The need to introduce
\marginpar{\color{blue} MDET}
some sort of multideterminantal (MDET) nature into DFT seems manifest 
if we wish to over come the effective violation of NVR problem.  
(See Ref.~\cite{THS+22} for a lengthy recent discussion of current
issues in DFT.)

A common way to try to incorporate some MDET character into DFT is
to use symmetry breaking of the Kohn-Sham orbitals, with emphasis
on different orbitals for different spin even in closed-shell systems.
This is one way to obtain reasonable MDET-like PESs at the expense of
other properties such as having the correct spin-densities.  It is also
difficult when using grid-based methods, where the underlying grid does
not necessarily reflect the symmetry of the molecule, to prevent symmetry
breaking.  However symmetry breaking suffers from at least two obvious problems:
(i) there may be (often are?) more than one way to break the symmetry of 
the wave function and different types of symmetry breaking might give the
lowest energy for different atomic configurations and (ii) our experience
with excited states indicates that not only are states more difficult to
assign once symmetry is broken but calculated excitation spectra often 
simply become unreliable.  Hence the need for a MDET extension without 
symmetry breaking.

\marginpar{\color{blue} SI}
[A brief review of some other MDET extensions of DFT has been included
in the Supplementary Information (SI) associated with this article.]

Our principle interest is in DFT tools that will help us treat the
electronic excited states of molecules.  Excited states, even more
than ground states, are often fundamentally multideterminantal.  The
two most popular DFT methods for treating excited states are TD-DFT
(see, e.g., \cite{CH12}) and the older Zieger-Rauk-Baerends-Daul MSM-DFT
\cite{ZRB77,D94}.  Each of these methods has its pros and cons and, 
although the fundamental equations look very different, tend to give
very similar answers in cases where both methods are applicable.
TD-DFT obtains information about excited states by looking at the 
response of the charge density to an external applied electric field.
In its usual formulation, it requires a single-determinantal reference
state, so that it is only a MDET method for excited states.  However
spin-flip TD-DFT using a triplet reference state allows some incorporation
of MDET character in the ground state.  Much of the popularity of 
TD-DFT resides in the fact that it is easily automated.  However it
has difficulty teaing charge transfer and double excitations.
MSM-DFT is the older method.  It assumes that DFT provides a good
description of dynamical correlation in states well described by 
a single-determinantal wave function and then uses symmetry arguments
to obtain information about static correlation that gives rise to 
multiplet states.  [The terms ``dynamic'' and ``static'' are imprecise
``fuzzy'' but still very useful terms.  In DFT, dynamic correlation is 
often associated with atoms and static correlationn is associated with 
quasidegeneracies.  In WFT, dynamic correlation is accounted for by 
many-body perturbation theory
\marginpar{\color{blue} MBPT}
(MBPT) supposing that the unperturbed wave function is a good starting 
point.  If this unperturbed wave function needs to be of MDET form, then
we speak of static correlation.]
MSM-DFT is more difficult to automate, being more strongly based upon
physical and chemical intuition.  In contrast to TD-DFT, MSM-DFT
sometimes provides an easy way to handle double excitations and handles
charge transfer excitations better than does TD-DFT.
The present paper focuses on MSM-DFT.  (Attempts to unite TD-DFT and
MSM-DFT are briefly reviewed in the SI.)

This article presents Feynman diagrams for analyzing terms in MSM-DFT.
It is important to realize that our approach is different from other
combinations of Feynman diagrams and DFT.  It is now common in theoretical
solid-state physics to do MBPT using DFT orbitals (e.g., Ref.~\cite{ORR02}
for a review).  Feynman diagrams have also been used to analyze TD-DFT
\cite{CH15} and to add MBPT corrections to in dressed TD-DFT \cite{HIRC11}.
In contrast to these approaches, the present approach is very simple
and (hopefully) transparent.  We use the simplest type of Feynman
diagrams---ones intended as an elementary baby step for pedagogical 
purposes in in Shavitt and Bartlett's book (Chapter 4 of Ref.~\cite{SB09},
pp.~90-129).  Our goal is to seek analogies and patterns which will
In particular, we will show that some double excitation terms may be
calculated using only single excitations.

This article is organized in the following manner: The next section
covers the basic theory used in the rest of the paper.  
Subsection~\ref{sec:MSM} is review, but needed to establish
notation and make sense of the otther subsections. 
A description of the WFT diagrams that we are using in this paper
has been included in the SI.  
The reader who is not familiar with this type of Feynman diagrams
is strongly urged to see the appropriate part of the SI.
Subsections~\ref{sec:SITdiags} and \ref{sec:MSMdiags} are new
and hence are included in the main text.
The exchange-only anzatz (EXAN) in Subsec.~\ref{sec:EXAN} is not
particulary new but has a special importance in the context of the
present paper.  Section~\ref{sec:details} presents computational
details.  We then go on to present the results of some example
calculations Sec.~\ref{sec:results}, notably for H$_2$ (Subsec.~\ref{sec:H2})
and for O$_2$ (Subsec.~\ref{sec:O2}).  
An example calculation for LiH is presented in the SI.
We will see that, while MSM-DFT terms cannot be replaced by their
corresponding MSM-WFT terms, MSM-DFT terms are roughly equal when they
correspond to MSM-WFT terms that are equal.  This is a symmetry-independent
result.  Section~\ref{sec:conclude} concludes.  

\section{Diagrammatic MSM-DFT}
\label{sec:theory}

The objective of this section is to present a diagrammatic representation 
for MSM-DFT which closely parallels diagrammatic wave function theory.  We
will proceed by steps.  We will first review the MSM with emphasis on the
usual triplet/open-shell singlet application.  A brief but essential 
review of a particular type of 
WFT diagrams introduced 
in Shavitt and Bartlett's book primarily for pedagogical purposes 
(chapter 4 of Ref.~\cite{SB09}, pp.~90-129),
but which suits our present needs very well,
is given in the SI and should
be studied by reader's not already familiar with this type of diagram. 
In a second step, we will modify these diagrams to include 
self-interaction terms (SITs).
\marginpar{\color{blue} SIT}
This is a crucial step because, although these SITs cancel out in WFT,
they do not cancel out in DFT.   Finally we will show how expansion of 
MSM-DFT expressions around a reference state lead to diagrams in close 
parallel with the WFT diagrams already presented.

\subsection{Multiplet Sum Method}
\label{sec:MSM}

Ziegler, Rauk, and Baerends introduced the MSM-DFT to handle the problem of
how to calculate multiplet splittings in excitation spectra using DFT
\cite{ZRB77}.  Their particular emphasis was on the calculation of the
energies of open-shell singlet excited states in closed shell molecules.  
They assumed that DFT gave a good description of states well described
by a single determinantal wave function but is not able to describe
states requiring a linear combination of quasidegenerate wave functions.
That is, DFT is able to describe dynamic correlation well but fails at
describing static correlation.  On the other hand, WFT is able to describe
static correlation with a wave function consisting of a few determinants
but often requires millions of terms to obtain quantitative results for
dynamic correlation.  It makes sense to seek the best of both worlds 
by combining DFT and WFT in such a way that DFT captures most of the
dynamic correlation while WFT catches the key static correlation.
We will illustrate the basic arguments here for the two-orbital two-electron
model (TOTEM).  
The TOTEM is adequate for treating both the problem of
the ground and lower states of H$_2$, the lower diabatic states of LiH, and the 
ground and lower states of O$_2$.  We will review these three applications 
in particular here.  Note that Claude Daul extended MSM-DFT to the more 
complicated problem of spatial multiplet splittings in transition metal 
complexes \cite{D94}.

\begin{figure}
\begin{center}
\includegraphics[width=0.7\textwidth]{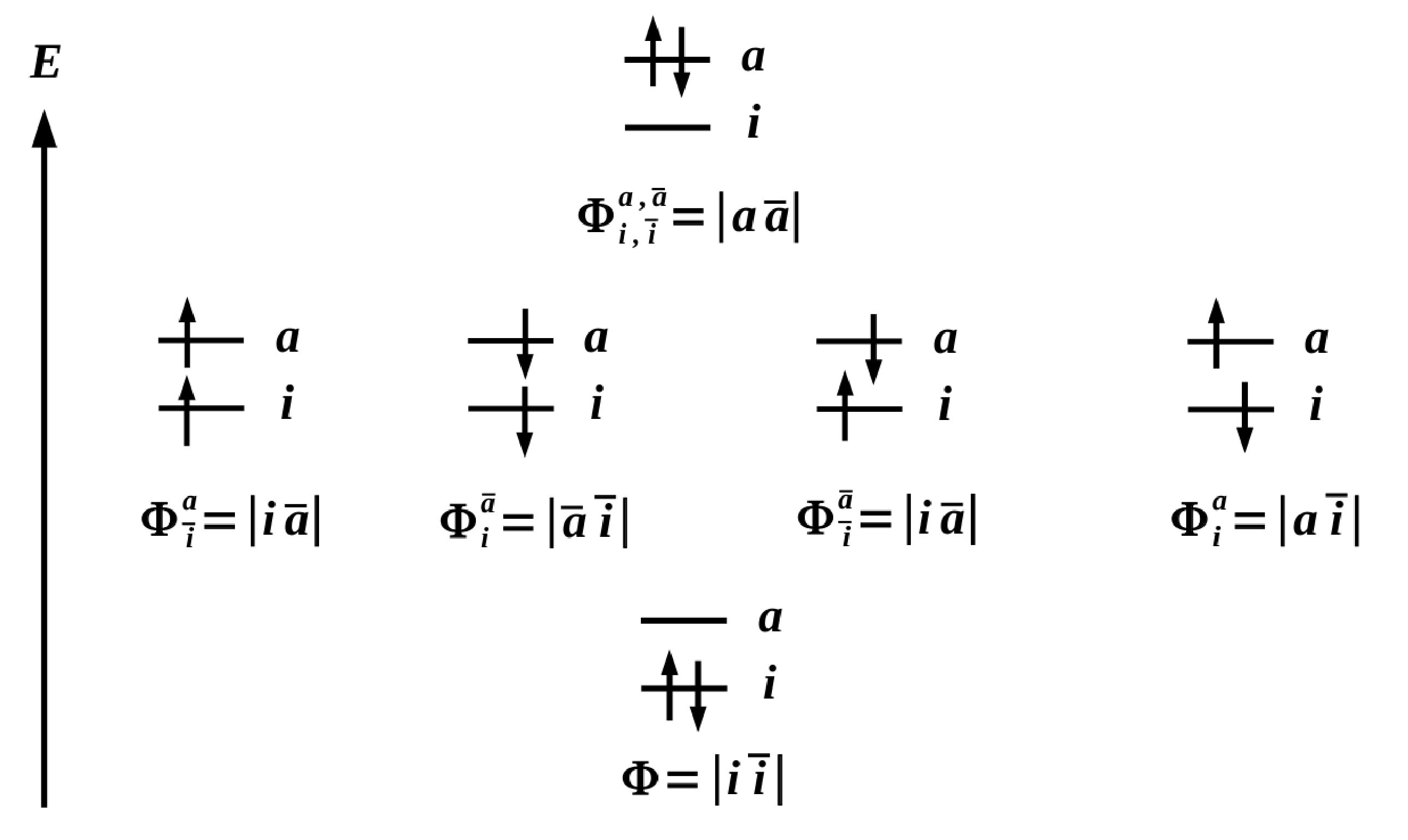}
\end{center}
\caption{
TOTEM state diagram when neglecting electron repulsions.
\label{fig:TOTEM1}
}
\end{figure}
We will mostly use orthonormal spin-orbitals when describing the 
formalism in this paper.  These spin-orbitals are divided into
occupied, unoccupied (virtual), and orbitals that are free to be either.
We will make heavy use of the index convention,
\begin{equation}
  \underbrace{a,b,c,\cdots,g,h}_{\text{unoccupied}}
  \underbrace{i,j,k,l,m,n}_{\text{occupied}}
  \underbrace{o,p,q,\cdots,x,yz}_{\text{free}}
  \, .
  \label{eq:theory.1}
\end{equation}
When necessary, spin will be indicated by an overbar for spin up and
the absence of an overbar for spin down.  We also use the common convention,
\begin{equation}
  \vert i_1, i_2, \cdots , i_N \vert = \frac{1}{\sqrt{N!}}
  \left| \begin{array}{cccc}
     \psi_{i_1}(1) & \psi_{i_2}(1) & \cdots & \psi_{i_N}(1) \\
     \psi_{i_1}(2) & \psi_{i_2}(2) & \cdots & \psi_{i_N}(2) \\
     \vdots        & \vdots & \ddots & \vdots \\
     \psi_{i_1}(N) & \psi_{i_2}(N) & \cdots & \psi_{i_N}(N) 
  \end{array} \right| \, ,
  \label{eq:theory.2}
\end{equation}
for a single determinant, where $(i)$ stands for the space and spin 
coordinates of electron $i$.  In our TOTEM, we have just one occupied
spatial orbital $i$ and one unoccupied orbital $a$ in the ground state.
If we assume that the orbital energy of $a$ is greater than that of $i$
and neglect electron repulsions, then we get the 
state diagram of {\bf Fig.~\ref{fig:TOTEM1}}, consisting
of six single-determinantal states.  

In the presence, or absence, of the electron repulsion, the spin operators
$\hat{S}^2$ and $\hat{S}_z$ commute with the hamiltonian $\hat{H}$.  Group 
theory then tells us that the eigenfunctions of the hamiltonian may be chosen 
as simulataneous eigenfunctions of these three operators, meaning that 
the eigenfunctions may be labeled using the spin quantum numbers $(S,M)$
and that there will be $2S+1$ degenerate functions for each value of $S$.
However the single-determinantal solutions in Fig.~\ref{fig:TOTEM1}
are not all spin eigenfunctions.  In fact, non-spin eigenfunctions are 
still a possible valid choice of eigenfunctions of the hamiltonian when 
electron repulsion is neglected, but only because we are allowed to take
linear combinations of degenerate states.  Introducing electron repulsions
will lift the degeneracy of the four states in the middle of 
Fig.~\ref{fig:TOTEM1} and these states will be labeled by $S$ and $M$.

\begin{figure}
\begin{center}
\includegraphics[width=0.7\textwidth]{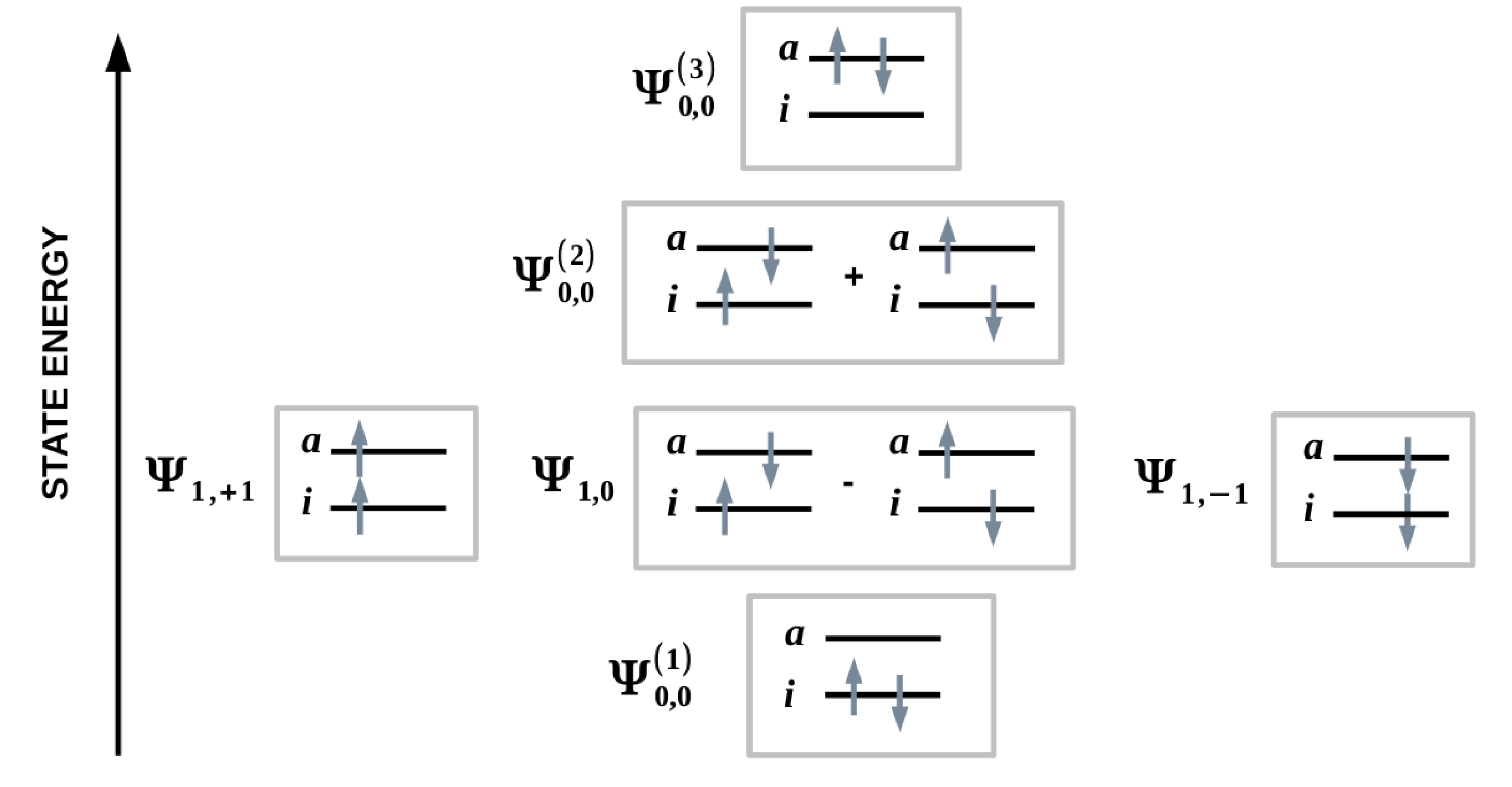}
\end{center}
\caption{
Generic TOTEM state diagram when electron repulsions are included.
The wave functions $\Psi_{S,M}$ are labeled using the usual quantum numbers
for the total spin ($S$) and for its projection on the $z$-axis ($M$).
\label{fig:genericstatediagram}
}
\end{figure}
There are many different ways to deduce the spin eigenfunctions $\Psi_{S,M}$,
but we wish to keep our presentation as simple as possible.  So we will just 
note that the two-electron case is a special case where the spatial and spin 
parts of the spin eigenfunctions factor.  This leads to the following spin 
eigenfunctions:
\begin{eqnarray}
  \Psi_{1,+1} & = & \vert i, a \vert \nonumber \\
              & = & \left[ \frac{1}{\sqrt{2}} 
              \left( i(1) a(2) - a(1) i(2) \right) \right]
              \left( \alpha(1) \alpha(2) \right) \nonumber \\
  \Psi_{1,0}  & = & \frac{1}{\sqrt{2}} \left( \vert i, {\bar a} \vert
              - \vert a, {\bar i} \vert \right) \nonumber \\
               & = & \left[ \frac{1}{\sqrt{2}} 
              \left( i(1) a(2) - a(1) i(2) \right) \right]
              \left[ \frac{1}{\sqrt{2}} \left( \alpha(1) \beta(2) +
              \beta(1) \alpha(2) \right) \right] \nonumber \\
  \Psi_{1,-1} & = & \vert {\bar a}, {\bar i} \vert \nonumber \\
              & = & \left[ \frac{1}{\sqrt{2}} 
              \left( i(1) a(2) - a(1) i(2) \right) \right]
              \left( \beta(1) \beta(2) \right) \nonumber \\
  \Psi_{0,0}  & = & \frac{1}{\sqrt{2}} \left( \vert i, {\bar a} \vert
              + \vert a, {\bar i} \vert \right) \nonumber \\
               & = & \left[ \frac{1}{\sqrt{2}} 
              \left( i(1) a(2) + a(1) i(2) \right) \right]
              \left[ \frac{1}{\sqrt{2}} \left( \alpha(1) \beta(2) -
              \beta(1) \alpha(2) \right) \right] \, . 
  \label{eq:theory.3}
\end{eqnarray}
As the Pauli principle tells us that the total wave function must be
antisymmetric, there are only two possibilities.  In the first possibility,
the spatial part of the wave function is antisymmetric and the spin part
is symmetric.  These are the triplet ($S=1$) wave functions.  They are
energetically degenerate for a spin-free hamiltonian.  The second possibility
is that the spatial part of the wave function is symmetric and the 
spin part is antisymmetric.  As there is only one way to do this, there is
only one singlet ($S=0$) wave function.  Furthermore antisymmetry also 
means that electrons with the same spin avoid each other in space (Fermi hole)
and so have a smaller electron repulsion energy than do electrons with the
opposite spins because same spin electrons are, on average, farther apart.  
This is a common explanation of why triplet energies are
usually lower than the corresponding singlet energies.  This gives us the
new generic state diagram given in {\bf Fig.~\ref{fig:genericstatediagram}}.
In principle, configuration mixing can occur between the three singlet
wave functions, $\Psi_{0,0}^{(1)}$, $\Psi_{0,0}^{(2)}$, and $\Psi_{0,0}^{(3)}$.
The MSM assumes that we can neglect mixing of $\Psi_{0,0}^{(2)}$ with 
the other two singlet wave functions, either because of other symmetries 
(the case of H$_2$) or because of Brillouin's theorem (see the section on
WFT diagram in the SI).
Hence we need only be concerned with the open-shell singlet
$\Psi_{0,0}^{(2)}$ and the three triplet states $\Psi_{1,+1}$, $\Psi_{1,0}$,
and $\Psi_{1,-1}$.  In fact, as only mixing between states with the same $M$
\marginpar{\color{blue} CI}
quantum number is allowed, we have only to solve the small configuration
interaction (CI) problem,
\begin{equation}
   \left[
       \begin{array}{cc} E[\vert a,{\bar i} \vert] & A \\
       A & E[\vert i, {\bar a} \vert] 
       \end{array}
   \right] 
   \left( 
       \begin{array}{c} C_i^a \\ C_{\bar i}^{\bar a} 
       \end{array}
   \right)
   = E
   \left( 
       \begin{array}{c} C_i^a \\ C_{\bar i}^{\bar a} 
       \end{array}
   \right) \, .
   \label{eq:theory.4}
\end{equation}
As $E_M = E[\vert a,{\bar i} \vert]=E[\vert i, {\bar a} \vert]$, this has two 
simple solutions, namely
\begin{eqnarray}
   E_S & = & E_M + A \nonumber \\ 
   E_T & = & E_M - A \, .
   \label{eq:theory.5}
\end{eqnarray}
Here $E_M$ stands for the energy of a single-determinantal mixed symmetry
state (i.e., one which is not a spin eigenfunction), $E_S$ stands for the
open-shell singlet energy, $E_T$ stands for the open-shell triplet energy,
and $A$ is assumed to be real and positive.  We will examine the nature of $A$
in greater detail in the subsections below.

For now, let us focus on trying to get the best of two worlds, namely 
taking dynamical correlation from DFT and static correlation from WFT.
A na\"{\i}ve way to do this is to simply replace $E_M$ with the 
single-determinantal energy calculated in DFT.  As will be illustrated
in Sec.~\ref{sec:results}, this does not work because the three triplet
states no longer have the same energy: $E[\vert i, a \vert]=E[\vert {\bar i},
{\bar a} \vert] \neq E[\vert i, {\bar a} \vert] - A$.  The MSM solution
to this problem is to rechoose the matrix element $A$ so as to enforce
degeneracy of the three triplet states.  Hence,
\begin{equation}
  A =  E[\vert i, {\bar a} \vert] - E[\vert i, a \vert] \, ,
  \label{eq:theory.6}
\end{equation}
in the MSM.  More generally, the MSM consists of finding as many symmetry
conditions as possible linking off-diagonal energies with diagonal 
single-determinantal energies and then deduce formulae for the off-diagonal
matrix elements.  For example, both spin and spatial symmetry was taken
into account in applying the MSM to the low-lying excited states of O$_2$
\cite{PEMC21} as will be discussed in more detail in Sec.~\ref{sec:results}.

In some cases, there may not be enough symmetry to do this,
so a few matrix elements are calculated explicitly {\em as they would be
in WFT}.  We will see in Sec.~\ref{sec:results} that this latter approach
is dangerous and should be avoided.  In fact, the objective of the diagrammatic
approach introduced in this article is to provide a tool for making better
guesses as to how to calculate off-diagonal matrix elements in the absence
of sufficient symmetry conditions.  Another point is that the MSM implicitly
assumes using the {\em same} MOs in constructing all of the states.  This
means first deciding on a reference configuration (preferably some type of
state average without symmetry breaking) which provides a unique set of MOs
to be used for calculating all of the MOs.  These MOs are then used throughout
the rest of the calculations at the price of possibly missing important
relaxation effects.

\subsection{SIT Diagrams}
\label{sec:SITdiags}

{\em We use a simple type of WFT Feynman diagrams in this article.
The reader not already familiar with this type of diagram is strongly 
urged to take a moment to review the explanation given in the SI before
continuing to read the rest of this article.}

Our interest is not WFT, but DFT.  In particular, for simplicity, we
are interested in Kohn-Sham DFT where the xc functional depends only
upon the charge density and we will not talk about hybrid functionals
(i.e., generalized Kohn-Sham DFT) although it seems trivial to modify 
our diagrammatic formalism to take hybrid functionals into account.  
Kohn-Sham DFT resembles Hartree theory more than Hartree-Fock because
of the use of the classical Hartree energy,
\begin{equation}
  E_H[\rho] = \frac{1}{2} \int \int \frac{\rho(1) \rho(2)}{r_{1,2}} \, d1 d2
  \, .
  \label{eq:theory.14}
\end{equation}
This necessarily implies the presence of SITs, both self-interaction
errors but also self-interaction corrections.  In fact, SITs of various
sorts are common in diagrammatic WFT where they go by the name of
\marginpar{\color{blue} EPV}
exclusion principle violating (EPV) terms and are recognizable by
the presence of the same hole indices or the same particle indices
at the same time (horizontal level in our diagrams).  EPV terms are
tolerated because they cancel each other, but this will no longer
be the case in diagrammatic DFT.  Making such terms expicit is also
helpful because it forces us to deduce diagrams algebraically without
second quantization and Wick's theorem.  The result will be new types
of diagrams not usually seen in diagrammatic WFT.

\begin{figure}
\begin{center}
\includegraphics[width=0.7\textwidth]{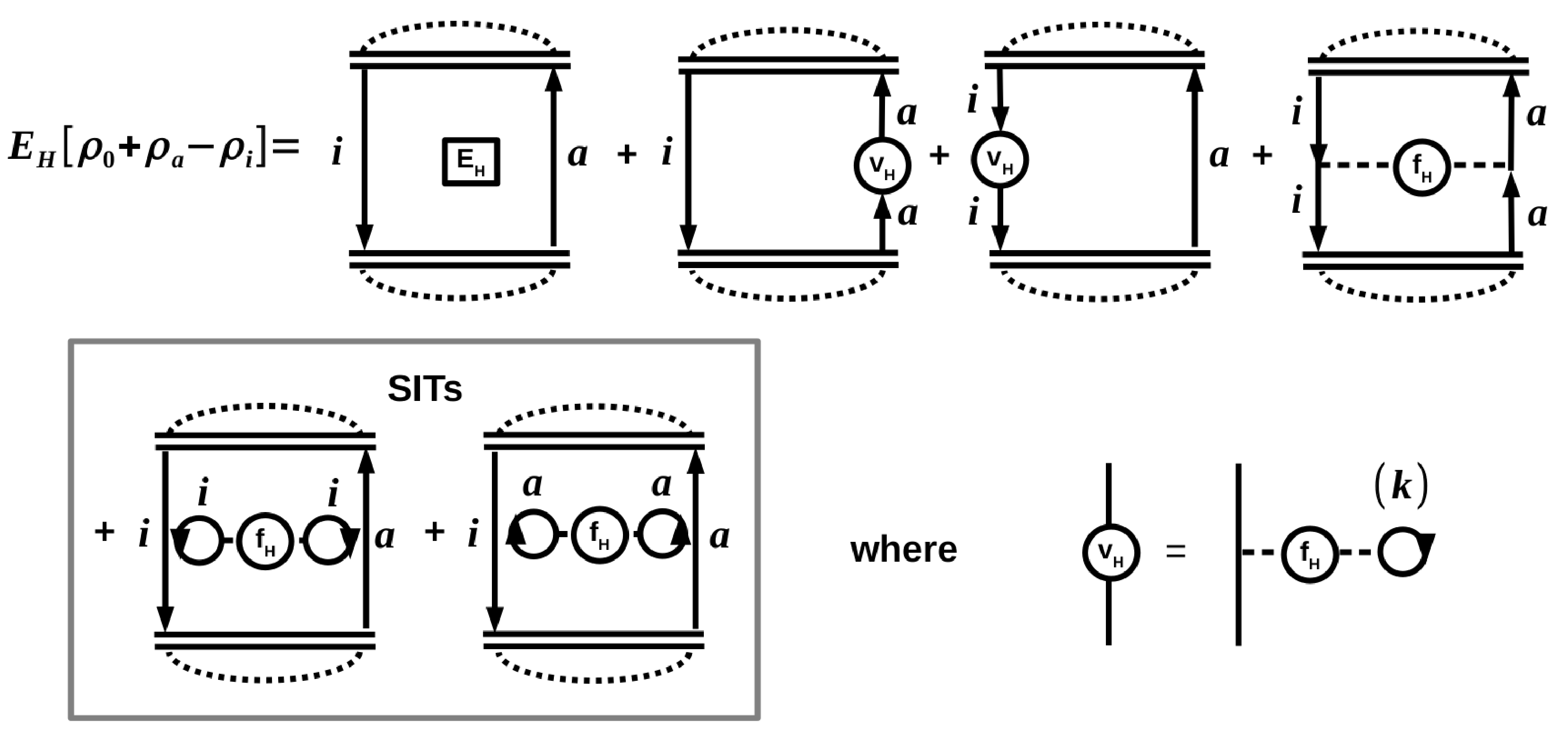}
\end{center}
\caption{
Diagrammatic expansion the Hartree part of $E[\vert i,a \vert]$.  
Here $\rho_0$ is the total charge density, $\rho_a = \vert \psi_a \vert^2$,
and $\rho_i = \vert \psi_i \vert^2$.  The SITs are a new type of diagram.  
Another point worth emphasizing is that the matrix element parts 
$E_H = E_H[\rho_0]$ and $v_H=v_H[\rho_0]$ depend upon the reference
density, although other choices are possible.  The kernel 
$f_H(1,2) = 1/r_{1,2}$ is independent of the charge density.
\label{fig:SIT1}
}
\end{figure}
Consider the specific case of evaluating $E[\vert i, a \vert]$.  This
will include the Hartree term,
\begin{eqnarray}
  E_H[\rho_0 + \rho_a - \rho_i] & = & \frac{1}{2} \int \int \left(\rho_0(1) + \rho_a(1)
  - \rho_i(1) \right) \frac{1}{r_{1,2}} \left(\rho_0(2) + \rho_a(2)
  - \rho_i(2) \right) \nonumber \\
  & = & \frac{1}{2} \int \int \rho_0(1) \frac{1}{r_{1,2}} \rho_0(2) \, d1 d2
  \nonumber \\
  & + & \int \int \rho_0(1) \frac{1}{r_{1,2}} \rho_a(2) \, d1 d2
  - \int \int \rho_0(1) \frac{1}{r_{1,2}} \rho_i(2) \, d1 d2
  \nonumber \\
  & - &  \int \int \rho_i(1) \frac{1}{r_{1,2}} \rho_a(2) \, d1 d2
  \nonumber \\
  & + & \frac{1}{2} \int \int \rho_i(1) \frac{1}{r_{1,2}} \rho_i(2) \, d1 d2
  + \frac{1}{2} \int \int \rho_a(1) \frac{1}{r_{1,2}} \rho_a(2) \, d1 d2 
  \nonumber \\
  & = & E_H[\rho_0] + \int v_H[\rho_0](1) \rho_a(1) \, d1 - \int v_H[\rho_0]
  \rho_i(1) \, d1 \nonumber \\
  & - & (ii \vert f_H \vert aa) + \frac{1}{2} (ii \vert f_H \vert ii) + \frac{1}{2} (aa \vert f_H \vert aa)
  \, .
  \label{eq:theory.15}
\end{eqnarray}
The corresponding diagrams are shown in {\bf Fig.~\ref{fig:SIT1}}. 
As none of these diagrams arise from Wick's theorem, it might be assumed
that some ``guessing'' occurs at this stage in writing down the diagrams.
However there is actually little or no leeway in determining how to draw
them.  The only real questions might arise in the context of a 
{\em new type of diagram}, namely the SIT diagrams.  We have not found
any other way to draw them except as the Hartree interactions that they 
represent.  There is a price to pay for this, namely that we need to 
add labels to indicate the specific orbitals involved and we violate 
{\em Rule 3} of the WF diagrams (see the SI) by using an unoccupied 
orbital index in loops that begin and end at the same time. 

\begin{figure}
\begin{center}
\includegraphics[width=0.7\textwidth]{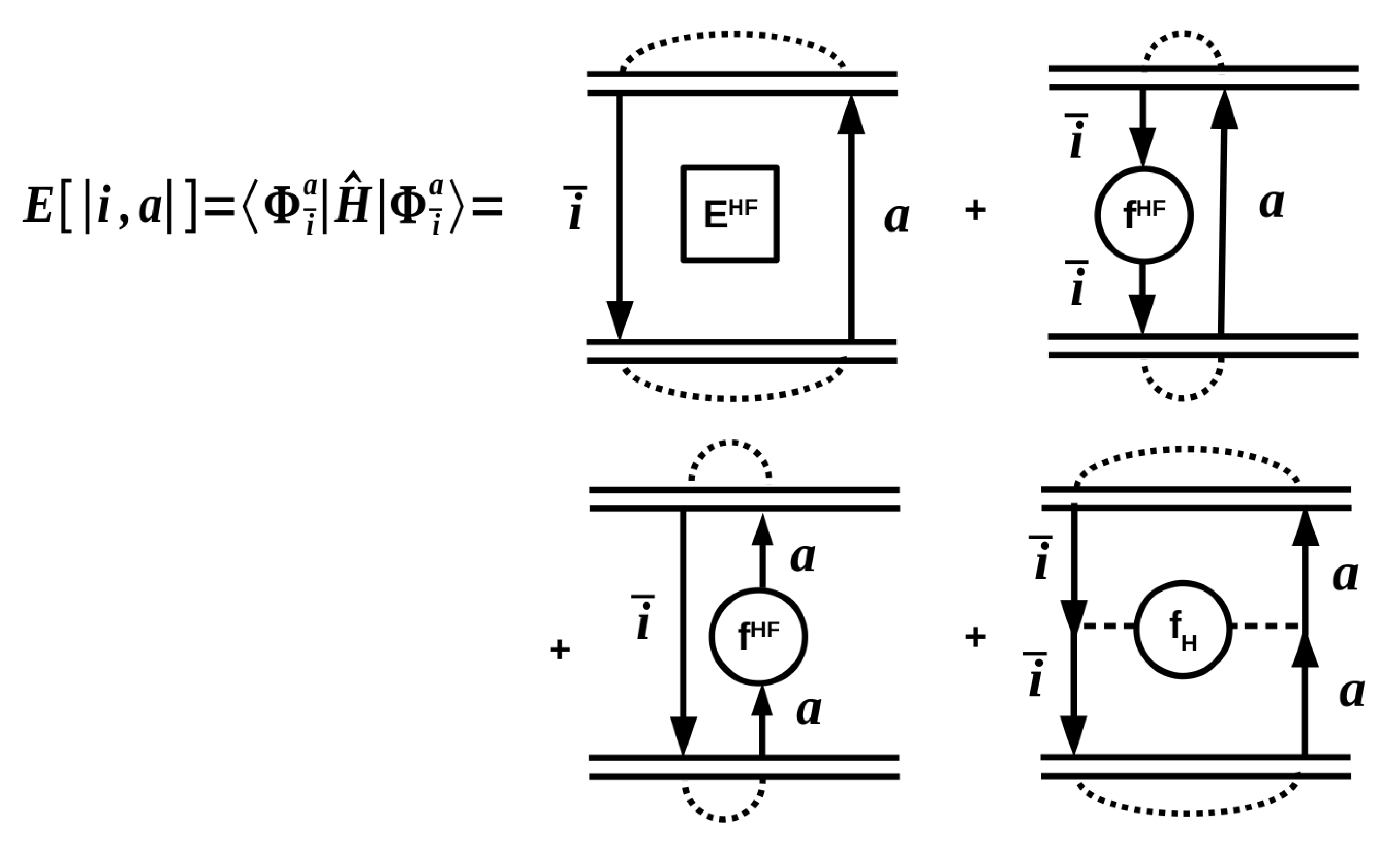}
\end{center}
\caption{
Diagrammatic evaluation of $E[\vert i,a \vert]$.
\label{fig:WFTdiags5}
}
\end{figure}
\begin{figure}
\begin{center}
\includegraphics[width=0.4\textwidth]{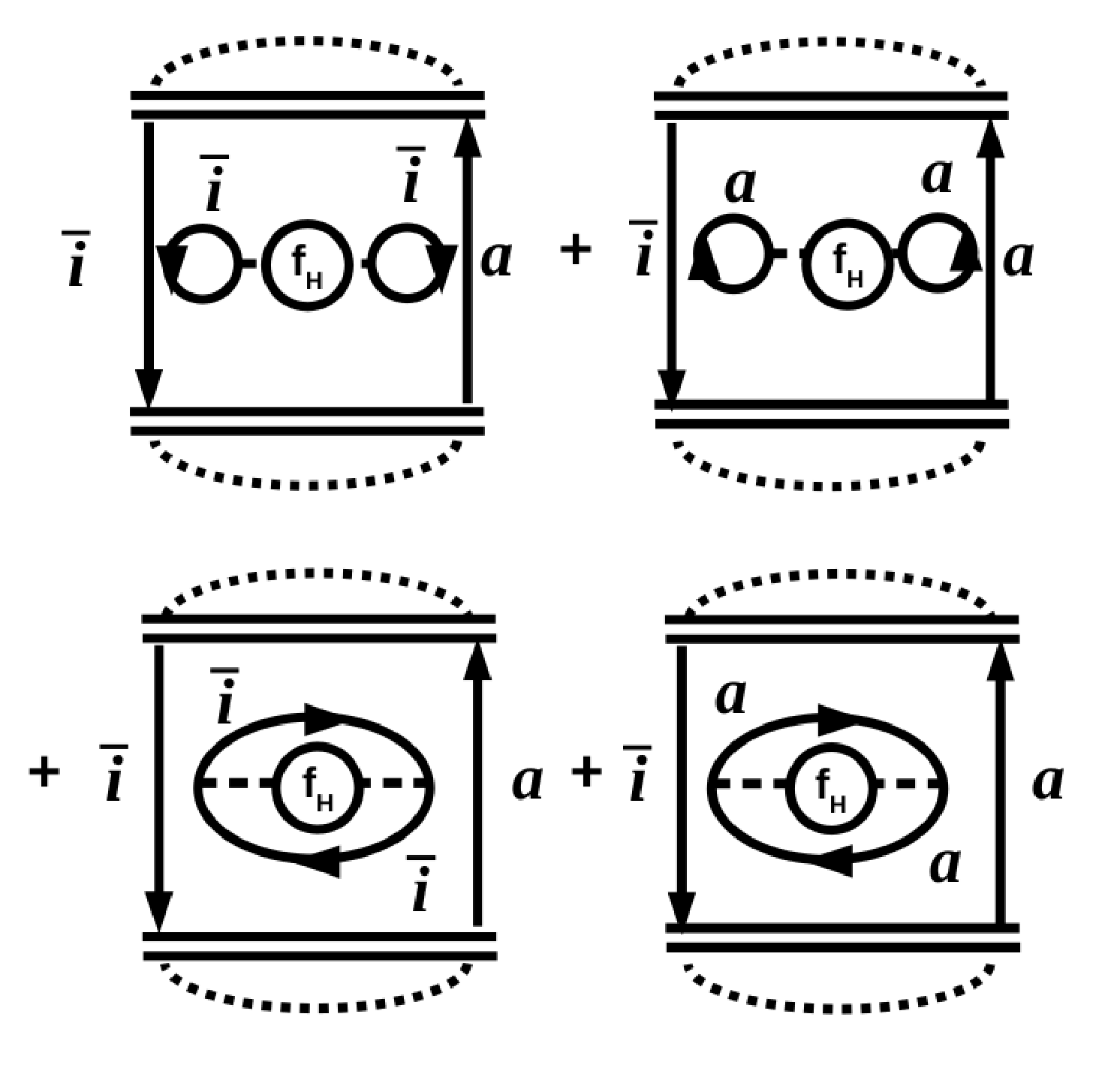}
\end{center}
\caption{
Explicit Hartree (direct) and exchange SIT diagrams implicit in 
{\bf Fig.~\ref{fig:WFTdiags5}} for $E[\vert i,a \vert]$.
\label{fig:SIT2}
}
\end{figure}
\begin{figure}
\begin{center}
\includegraphics[width=0.7\textwidth]{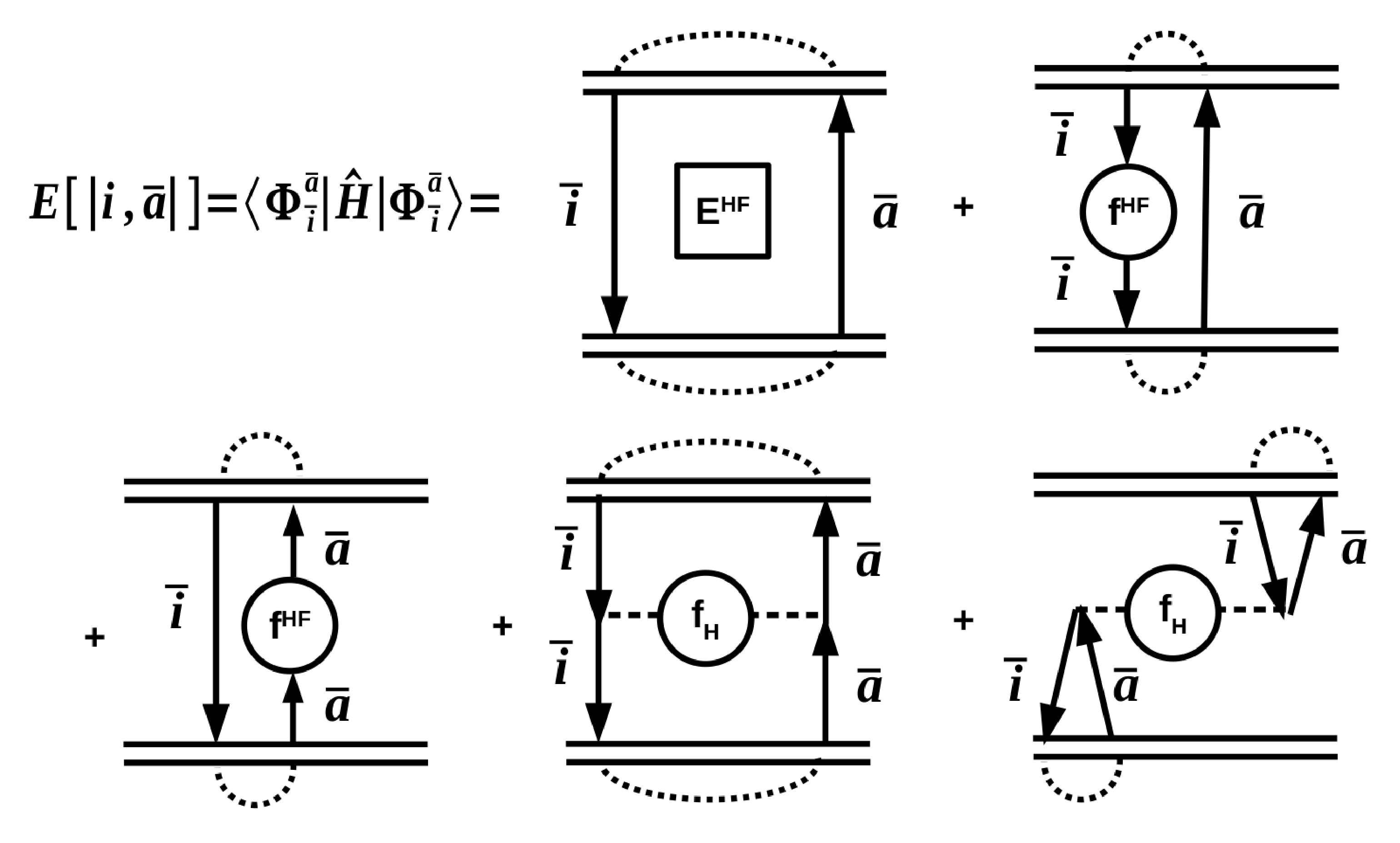}
\end{center}
\caption{
Diagrammatic evaluation of $E[\vert i,{\bar a} \vert]$.
\label{fig:WFTdiags6}
}
\end{figure}
\begin{figure}
\begin{center}
\includegraphics[width=0.4\textwidth]{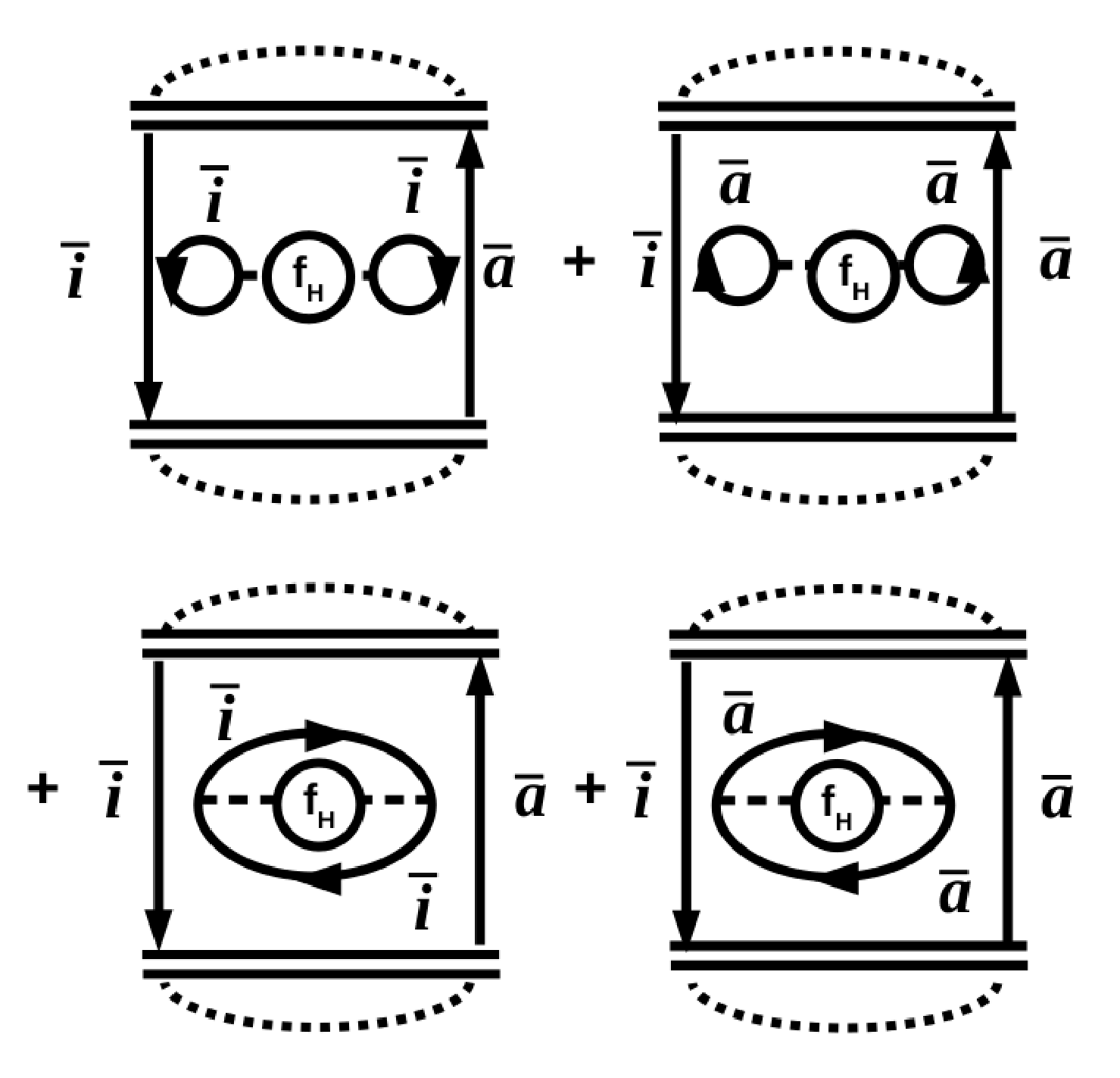}
\end{center}
\caption{
Explicit Hartree (direct) and exchange SIT diagrams implicit in 
{\bf Fig.~\ref{fig:WFTdiags6}} for $E[\vert i,{\bar a} \vert]$.
\label{fig:SIT3}
}
\end{figure}
We can also follow the same procedure for the HF exchange energies, but
using the one-electron reduced density matrix instead of the charge density.
This gives the SITs shown in {\bf Fig.~\ref{fig:SIT2}} and 
{\bf Fig.~\ref{fig:SIT3}}.  Note how these cancel out when the diagrams 
are evaluated according to the five rules given above.  This will no
longer be the case in the MSM-DFT diagrams.

\subsection{MSM-DFT Diagrams}
\label{sec:MSMdiags}

MSM-DFT diagrams are just like MSM-HF except that we have to keep the
Hartree SITs and the HF exchange terms are replaced with DFT xc terms.
The xc terms include the xc energy $E_{xc}[\rho^\uparrow,\rho^\downarrow]$,
the xc potential $v_{xc}^\tau[\rho^\uparrow,\rho^\downarrow](1)
= \delta E_{xc}[\rho^\uparrow,\rho^\downarrow]/\delta \rho^\tau(1)$,
the xc kernel $f_{xc}^{\sigma,\tau}[\rho^\uparrow,\rho^\downarrow] = \delta^2 
E_{xc}[\rho^\uparrow,\rho^\downarrow]/\delta \rho^\sigma(1) 
\delta \rho^\tau(2)$, and higher-order terms (HOT) such as
\marginpar{\color{blue} HOT}
$g_{xc}^{\xi, \sigma, \tau}[\rho^\uparrow,\rho^\downarrow] =
\delta^3 E_{xc}[\rho^\uparrow,\rho^\downarrow]/\delta \rho^\xi(1) 
\delta \rho^\sigma(2) \rho^\tau (3)$.  We need to do an expansion around
the zero-order densities $\rho_0^\uparrow$ and $\rho_0^\downarrow$.
This is possible once it is understood that a functional is just like
a function of a vector except that instead of the vector having
discrete indices for its components, the vector becomes a function
(such as $\rho(x)$) with a continuous index ($x$).  This allows us to
develop a calculus of variations \cite{GF63} in close parallel with 
the multivariable calculus.  In particular, we can carry out the 
calculus of variations analogue of a Taylor's expansion.  

\begin{figure}
\begin{center}
\includegraphics[width=0.6\textwidth]{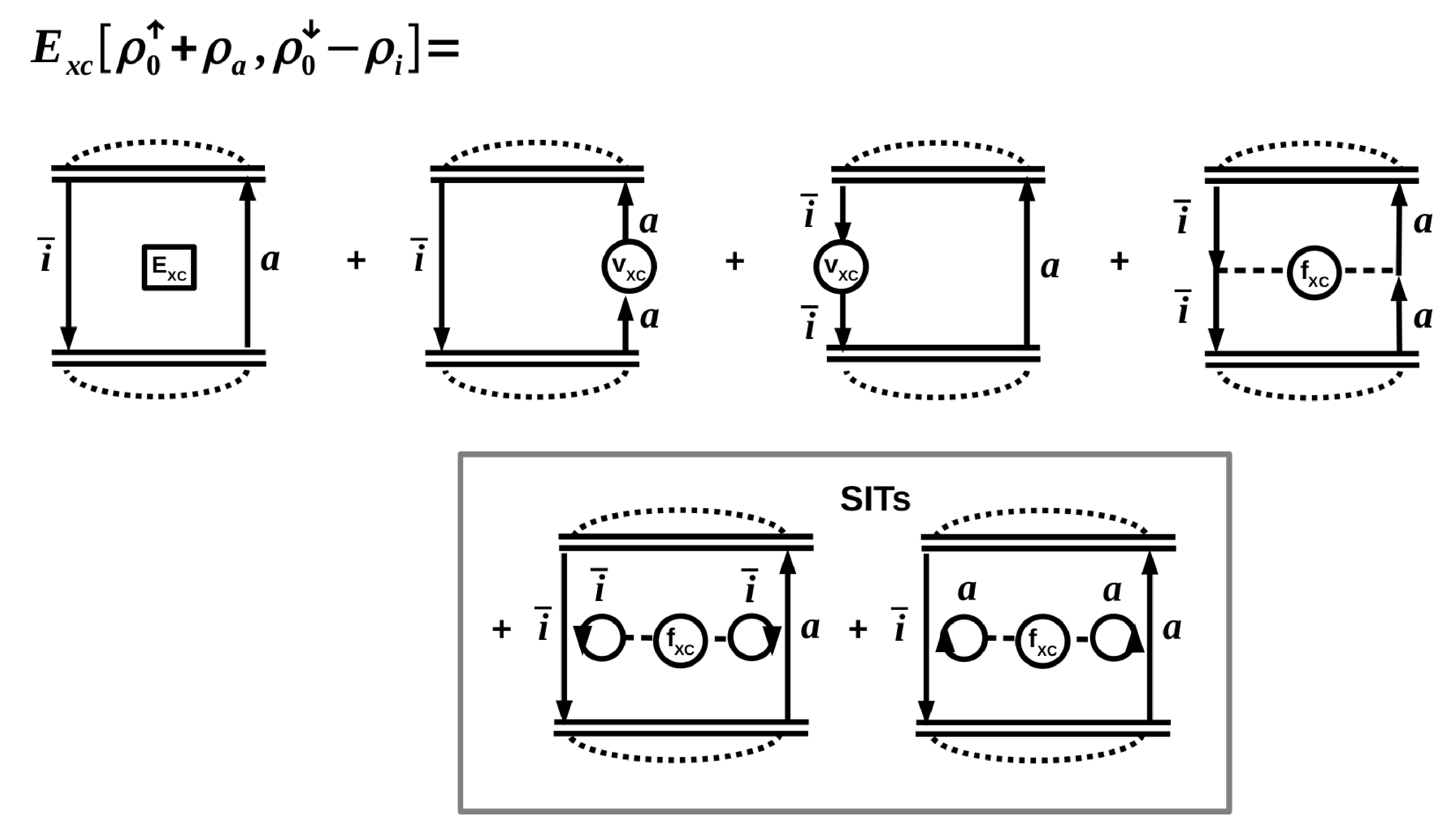}
\end{center}
\caption{
Explicit xc diagrams associated with $E[\vert i,a \vert]$.
The matrix element symbols in the diagram have the same arguments,
namely $[\rho_0^\uparrow, \rho_0^\downarrow]$.
Notice that the spin indices on the xc potentials and kernels
need not be specified as they are determined by the spins of 
incoming and outgoing lines.  
\label{fig:xc1}
}
\end{figure}
We will just give one example here and explain the relation with diagrams.
Thus we consider the excitation of an electron from an occupied
spin-down orbital $i$ to an unoccupied spin-up orbital $a$.  The 
spin-down density is reduced by an amount $-\rho_i = \vert \psi_i \vert^2$
while the spin-up density is increased by an amount 
$\rho_a = \vert \psi_a \vert^2$.  We may then expand,
\begin{eqnarray}
  E_{xc}[\rho_0^\uparrow+\rho_a, \rho_0^\downarrow-\rho_i] 
  & = & E_{xc}[[\rho_0^\uparrow, \rho_0^\downarrow]
  \nonumber \\
  & + & \int \frac{\delta E_{xc}[\rho_0^\uparrow, \rho_0^\downarrow]}{\delta 
  \rho_0^\uparrow(1)} \rho_a(1) \, d1 
  - \int \frac{\delta E_{xc}[\rho_0^\uparrow, \rho_0^\downarrow]}{\delta 
  \rho_0^\downarrow(1)} \rho_i(1) \, d1 \nonumber \\
  & - & \int \int \rho_i(1) \frac{\delta^2 E_{xc}[\rho_0^\uparrow,
  \rho_0^\downarrow]}{\delta \rho^\downarrow(1) \delta \rho^\uparrow(2)}
  \rho_a(2) \, d1 d2 \nonumber \\
  & + & \frac{1}{2} \int \int \rho_i(1) \frac{\delta^2 E_{xc}[\rho_0^\uparrow,
  \rho_0^\downarrow]}{\delta \rho^\downarrow(1) \rho^\downarrow(2)}
  \rho_i(2) \, d1 d2 
  + \frac{1}{2} \int \int \rho_a(1) \frac{\delta^2 E_{xc}[\rho_0^\uparrow,
  \rho_0^\downarrow]}{\delta \rho^\uparrow(1) \rho^\uparrow(2)}
  \rho_a(2) \, d1 d2
  \nonumber \\
  & + & \mbox{ HOT} \nonumber \\
  & = & E_{xc}[\rho_0^\uparrow, \rho_0^\downarrow]
  \nonumber \\
  & + & \int v_{xc}^\uparrow(1)[\rho_0^\uparrow, \rho_0^\downarrow] \rho_a(1) \, d1
  - \int v_{xc}^\downarrow(2)[\rho_0^\uparrow, \rho_0^\downarrow] \rho_i(1) \, d2 
  \nonumber \\
  & - & (ii \vert f_{xc}^{\uparrow,\downarrow}[\rho_0^\uparrow, \rho_0^\downarrow] \vert aa) 
  \nonumber \\
  & + & \frac{1}{2} (ii \vert f_{xc}^{\downarrow, \downarrow}[\rho_0^\uparrow, \rho_0^\downarrow] \vert ii)
  + \frac{1}{2} (aa \vert f_{xc}^{\uparrow, \uparrow}[\rho_0^\uparrow, \rho_0^\downarrow] \vert aa)
  \nonumber \\
  & + & \mbox{ HOT} \, .
  \label{eq:theory.16}
\end{eqnarray}
For the Hartree terms, the series trucates naturally after the second-order
term as all HOT are zero.  This is not true for xc terms, but nevertheless
we will confine our development to no higher than the xc kernel.
Equation~[\ref{eq:theory.16}] are shown in {\bf Fig.~\ref{fig:xc1}}.
Notice that only direct diagrams can appear in MSM-DFT as DFT is more 
Hartree-like than HF-like.  Also $v_{xc}^\uparrow[\rho_0^\uparrow, \rho_0^\downarrow]=v_{xc}^\downarrow[\rho_0^\uparrow, \rho_0^\downarrow]$,
$f_{xc}^{\uparrow,\uparrow}[\rho_0^\uparrow, \rho_0^\downarrow]=f_{xc}^{\downarrow,\downarrow}[\rho_0^\uparrow, \rho_0^\downarrow]$, and
$f_{xc}^{\uparrow,\downarrow}[\rho_0^\uparrow, \rho_0^\downarrow]=f_{xc}^{\downarrow,\uparrow}[\rho_0^\uparrow, \rho_0^\downarrow]$, when
$\rho_0^\uparrow=\rho_0^\downarrow$ as is normal for closed-shell molecules
without any symmetry breaking.

\begin{figure}
\begin{center}
\includegraphics[width=0.8\textwidth]{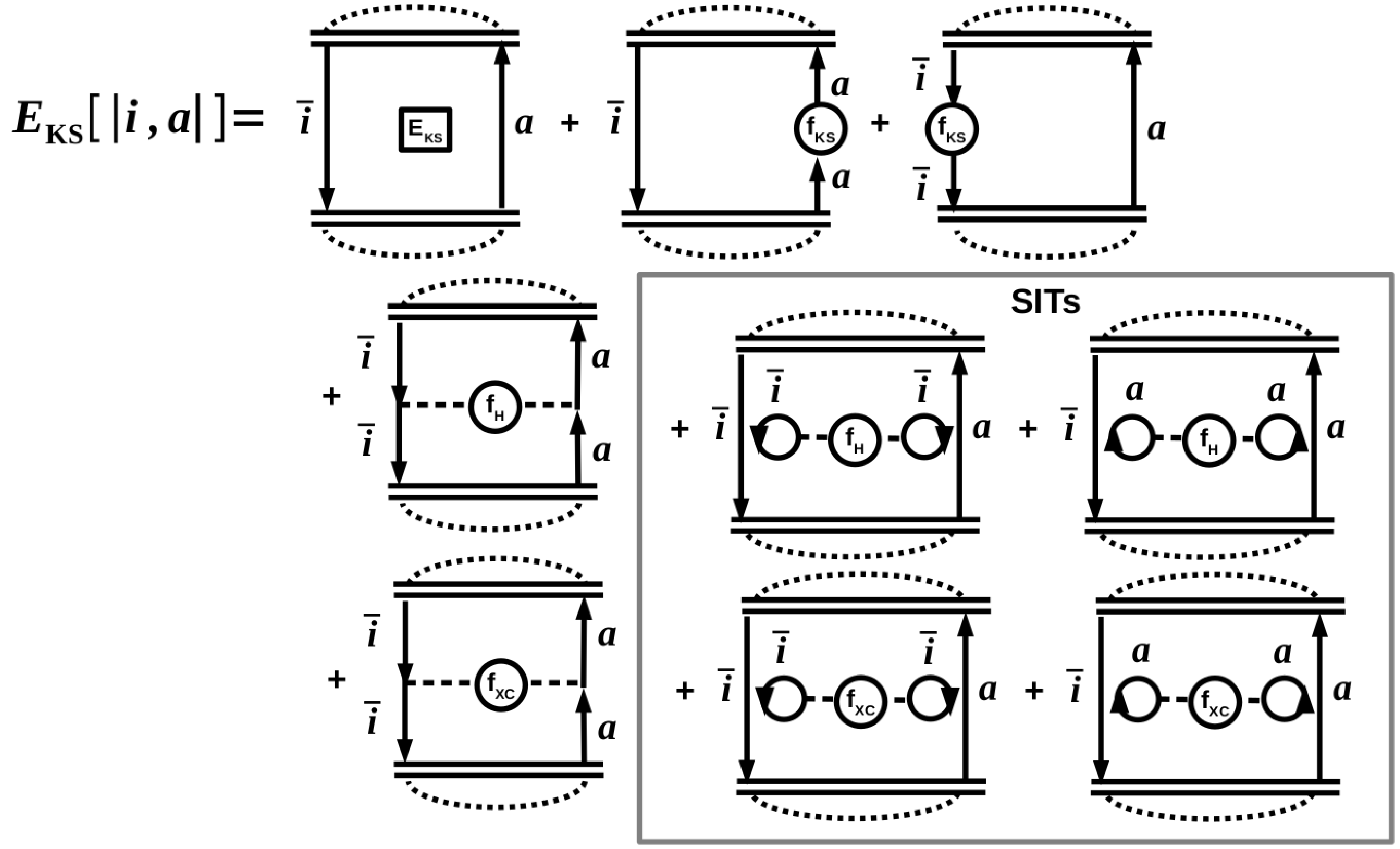}
\end{center}
\caption{
Explicit DFT diagrams associated with $E[\vert i,a \vert]$.
The xc matrix element symbols in the diagram have the same arguments,
namely $[\rho_0^\uparrow, \rho_0^\downarrow]$.
\label{fig:xc2}
}
\end{figure}
\begin{figure}
\begin{center}
\includegraphics[width=0.8\textwidth]{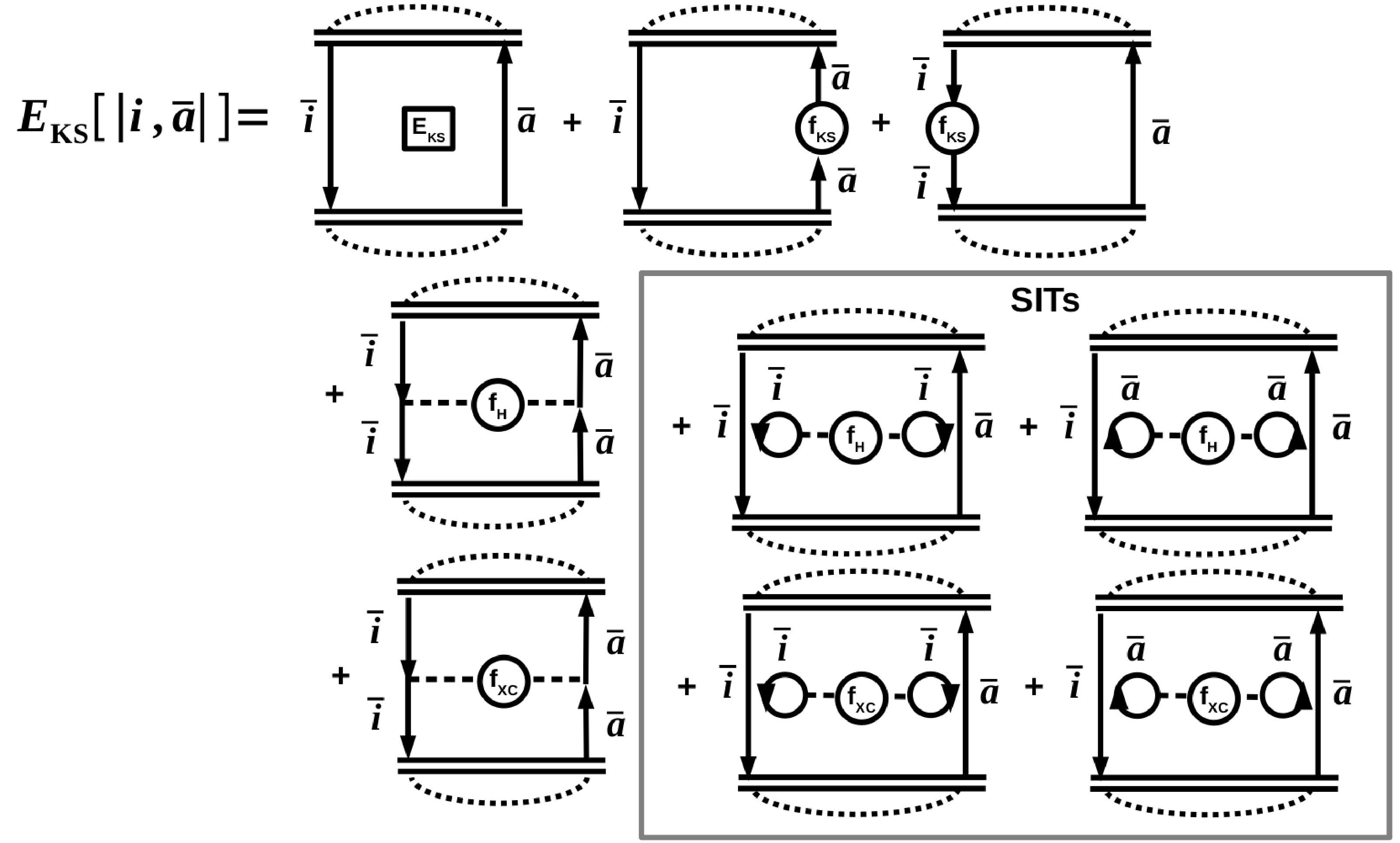}
\end{center}
\caption{
Explicit DFT diagrams associated with $E[\vert i, {\bar a} \vert]$.
The xc matrix element symbols in the diagram have the same arguments,
namely $[\rho_0^\uparrow, \rho_0^\downarrow]$.
\label{fig:xc3}
}
\end{figure}
\begin{figure}
\begin{center}
\includegraphics[width=0.8\textwidth]{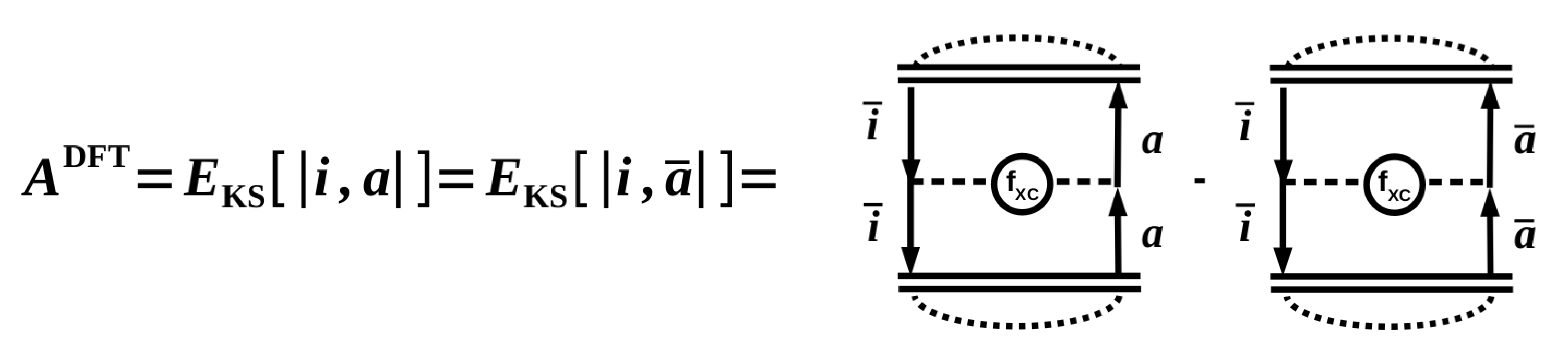}
\end{center}
\caption{
Diagrammatic representation of the $A$ matrix element in MSM-DFT.
The xc matrix element symbols in the diagram have the same arguments,
namely $[\rho_0^\uparrow, \rho_0^\downarrow]$.
\label{fig:xc4}
}
\end{figure}
We now wish to derive the diagrams for the $A$ matrix element in MSM-DFT.
To do this, we need the diagrammatic expansions for $E[\vert i,a \vert]$
and $E[\vert i,{\bar a} \vert]$ which are shown respectively in
{\bf Figs.~\ref{fig:xc2}} and {\bf \ref{fig:xc3}}.  Then, from 
Eq.~\ref{eq:theory.6} (represented diagrammatically in 
{\bf Fig.~\ref{fig:xc4}}),
\begin{equation}
  A^{\mbox{DFT}} = (ii\vert 
  f_{xc}^{\downarrow,\downarrow}[\rho_0^\downarrow,\rho_0^\uparrow]
  \vert aa)
  -(ii \vert f_{xc}^{\downarrow,\uparrow}[\rho_0^\downarrow,\rho_0^\uparrow]
  \vert aa ) \, .
  \label{eq:theory.17}
\end{equation}

\subsection{Exchange-Only Ansatz}
\label{sec:EXAN}

\begin{figure}
\begin{center}
\includegraphics[width=0.8\textwidth]{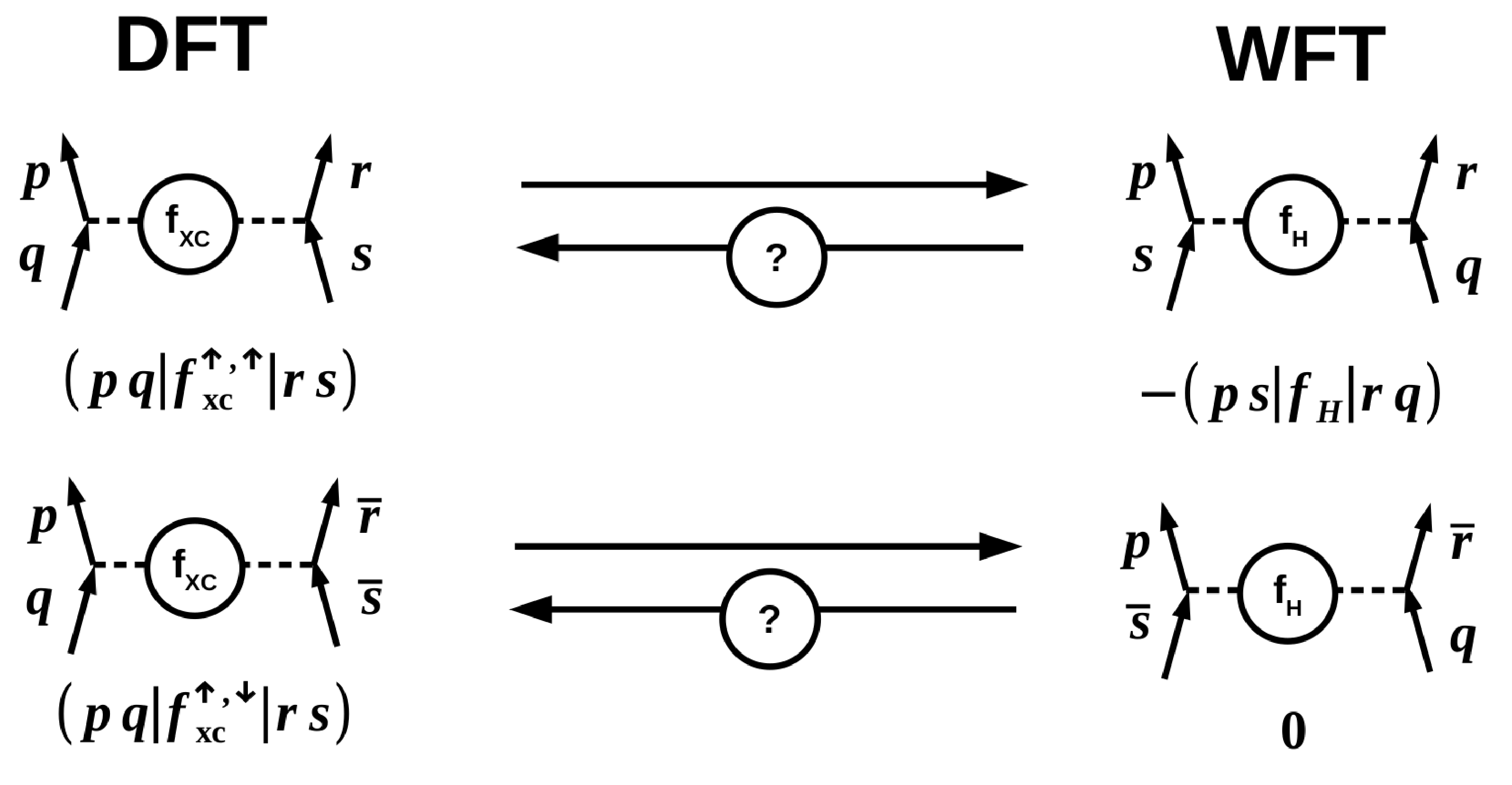}
\end{center}
\caption{
The exchange-only ansatz (EXAN).
\label{fig:EXAN}
}
\end{figure}
Up to this point, the only thing that is really new is the introduction of
diagrams as a tool.  It is also possible to check the MSM-DFT diagrams
using the EXAN. 
This neglects antiparallel spin 
correlation and assumes that parallel spin correlation may be replaced
by an exchange-type integral.  The exact rules are given in {\bf
Fig.~\ref{fig:EXAN}}.  It allows us to take both the MSM-DFT and 
TD-DFT in the TDA and
transform them into the corresponding HF-DFT or TD-HF equations.
Note that this is not done by replacing DFT terms with terms of the same
magnitude, but rather just shows parallels between the structure of the
two theories.  {\em If} we could go backwards and use the EXAN to make 
educated guesses for terms in MSM-DFT (or TD-DFT) from WFT, notably
in the absence of symmetry arguments, then we would be able to show the 
usefulness of the diagrammatic analysis.  The general formula for the
WFT $\rightarrow$ DFT mapping might be,
\begin{eqnarray}
  m (pr \vert f_H \vert qs ) & \rightarrow & 
  -m(pq \vert f_{xc}^{\uparrow,\uparrow} \vert rs) 
  + n \left[ (pr \vert f_H \vert qs ) + (pq \vert f_{xc}^{\uparrow,\uparrow} \vert rs) \right] \nonumber \\
  & + & \mbox{ SITs} + \mbox{ terms involving $f_{xc}^{\uparrow,\downarrow}$ }
  \, ,
  \label{eq:EXAN}
\end{eqnarray}
assuming that $\rho_0^\uparrow = \rho_0^\downarrow$ and presuming that
$n$ and $m$ are integers.  The simplest choice would then be to drop the SITs
and the $f_{xc}^{\uparrow,\downarrow}$ terms and to set $n=0$, but the simplest
choice is not always the best choice.  {\em We will show in 
Sec.~\ref{sec:results} that the diagrammatic analysis sometimes provides
a guide as to how best to make our educated guesses.}

%
%

\section{Computational Details}
\label{sec:details}

Calculations were done with {\sc deMon2k} \cite{deMon2k}.  Historically 
{\sc deMon} ({\em densit\'e de Montr\'eal}, so called because it was 
developed at the {\em Universit\'e de Montr\'eal}) was one of the first 
\marginpar{\color{blue} GTO}
DFT programs to use gaussian-type orbitals (GTOs) and so to be able to 
profit from many well-established quantum chemistry algorithms.  In 
addition, an auxiliary basis set is used to reduce the formal scaling 
of the program from ${\cal O}(N^4)$ (because of 4-center integrals) 
to ${\cal O}(N^3)$.  Prescreening and other techniques further increase 
the efficiency of the program.  The {\tt FOCK} keyword in {\sc deMon2k} 
provides an auxiliary-function calculation of HF exchange so
that we also have access to a good approximation to Hartree-Fock and to 
hybrid functionals.  The calculations reported here were begun at the 7th
African School on Electronic Structure Methods and Applications (ASESMA)
\cite{ASESMA} with the intent of showing that research quality work could
be done with limited computational resources provided one has a very good
understanding of the underlying formal theory.  We used a freely 
downloadable version of {\sc deMon2k} that runs on all the {\sc Linux} 
systems that we have tried.
This may be used for teaching purposes as a first workbook has been written
with this exactly this in mind \cite{C22,OPEC23}.  The calculations reported
here were done on our portable computers.

All calculations used a valence double zeta plus polarization ({\tt DZVP}) 
orbital basis set and the {\tt GEN-A2} auxiliary basis set.  Only two 
functionals were used, namely the local (spin) density approximation
\marginpar{\color{blue} LDA, VWN}
(LDA) using the Vosko-Wilk-Nusair (VWN) parameterization \cite{VWN80}
of Ceperley and Alder's quantum Monte Carlo results \cite{CA80} for
the homogeneous electron gas
[confusingly denoted VWN5 in the popular {\em Gaussian} program where
VWN is (mistakenly) used to designate a parameterization of random-phase
\marginpar{\color{blue} RPA}
approximation (RPA) reaults in the Vosko-Wilk-Nusair article \cite{VWN80}]. 

Our MSM-HF and MSM-VWN calculations consisted of a two-step procedure.  
In the first step, a reference state was imposed
using the keyword {\tt MOMODIFY} which allows the fixing of the occupation
numbers of specific orbitals.  As we wished to avoid problems with symmetry
breaking in this part of the calculation, we used a high quality grid
\marginpar{\color{blue} SCF}
({\tt GRID FIXED REFERENCE}) and tight self-consistent field (SCF) 
convergence ({\tt SCFTYPE UKS TOL=1.E-8}).  In addition, we imposed 
the same occupation numbers 
\marginpar{\color{blue} MO}
for both spin up and spin down molecular orbitals (MOs), also with a view
towards avoiding problems due to symmetry breaking.  
Even so, we had difficulties with symmetry breaking in our O$_2$ calculations
and so decided to use HF reference state MOs for this molecule.
The MSM-DFT concept of a reference state is a little
\marginpar{\color{blue} CASSCF}
bit like the concept of {\em state-averaging} in complete active space (CAS) SCF
calculations in that it will impose the use of the same orbitals for all
our MSM-DFT states.  We set $n_\uparrow=n_\downarrow=1/2$ within the
two-orbital active space and otherwise $n=1$ for occupied orbitals
or $n=0$ for unoccupied orbitals in all our examples.
This first step of the calculations generated 
restart files that were then used in the second step
of the calculations.  Calculations in the second step read the restart files 
generated in the first step and calculate the energy with different orbital 
occupations {\em without} any SCF iterations ({\tt SCFTYPE UKS MAX=0}).
MO occupations are controlled with the {\tt MOMODIFY} keyword.

\section{Exploratory Examples}
\label{sec:results}

The diagrammatic MSM-DFT analysis introduced in Section~\ref{sec:theory}
reduces to the corresponding WFT diagrams under the EXAN, and the EXAN also
places some constraints on how we are allowed to add new terms to MSM-DFT,
other than those imposed by symmetry.  However the EXAN does not provide
a unique recipe for determining MSM-DFT terms in the absence of symmetry.
In this section, we will use some specific physical examples to which we
may apply the TOTEM model developed in Sec.~\ref{sec:theory} to illustrate
these points.  These are the two cases of H$_2$ and O$_2$. The case of 
LiH is treated in the SI.

We will carry out both MSM-HF (really MSM-{\tt FOCK}) and MSM-DFT 
(meaning MSM-VWN) calculations and compare the results against those 
from high-quality reference work
\marginpar{\color{blue} EXACT}
which we will denote as EXACT.  Of course, the EXACT calculations are
not perfectly exact, but they are much closer to exact results than our
our MSM calculations that they make an excellent comparison.  These objective
of these first calculations is a reality check and review of the MSM but
is not really anything new.  But we then go on to use the power of
diagrammatic MSM to make intelligent guesses as to the value of matrix
elements {\em not} determined by symmetry or determined by symmetry but
which the diagrammatic approach tells us might be approximated in an
alternative way.  This is new and is intended to show the power of the
the diagrammatic analysis as an analytic tool.

\subsection{H$_2$}
\label{sec:H2}

\begin{figure}
\begin{center}
\includegraphics[width=0.8\textwidth]{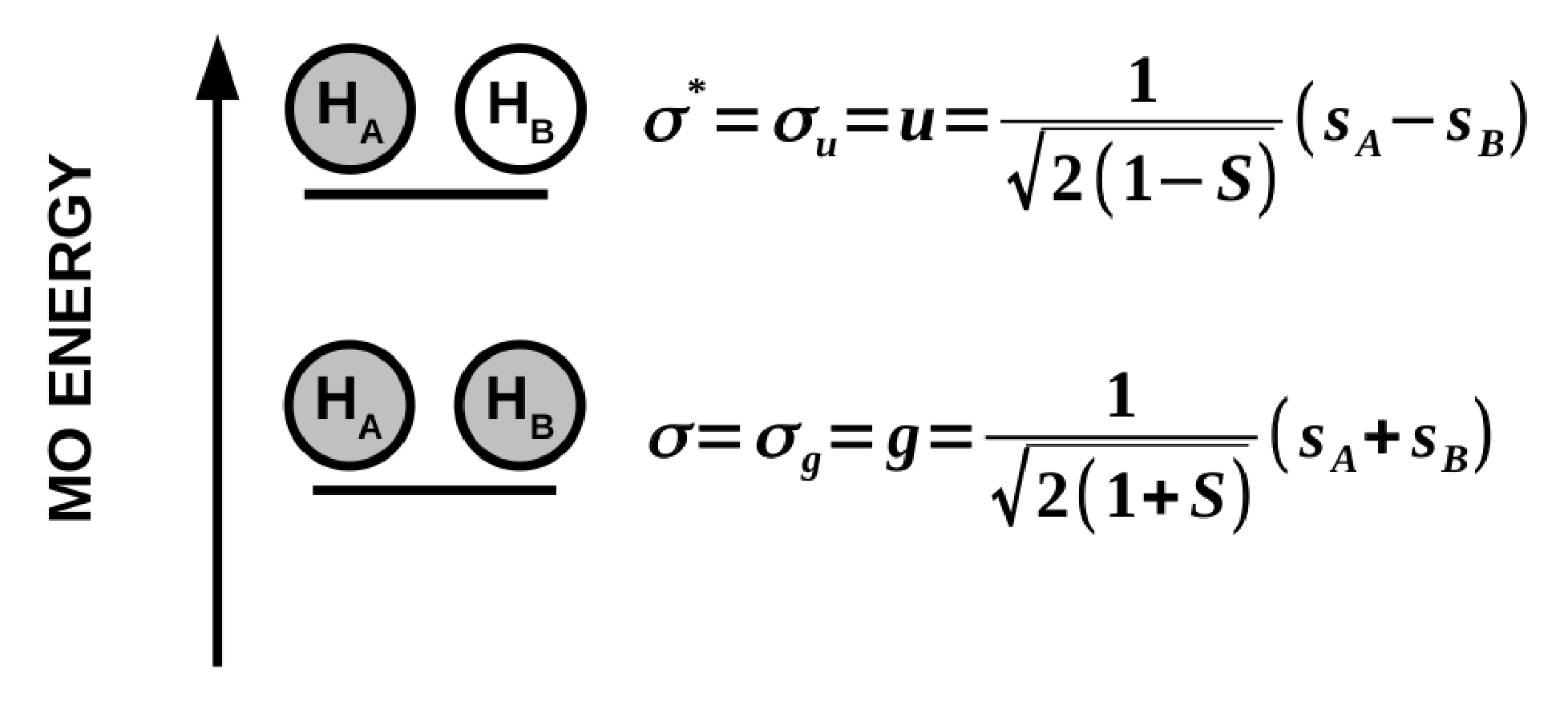}
\end{center}
\caption{
Textbook-like MO diagram for H$_2$ defining the notation used in this
article.  In particular, $g$ stands for the German
word {\em gerade} for even symmetry with respect to exchanging the
A and B labels while $u$ stands for the German word {\em ungerade}
for odd symmetry with respect to exchanging the A and B labels.
Also $S=\langle g \vert u \rangle$ is the overlap integral.  We
assume real orbitals.
\label{fig:H2MOdiagram}
}
\end{figure}
H$_2$ is the ``fruitfly''of quantum chemistry in so far as it is
the simplest molecule (after H$_2^+$) and so is often used to illustrate 
various level of theory.  We will not be different in this respect.
{\bf Figure~\ref{fig:H2MOdiagram}} shows the usual linear combination
\marginpar{\color{blue} LCAO}
of (valence) atomic orbitals (LCAO) description usually used as a first
approximation for describing the MOs of H$_2$ and is primarily shown
as a quick way to introduce our notation.  This is a TOTEM identical
to that introduced in Fig.~\ref{fig:genericstatediagram}, except 
that $i$ has been replaced by $g$ and $a$ has been replaced by $u$ so 
that there is additional symmetry.
\begin{figure}
\begin{center}
\includegraphics[width=0.8\textwidth]{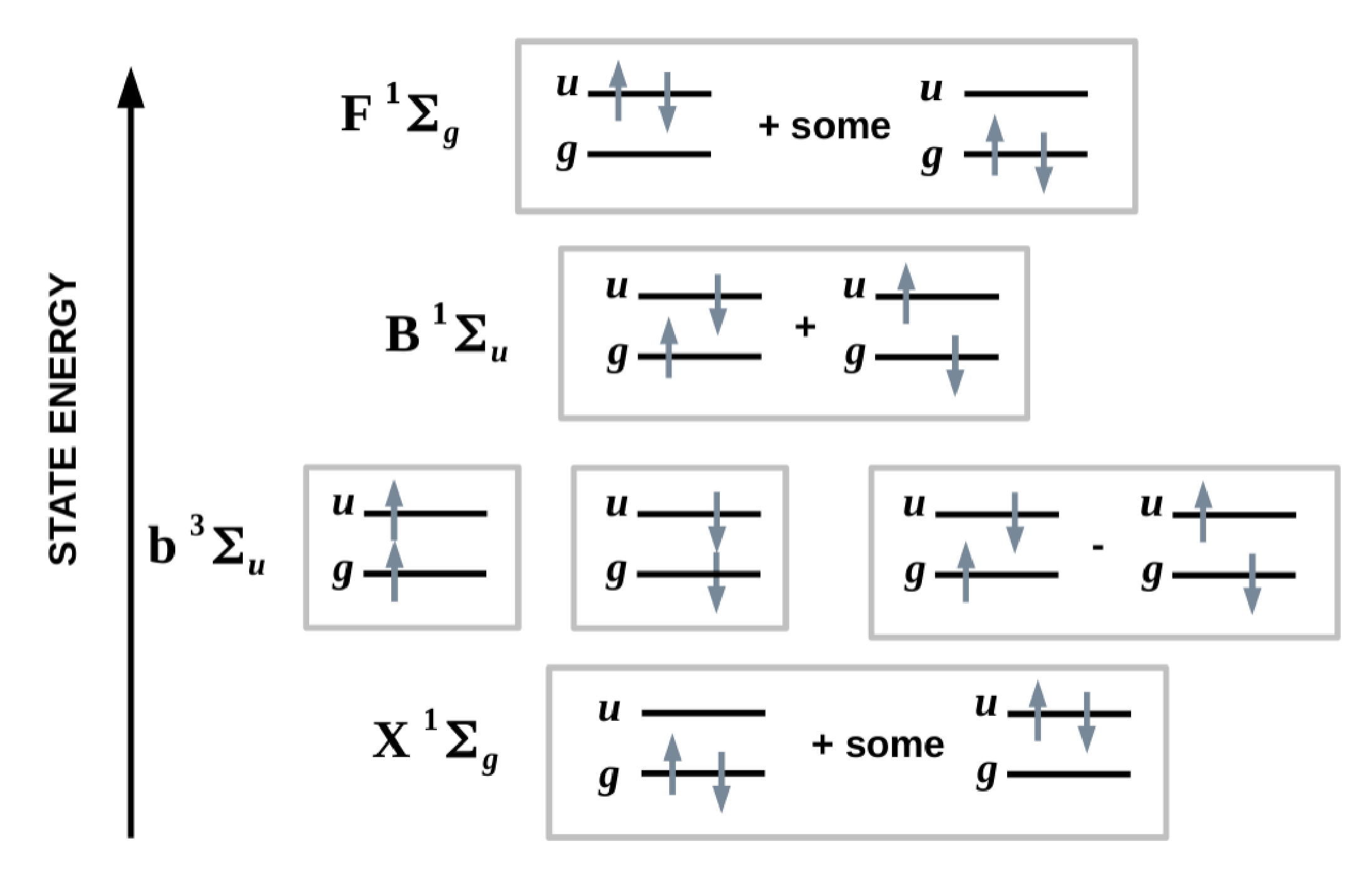}
\end{center}
\caption{
Diagram showing the six states which may be constructed by placing 
two electrons in the two MOs of Fig.~\ref{fig:H2MOdiagram}.
\label{fig:H2statediagram}
}
\end{figure}
This gives rise to four states, namely the ground $X \,^1\Sigma_g^+$
state, the $b \,^3\Sigma_u^+$ triplet state, the $B \,^1\Sigma_g^+$ 
open-shell singlet state, and the $F \,^1\Sigma_g^+$ doubly-excited
singlet state ({\bf Fig.~\ref{fig:H2statediagram}}).  We will habitually
drop the "+" part of the symmetry label as it is not important in the
present context.

\begin{figure}
\begin{center}
\includegraphics[width=0.5\textwidth]{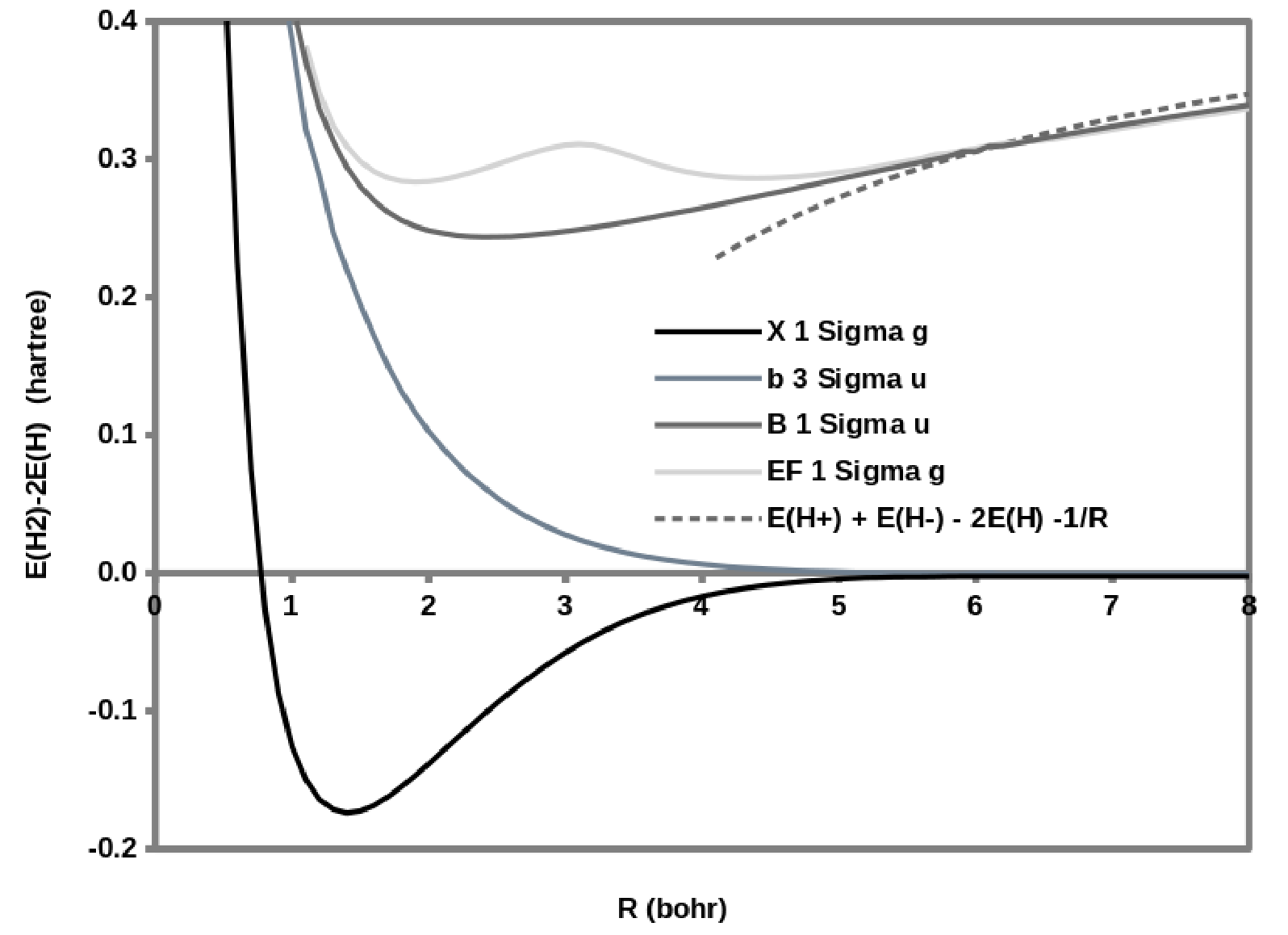}
\end{center}
\caption{
EXACT H$_2$ PECs obtained by Ko{\l}os and Wolniewicz by calculations
using basis functions with an explicit $r_{1,2}$ dependence.  See text.
We used the value of 0.754189 eV for the electron affinity of the hydrogen
atom determined by laser photoelectron spectroscopy \cite{LML91} to construct
the dashed asymptotic estimate for the long distance asymptote in the
[H$^+$ H:$^-$ $\leftrightarrow$ H:$^-$ + H$^+$] limit.
\label{fig:exactH2}
}
\end{figure}
The 1971 review written by Sharp provides a nice picture of what was known
\marginpar{\color{blue} PEC}
about the various potential energy curves (PECs) of H$_2$ and its ions at that time \cite{S71}.
Our TOTEM can only hope to describe some of these states, namely the EXACT PECs
shown in {\bf Fig.~\ref{fig:exactH2}}.  These are PECs from
the well-known excellent calculations of Ko{\l}os and Wolniewicz obtained
using explicit $r_{1,2}$ basis functions 
\cite{KW65,KW66,KW68,KW69,KR76,K76,KR77,WD88,SW02,WS03}.  
In practice we use numbers obtained by digitizing Fig.~1 of
Ref.~\cite{LGC13} which provides curves which are already much more accurate
than what we expect from our MS-DFT calculations. Note that the $E,F$
$^1\Sigma_g$ curve has a double minimum arising from an avoided crossing
between the $E$ $^1\Sigma_g(1\sigma_g,2\sigma_g)$ diabatic curve (left-hand
minimum) and the $F$ $^1\Sigma_g(1\sigma_u^2)$ (right-hand minimum).

\begin{figure}
\begin{center}
\includegraphics[width=0.9\textwidth]{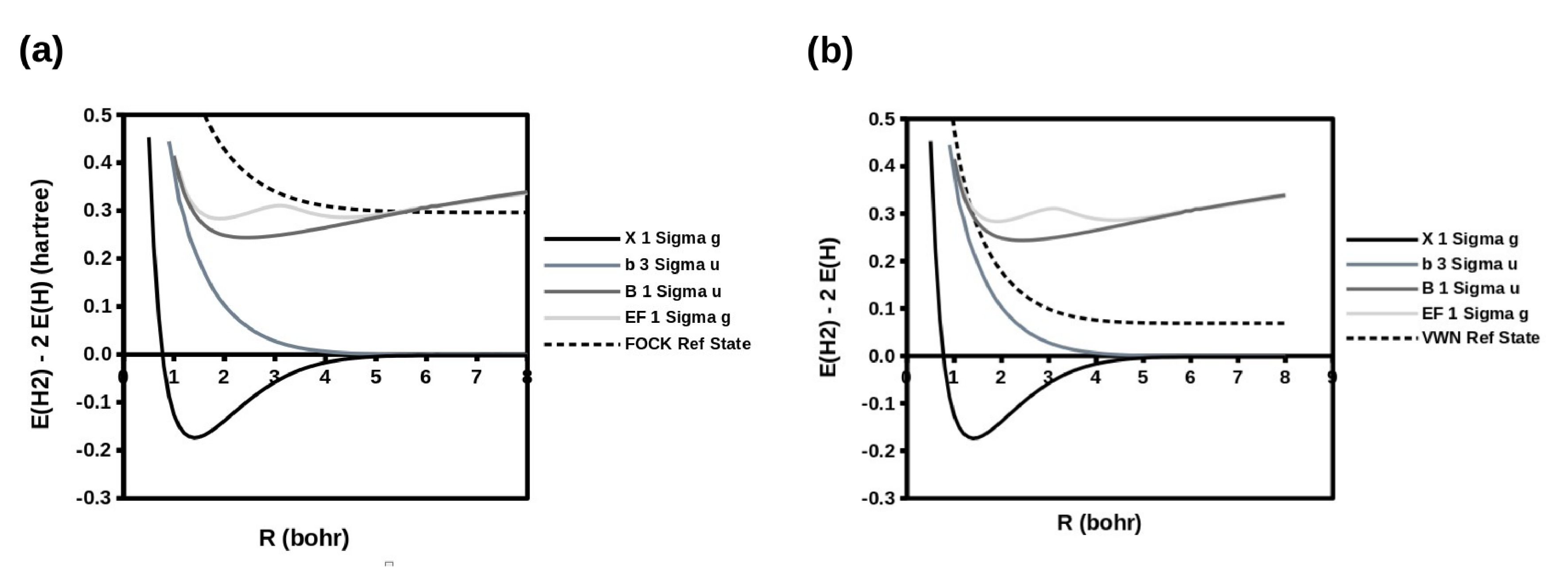}
\end{center}
\caption{
MSM reference state PECs: (a) MSM-HF and (b) MSM-VWN.
The energies of the separated atoms
have been calculated as twice the energy of the neutral hydrogen atom
calculated using identical conditions (atomic basis set, density fitting,
basis set, grid, convergence conditions) to those used in calculating
the H$_2$ molecule reference state.
\label{fig:H2ref}
}
\end{figure}
\begin{figure}
\begin{center}
\includegraphics[width=0.5\textwidth]{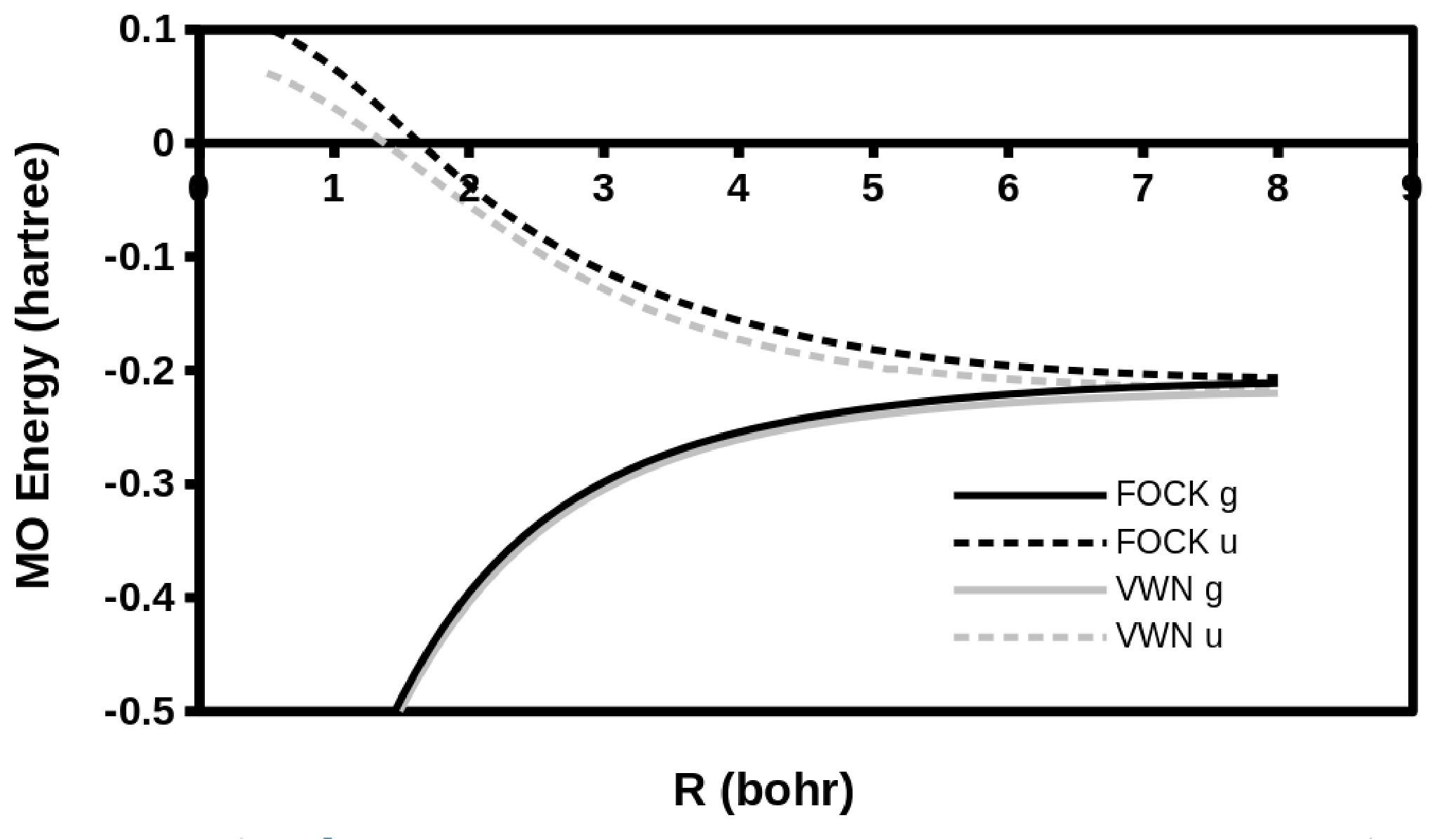}
\end{center}
\caption{
MO energies from the HF and VWN reference state calculations.
\label{fig:H2MOenergies}
}
\end{figure}
Our MSM-HF and MSM-VWN calculations use a reference calculation whose
PEC is shown in {\bf Fig.~\ref{fig:H2ref}} as a function of the bond 
distance $R$. Here, and throughout this subsection, we plot the H$_2$ 
energy relative to twice the energy of a single neutral hydrogen atom 
calculated using identical conditions (atomic basis set, density fitting,
basis set, grid, convergence conditions) to those used in calculating
the H$_2$ molecule reference state.  It is evident that the MSM-VWN PEC 
is much lower in energy than the MSM-HF PEC although, as 
{\bf Fig.~\ref{fig:H2MOenergies}} shows, the difference in the 
corresponding HF and VWN MO energy curves 
is much less dramatic.  Notable in the figure is that the $u$ MO energy 
is unbound in HF for bond lengths below about $R \approx \mbox{ 1.6 bohr}$ and
in HF for bond lengths below $R \approx \mbox{ 1.3 bohr}$ so that
the reference state cannot be considered to be converged with respect
to basis set expansion at short bond lengths.  This is an obvious difficulty 
with the MSM, but is not very important in the present application.

\begin{figure}
\begin{center}
\includegraphics[width=0.9\textwidth]{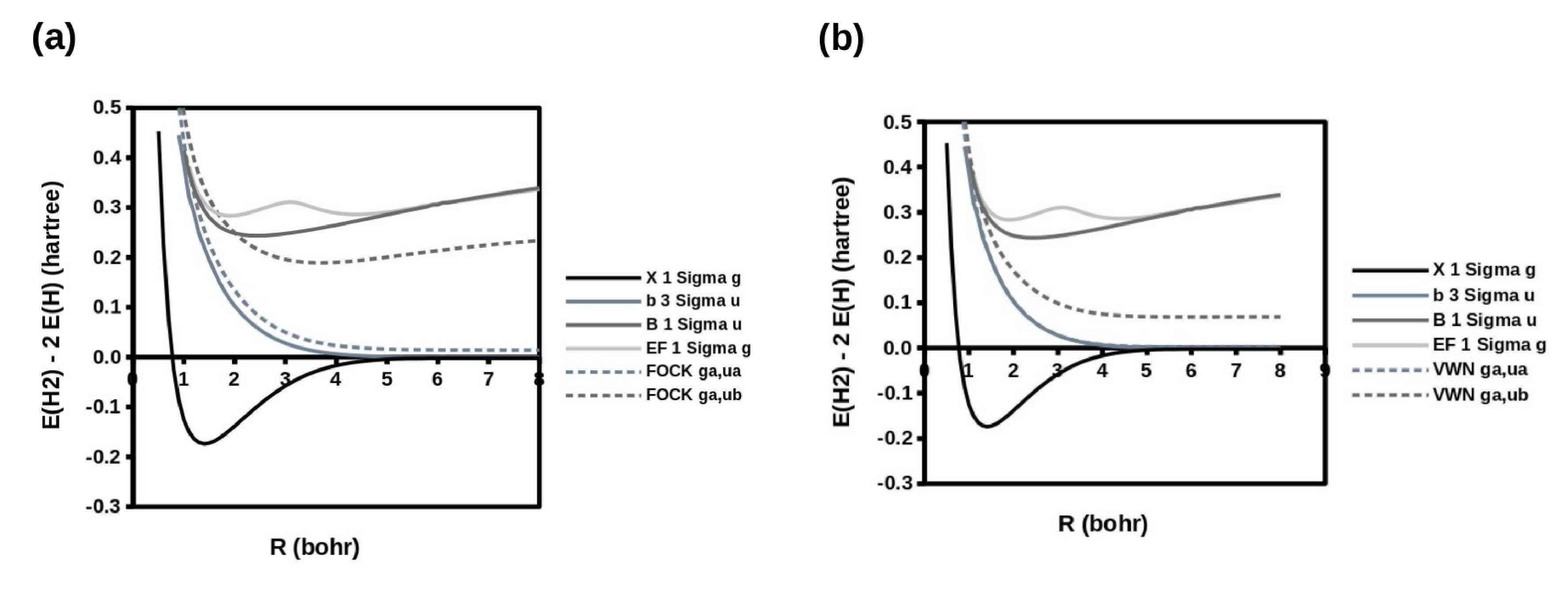}
\end{center}
\caption{
HF and VWN PECs for the triplet state determinant $\Psi_{1,1}=\vert g,u \vert$
and for the mixed-symmetry singly-excited determinant
$\Psi_M = \vert g, {\bar u} \vert$.  Note that the VWN triplet PEC is
indistinguishable from the EXACT triplet PEC.
\label{fig:H2TripletMixed}
}
\end{figure}
The first thing that needs to be done in the MSM is to calculate the energies
corresponding to a pure-symmetry triplet determinant 
$\Psi_{S,M} = \Psi_{1,\pm 1}$ and for a mixed-symmetry singly-excited 
determinant $\Psi_M$.  We have chosen to do this for 
$\Psi_{1,1} = \vert g, u \vert$ and for $\Psi_M = \vert g, {\bar u} \vert$.
The results are shown in {\bf Fig.~\ref{fig:H2TripletMixed}}.  Note that
we are use the unrelaxed orbitals from the reference calculation to evalute
the single determinant energies.  Only the spin-orbital occupations have
been changed using the {\tt MOMODIFY} keyword.  Relaxing the triplet HF orbitals
would lower the HF triplet PEC, bringing it closer to the EXACT triplet PEC.`
Interesting the VWN triplet PEC is indistinguishable from the EXACT triplet
PEC.  The HF mixed-symmetry PEC is not only higher in energy and closer to the
EXACT $B$ $^1\Sigma_u$ PEC than is the VWN mixed-symmetry PEC, but the
HF mixed-symmetry PEC has a minimum.  This is similar to the EXACT 
$B$ $^1\Sigma_u$ PEC but is lacking in the VWN mixed-symmetry PEC.  This is
a first indication of a failure of the MSM-VWN calculation to produce a
qualitatively correct result.

\begin{figure}
\begin{center}
\includegraphics[width=0.6\textwidth]{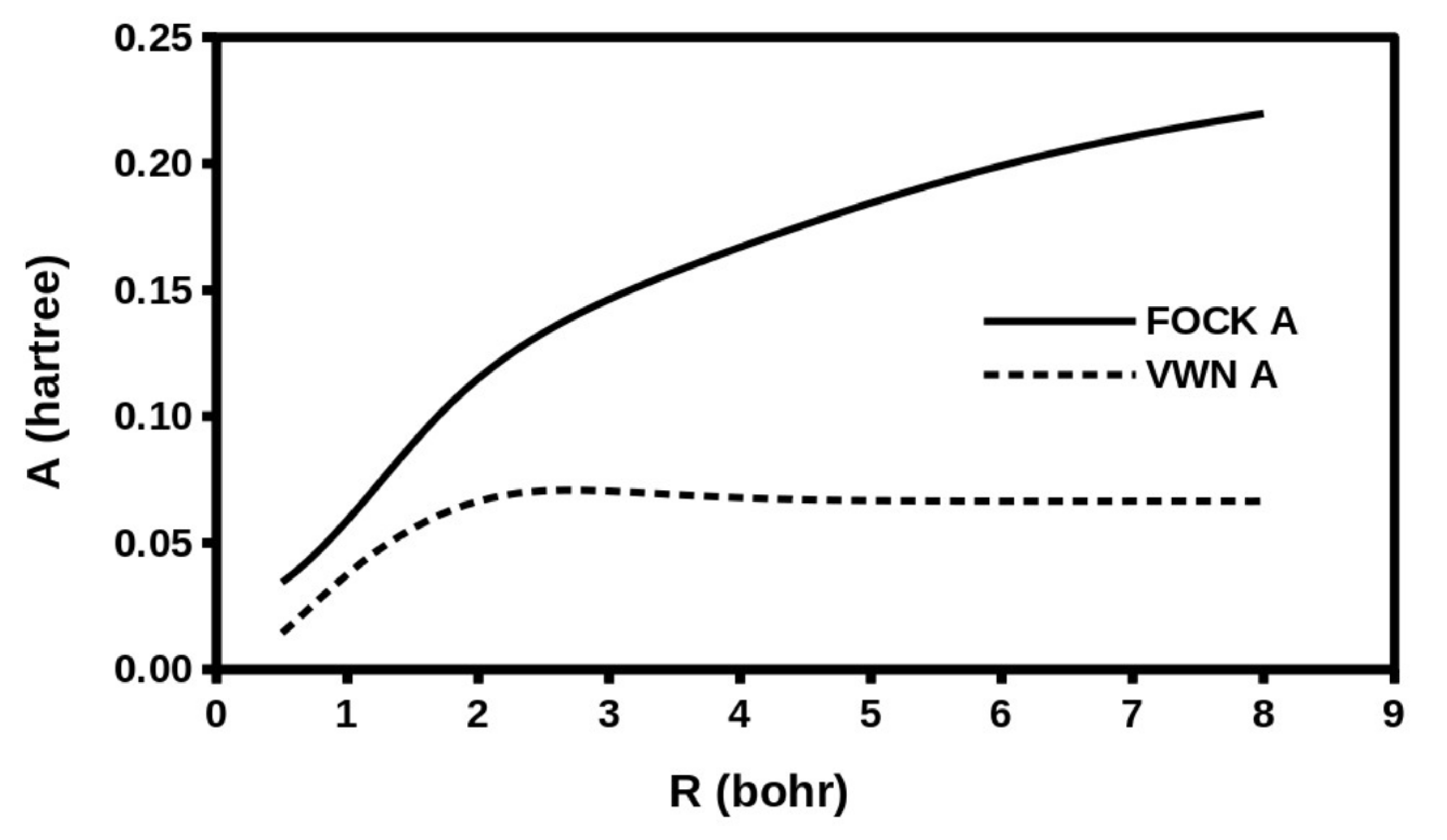}
\end{center}
\caption{
MSM $A=E[\vert g, {\bar u} \vert]-E[\vert g, u \vert]$ matrix element.
\label{fig:H2A}
}
\end{figure}
\begin{figure}
\begin{center}
\includegraphics[width=0.9\textwidth]{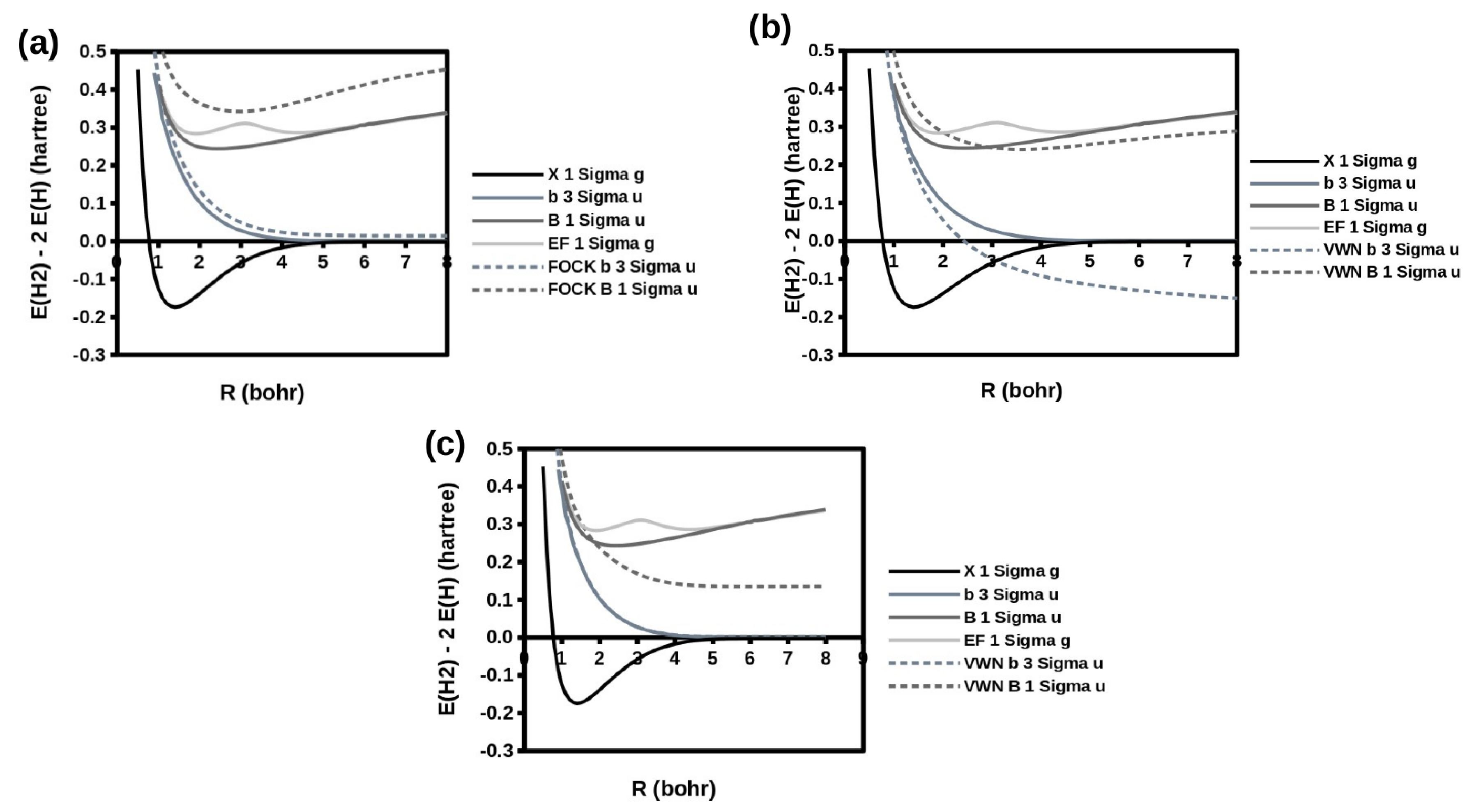}
\end{center}
\caption{
Three ways to calculate MSM $b$ $^3\Sigma_u$ and $B$ $^1\Sigma_u$ PECs:
(a) MSM-HF, (b) MSM-VWN using HF $A$, and (c) MSM-VWN (using VWN $A$).
\label{fig:H2MSMgu}
}
\end{figure}
According to the explanation given in Section~\ref{sec:theory}, these 
VWN single-determinant energies constitute a way to include dynamical 
correlation in the diagonal elements of the CI eigenvalue problem,
\begin{equation}
   \left[
       \begin{array}{cc} E[\vert u,{\bar g} \vert] & A \\
       A & E[\vert g, {\bar u} \vert] 
       \end{array}
   \right] 
   \left( 
       \begin{array}{c} C_g^u \\ C_{\bar g}^{\bar u} 
       \end{array}
   \right)
   = E
   \left( 
       \begin{array}{c} C_g^u \\ C_{\bar g}^{\bar u} 
       \end{array}
   \right) \, .
   \label{eq:results.1}
\end{equation}
The problem then is to introduce static correlation by chosing an appropriate
value of the matrix element $A$.  The default guess is to use WFT methods
to obtain,
\begin{equation}
   A^{\text{HF}} 
   = E_{\text{HF}}[\vert g, {\bar u} \vert]-E_{\text{HF}}[\vert g, u \vert] 
   = (gu \vert f_H \vert ug) \, .
   \label{eq:results.2}
\end{equation}
This matrix element is shown in {\bf Fig.~\ref{fig:H2A}} as a function of
distance.  This gives the MSM results shown in {\bf Fig.~\ref{fig:H2MSMgu}}.
By construction, the MSM-HF $b$ $^3\Sigma_u$ state made by subtracting
$A$ from the energy of the mixed symmetry state has exactly the same
energy as the HF $b$ $^3\Sigma_u$ state shown in Fig.~\ref{fig:H2TripletMixed}.
The MSM-HF $B$ $^3\Sigma_u$ PEC is a significant overestimate of the EXACT
$B$ $^3\Sigma_u$ PEC compared to what is obtained from the mixed-state
PEC (Fig.~\ref{fig:H2TripletMixed}) which matches the EXACT PEC remarkably
well for $R \leq \mbox{ 2 bohr}$.  However the MSM-HF $B$ $^3\Sigma_u$ PEC,
though overestimating the energy, is clearly a better global description as
it has a very similar overall shape to the corresponding EXACT PEC including
having a minimum in about the right place.  This state should also dissociate
into the ions whose energy, relative to the neutral atoms, is 0.57 hartree in
our HF calculations.

Figure~\ref{fig:H2MSMgu} confirms that using the HF $A$ matrix
element in the MSM-VWN calculation by subtracting
$A$ from the energy of the mixed symmetry state gives a very different
energy than that obtained from calculating the energies of the two single
determinantal $\Psi_{1,\pm 1}$ triplet states.  In particular, the dissociation
limit of the MSM-VWN $B$ $^1\Sigma_u$ PEC is disasterously low. 
On the other hand, the MSM-VWN $B$ $^1\Sigma_u$ PEC obtained in this way 
is (arguably) not that bad.  Its energy is about right and it has a minimum, 
just like the EXACT $B$ $^1\Sigma_u$ PEC.

\begin{figure}
\begin{center}
\includegraphics[width=0.9\textwidth]{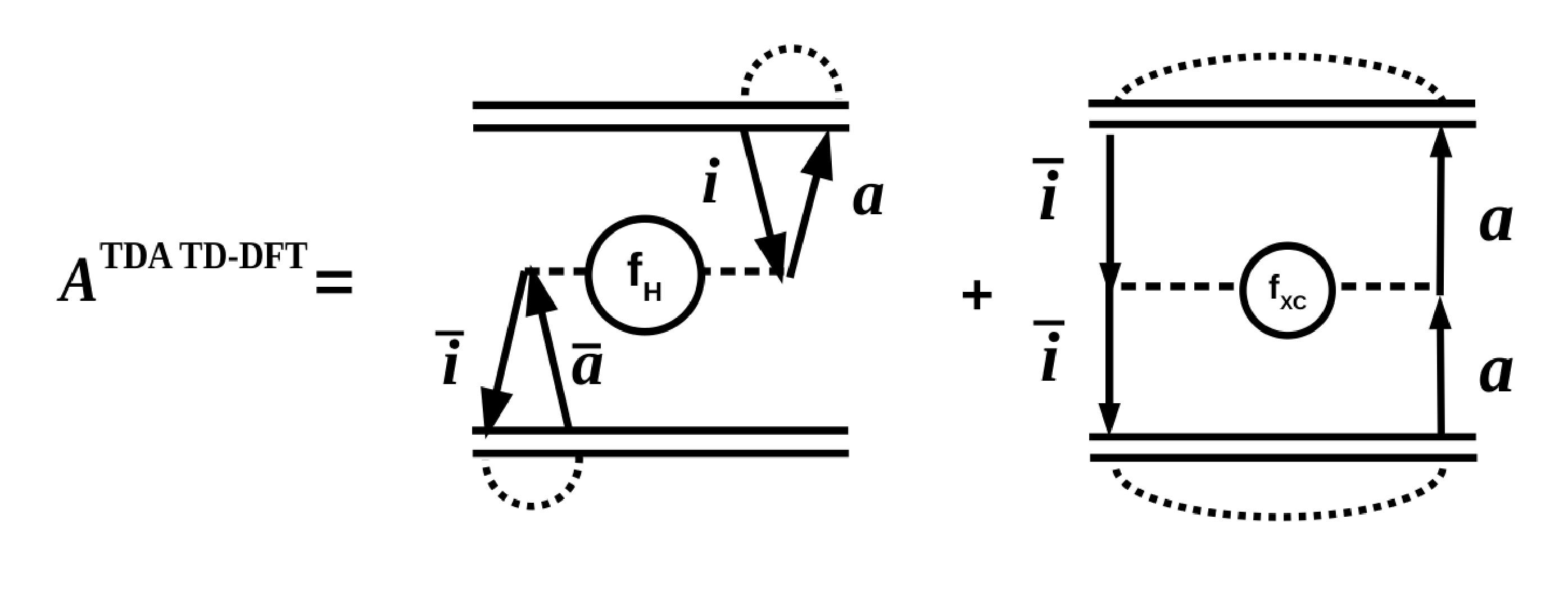}
\end{center}
\caption{
Diagrammatic representation of the $A$ matrix element in TDA TD-DFT.
The indices $i$ and $a$ have been used to facilitate comparison with
Figs.~\ref{fig:WFTdiags7} and \ref{fig:xc4}.
For H$_2$, replace $i$ with $g$ and $a$ with $u$.
\label{fig:TDDFTdiags}
}
\end{figure}

In order to fix the MSM-VWN triplet state problem, we must reset the $A$
matrix element to be,
\begin{equation}
   A^{\text{VWN}} 
   = E_{\text{VWN}}[\vert g, {\bar u} \vert]-E_{\text{VWN}}[\vert g, u \vert] 
   \approx (gg \vert f_{xc}^{\uparrow, \downarrow} 
   - f_{xc}^{\uparrow, \uparrow} \vert uu) \, .
   \label{eq:results.3}
\end{equation}
This gives the PECs in part (c) of Fig.~\ref{fig:H2MSMgu} where all the 
MSM-VWN triplet curves are now perfectly degenerate by construction.
The MSM-VWN $B$ $^1\Sigma_u$ PEC is matches the EXACT $B$ $^1\Sigma_u$ PEC
for $R \leq \mbox{ 2 bohr}$, but thereafter is a significant underestimate
of the EXACT PEC. This correlates very nicely with the fact that the 
VWN and HF values of $A$ are roughly similar for $R \leq \mbox{ 2 bohr}$.
Beyond this bond length, the HF $A$ increases while the VWN $A$ levels out
to a near constant value of about 0.066 hartree.  Perhaps surprisingly,
the MSM-VWN $B$ $^1\Sigma_u$ PEC is dissociating to an energy of about
0.135 hartree which is well below the expected VWN ionic dissociation energy of
0.515 hartree.  This strange result has been seen before where it was
also noted that a qualitatively correct $B$ $^1\Sigma_u$ PEC is obtained
\marginpar{\color{blue} TDA}
using TD-DFT in the Tamm-Dancoff approximation (TDA) 
\cite{CGG+00}.  The corresponding $A$ matrix element in TDA TD-DFT
is
\begin{equation}
   A^{\text{TDA TD-DFT}} 
   = (gu \vert f_H \vert ug) + (gg \vert f_{xc}^{\uparrow, \downarrow} \vert uu) \, ,
   \label{eq:results.4}
\end{equation}
represented diagrammatically in {\bf Fig.~\ref{fig:TDDFTdiags}}.   
This might suggest the superiority
of TDA TD-DFT over MS-DFT, but it should be noted that the two theories
give very similar results at the ground-state equilibrium bond length of
about 1.4 bohr, where the hypothesis of a single-determinantal ground state
wave function remains reasonable.

\begin{figure}
\begin{center}
\includegraphics[width=0.9\textwidth]{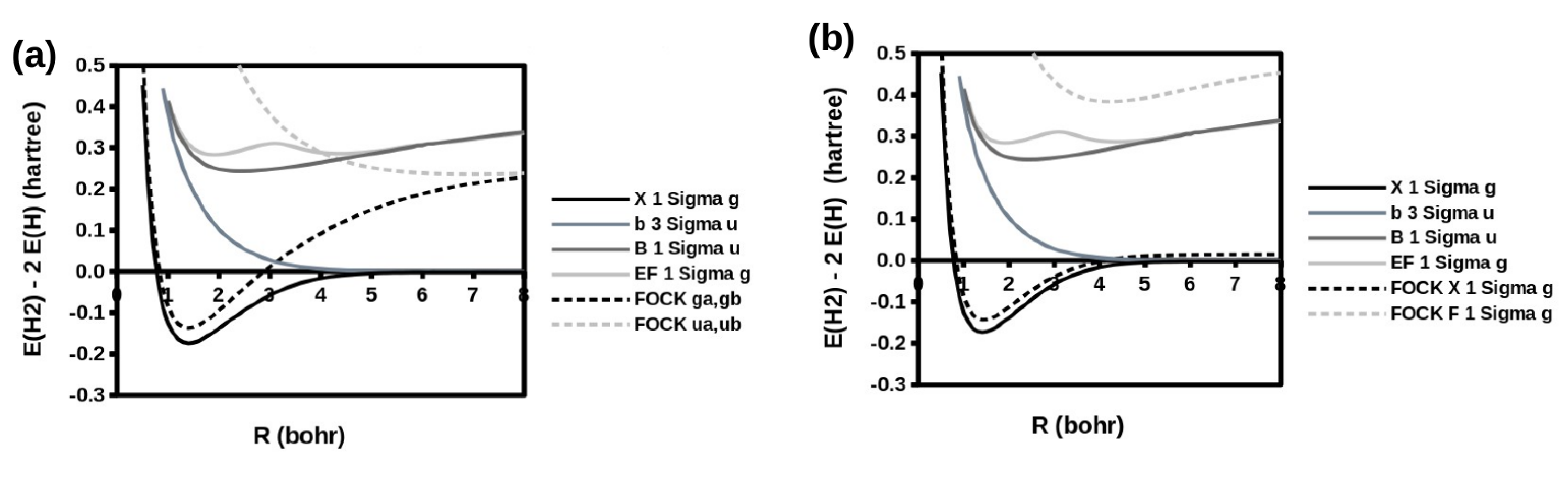}
\end{center}
\caption{
$X$ $^1\Sigma_g$ and $F$ $^1\Sigma_g$ HF PECs: (a) without $A$ and (b)
including $A$.
\label{fig:H2SigmagHF}
}
\end{figure}
\begin{figure}
\begin{center}
\includegraphics[width=0.9\textwidth]{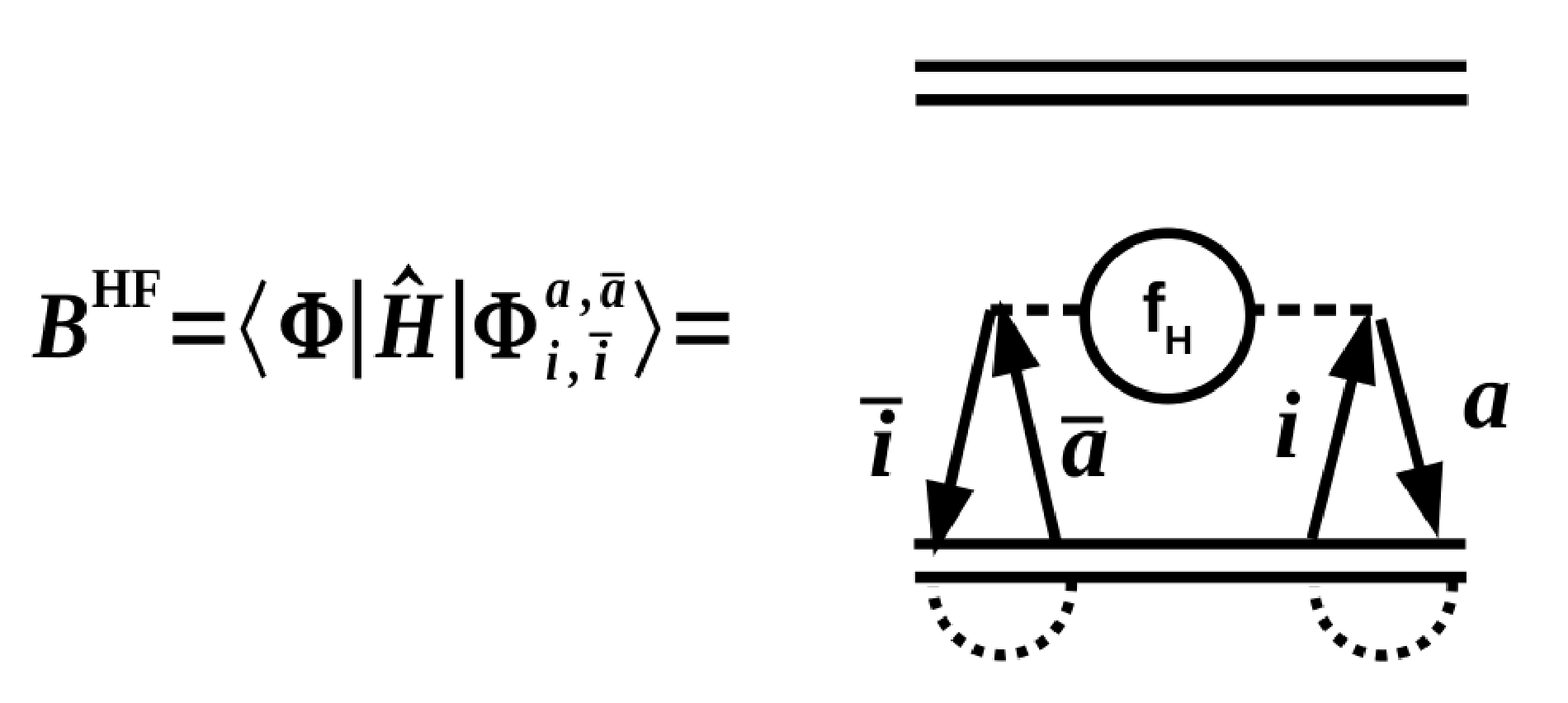}
\end{center}
\caption{
Diagrammatic evaluation of $B^{\mbox{HF}}=\langle \Phi \vert \hat{H} \vert
\Phi_{i,{\bar i}}^{a,{\bar a}} \rangle$.  In the particular case of H$_2$, then
$i$ and $a$ should be replaced respectively with $g$ and $u$.
\label{fig:WFTdiagB}
}
\end{figure}

Up to this point, there is nothing (or at least very little) that is new.
We have simply illustrated the use of MS-DFT for describing the triplet
and open-shell singlet PECs of H$_2$.  However, {\em we now look at something
new} inspired by diagrammatic MS-DFT.  Let us begin with WFT.  
{\bf Figure~\ref{fig:H2SigmagHF}}(a) shows the ground and doubly-excited
state PECs calculated within the single-determinantal approximation.
These two HF PECs are dissociating to the same wrong energy.  The reason
that this happens is well understood but may require review for some
\marginpar{\color{blue} VB \\ LDS}
readers as it is closely associated with the valence-bond (VB) method
which allows wave functions to be interpretted in terms of Lewis dot
structures (LDSs) and which is not as often taught as it once was \cite{C52}.  
At large $R$, the overlap matrix (Fig.~\ref{fig:H2MOdiagram}) goes to zero 
and the limiting form of the MOs is just,
\begin{eqnarray}
   g & = & \frac{1}{\sqrt{2}} \left( s_A - s_B \right) \nonumber \\
   u & = & \frac{1}{\sqrt{2}} \left( s_A + s_B \right) \, .
   \label{eq:results.5}
\end{eqnarray}
This allows us to calculate the large $R$ limit for the three triplet
$b$ $^3\Sigma_u$ wave functions $\Psi_{S,M}$,
\begin{eqnarray}
   \Psi_{1,1} & = & \vert g , u \vert \nonumber \\
    & \rightarrow & -\vert s_A , s_B \vert \nonumber \\
    & & \left[ \mbox{H$\uparrow$ H$\uparrow$} \right] \nonumber \\
   \Psi_{1,0} & = & \frac{1}{\sqrt{2}} \left( \vert g, {\bar u} \vert
   - \vert u , {\bar g} \vert \right) \nonumber \\
   & \rightarrow & -\frac{1}{\sqrt{2}} 
   \left( \vert s_A, {\bar s}_B \vert - \vert s_B, {\bar s_A} \right)
   \nonumber \\
   & & \left[ \mbox{H$\uparrow$ H$\downarrow$} \leftrightarrow
              \mbox{H$\downarrow$ H$\uparrow$} \right] \nonumber \\
   \Psi_{1,-1} & = & \vert {\bar g}, {\bar u} \vert \nonumber \\
   & \rightarrow & \vert {\bar s}_A , {\bar s}_B \vert \nonumber \\
   & & \left[ \mbox{H$\downarrow$ H$\downarrow$} \right] \, ,
   \label{eq:results.6}
\end{eqnarray}
where the corresponding LDSs are also shown.  This explains why 
the $b$ $^3\Sigma_u$ PEC dissociates to the neutral atom limit in 
Fig.~\ref{fig:exactH2}.  Similarly for the open-shell singlet $B$ $^1\Sigma_u$
wave function,
\begin{eqnarray}
   \Psi_{0,0} & = & \frac{1}{\sqrt{2}} \left( \vert g, {\bar u} \vert
   + \vert u , {\bar g} \vert \right) \nonumber \\
   & \rightarrow & \frac{1}{\sqrt{2}} 
   \left( \vert s_A, {\bar s}_A \vert - \vert s_B, {\bar s_B} \right)
   \nonumber \\
   & & \left[ \mbox{H:$^-$ H$^+$} \leftrightarrow
              \mbox{H$^+$ H$^-$} \right] \, ,
   \label{eq:results.7}
\end{eqnarray}
which explains why the $B$ $^1\Sigma_u$ PEC dissociates to the separated
ion limit in Fig.~\ref{fig:exactH2}.  However,
\begin{eqnarray}
   \vert g, {\bar g} \vert & \rightarrow & \frac{1}{\sqrt{2}}
   \left[ \frac{1}{\sqrt{2}} \left( \vert s_A , {\bar s}_A \vert + 
   \vert s_B, {\bar s}_B \vert \right) 
   + \frac{1}{\sqrt{2}} \left( \vert s_A , {\bar s}_B \vert + 
   \vert s_B, {\bar s}_A \vert \right) \right] \nonumber \\
   & & \left[ \mbox{H:$^-$ H$^+$} \leftrightarrow \mbox{H$^+$ H:$^-$} \right]
   + \left[ \mbox{H$\uparrow$ H$\downarrow$} \leftrightarrow 
   \mbox{H$\downarrow$ H$\uparrow$} \right] \nonumber \\
   \vert u, {\bar u} \vert & \rightarrow & \frac{1}{\sqrt{2}}
   \left[ \frac{1}{\sqrt{2}} \left( \vert s_A , {\bar s}_A \vert + 
   \vert s_B, {\bar s}_B \vert \right) 
   + \frac{1}{\sqrt{2}} \left( \vert s_A , {\bar s}_B \vert + 
   \vert s_B, {\bar s}_A \vert \right) \right] \nonumber \\
   & & \left[ \mbox{H:$^-$ H$^+$} \leftrightarrow \mbox{H$^+$ H:$^-$} \right]
   + \left[ \mbox{H$\uparrow$ H$\downarrow$} \leftrightarrow 
   \mbox{H$\downarrow$ H$\uparrow$} \right] \, ,
   \label{eq:results.8}
\end{eqnarray}
which shows that the ground and doubly-excited determinants dissociate
to the same incorrect limit which is a combination of ionic and of
neutral atom separation.  In the large $R$ limit, the $X$ $^1\Sigma_g$
wave function should have the form,
\begin{eqnarray}
  \frac{1}{\sqrt{2}} \left( \vert g, {\bar g} \vert - \vert u , {\bar u} \vert
  \right) & = & \frac{1}{\sqrt{2}} \left( \vert s_A, {\bar s}_B \vert
  + \vert s_B , {\bar s}_A \vert \right) \nonumber \\
  \left[ \mbox{H$\uparrow$ H$\downarrow$} \right.
  & \leftrightarrow &  \left. \mbox{H$\downarrow$ H$\uparrow$} \right] \, ,
  \label{eq:results.9}
\end{eqnarray}
and the $F$ $^1\Sigma_g$ wave function should have the form,
\begin{eqnarray}
  \frac{1}{\sqrt{2}} \left( \vert g, {\bar g} \vert + \vert u , {\bar u} \vert
  \right) & = & \frac{1}{\sqrt{2}} \left( \vert s_A, {\bar s}_B \vert
  - \vert s_B , {\bar s}_A \vert \right) \nonumber \\
  \left[ \mbox{H:$^-$ H$^+$} \right.
  & \leftrightarrow &  \left. \mbox{H$^+$ H:$^-$} \right] \, .
  \label{eq:results.10}
\end{eqnarray}
For smaller values of $R$, the first approximation is that these two
wave functions have the form,
\begin{equation}
  \Psi = C_0 \vert g, {\bar g} \vert + C_{g,{\bar g}}^{u,{\bar u}}
  \vert u, {\bar u} \vert \, ,
  \label{eq:results.11}
\end{equation}
and the coefficients and corresponding energies are determined by solving
the small CI problem,
\begin{equation}
   \left[
       \begin{array}{cc} E[\vert g,{\bar g} \vert] & B \\
       B & E[\vert u, {\bar u} \vert] 
       \end{array}
   \right] 
   \left( 
       \begin{array}{c} C_0 \\ C_{g, {\bar g}}^{u, {\bar u}}
       \end{array}
   \right)
   = E
   \left( 
       \begin{array}{c} C_0 \\ C_{g, {\bar g}}^{u, {\bar u}}
       \end{array}
   \right) \, .
   \label{eq:results.12}
\end{equation}
This is reminiscent of Eq.~[\ref{eq:results.1}], except that the two
diagonal elements are no longer equal, so that simple quadratic equation must
be solved to find the eigenvalues.  {\bf Figure~\ref{fig:WFTdiagB}} shows that 
\begin{equation}
  B = (ug \vert f_H \vert ug) \, ,
  \label{eq:results.13}
\end{equation}
which is the same as the $A$ matrix element already calculated as long as 
the MOs are real.  Solving the CI Eq.~[\ref{eq:results.12}], gives the graph
shown in part (b) of Fig.~\ref{fig:H2SigmagHF}.  Not only does the 
$X$ $^1\Sigma_g$ PEC dissociate correctly to the limit of separated
neutral atoms (within the accuracy of the frozen reference orbital approximation
being used) but the $F$ $^1\Sigma_g$ PEC has a minimum around the right
hand minimum of the $E,F$ $^1\Sigma_g$ PEC as it should and dissociates
correctly towards the HF ionic limit of 0.57 hartree.

The situation is very different in DFT. When the exact xc functional is 
used, there is no NVR problem and a single-determinantal wavefunction
suffices.  This is easily shown by setting the $g$ orbital equal to the
square root of half the exact charge density and then reconstructing the
non-interacting potential as,
\begin{eqnarray}
  \left( -\frac{1}{2} \nabla^2 + v_s \right) \psi_g & = & \epsilon_g \psi_g
  \nonumber \\
  \psi_g & = & \epsilon_g + \frac{1}{2} \nabla^2 \psi_g \, .
  \label{eq:results.14}
\end{eqnarray}
That $\psi_g$ is the ground state for this $v_s$ results from the absence
of nodes in $\psi_g$. However practical xc approximations result in symmetry
breaking beyond the Coulson-Fischer point (e.g., see Ref.~\cite{CGG+00}
\marginpar{\color{blue} DODS \\ SODS}
and references therein) where the different-orbitals-for-different-spins (DODS)
solution becomes lower in energy than the same-orbitals-for-different-spins
(SODS) solution. This implies the need to improve the wave function by taking
a linear combination of two determinants as in Eq.~[\ref{eq:results.11}].
Alternatively, allowing the symmetry breaking to occur and using the DODS
solution suffices for many cases, but {\em not} when excited states are to be 
calculated from the ground state solution (e.g., Ref.~\cite{PEMC21}) as
symmetry is needed to get even qualitatively correct PECs.  It is therefore
important to be able to formulate a multideterminantal DFT, using WFT to include
static correlation and DFT to include static correlation and trusting that
overcounting of correlation effects will not prose a serious problem in
practical applications as long as the number of determinants used in the
multideterminantal DFT is kept small.

\begin{figure}
\begin{center}
\includegraphics[width=0.9\textwidth]{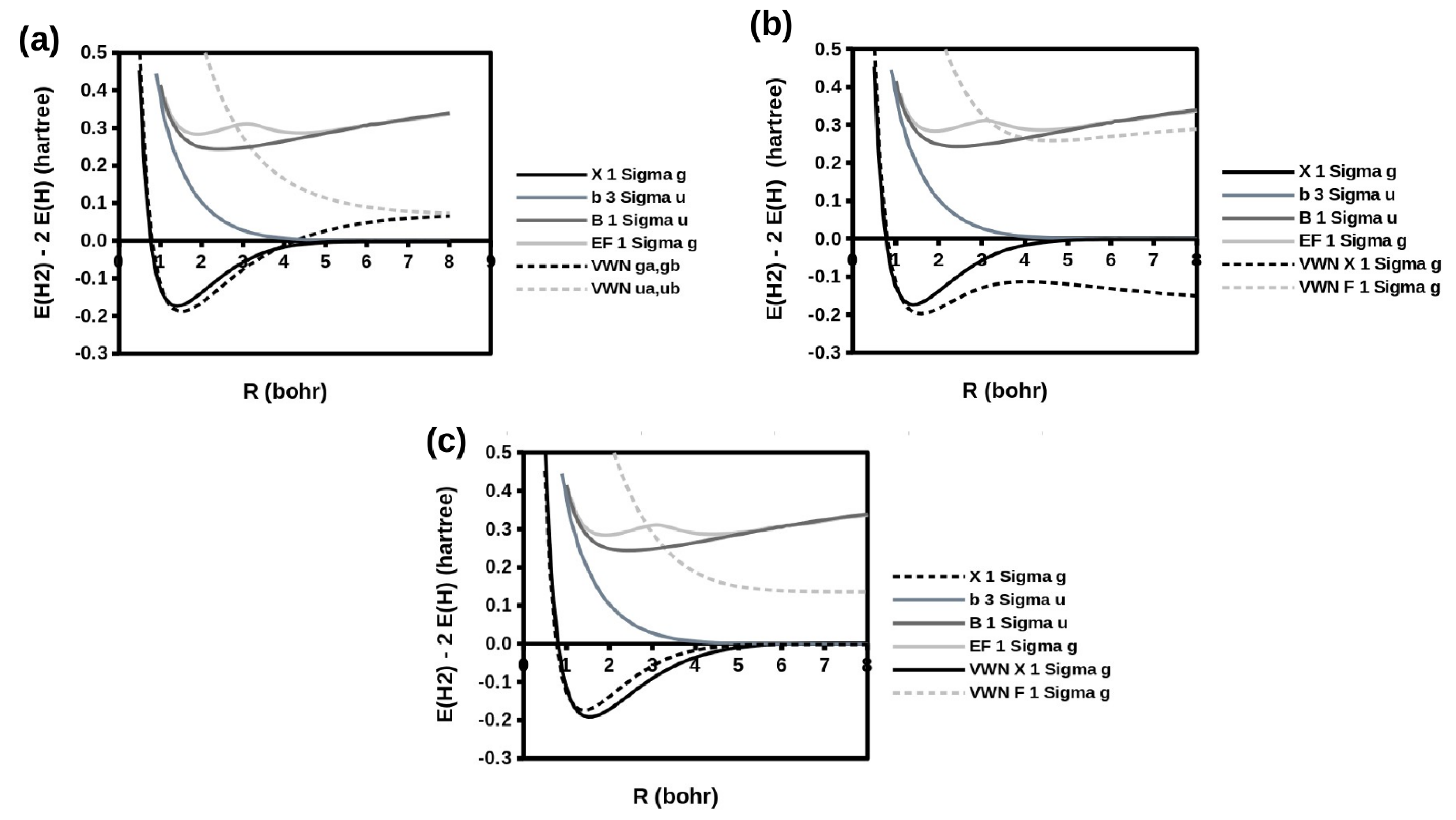}
\end{center}
\caption{
Three ways to calculate MSM $X$ $^1\Sigma_g$ and $F$ $^1\Sigma_g$ PECs:
(a) VWN single-determinantal energies, (b) VWN single-determinantal energies
corrected using the HF $B=A$, and (c) VWN single determinantal energies
obtained using $B =$ VWN $A$.
\label{fig:H2SigmagVWN}
}
\end{figure}
Inspired by MSM-DFT, we will use DFT to calculate the diagonal elements
in the CI Eq.~[\ref{eq:results.12}].  But what shall we use for $B$?
Unlike in MSM-DFT we have no symmetry to guide us! The two ``obvious'' 
choices for $B$ are either to use the HF $A$ from WFT or to use the
VWN $A$ obtained from MSM-DFT using symmetry and for a different purpose.
{\bf Figure~\ref{fig:H2SigmagVWN}} shows the results of these two choices.
The choice of using the HF $B=A$ to correct the single-determinental 
VWN energies looks like it might be reasonable for the $F$ $^1\Sigma_g$
excited state, but it is clearly wrong for the very important problem of
obtaining the correct dissociation of the $X$ $^1\Sigma_g$ ground state.
If $B$ may be replaced with $A$ in WFT, we might also suppose based upon
diagrammatic MS-DFT that $B$ might also be replaced with $A$ in DFT, provided
we use the correct MS-DFT $A$.  Remarkably Fig.~\ref{fig:H2SigmagVWN}(c)
shows that this is exactly correct for obtaining the correct dissociation
of the ground state.  {\em It is worth emphasizing that this an apparently
new result based upon an educated guess obtained from our diagrammatic analysis
of MS-DFT.  If this holds in a more general context, then we shall have 
produced something new and unexpected---namely a parameter-free 
multideterminantal DFT for ground state potential energy surfaces.}
The corresponding $F$ $^1\Sigma_g$ PEC is less impressive, suggesting that
we should use an alternative route to obtain this excited-state PEC, such
as doing response theory on our new multideterminantal DFT ground state
energy expression (something whose details should not be too difficult to
work out.) However the $F$ $^1\Sigma_g$ and $B$ $^1\Sigma_u$ PECs {\em are}
fully consistent in that they are both converging to the same value of 
about 0.135 hartree, instead of the expected DFT value of 0.514 hartree 
for dissociation into separated ions.  Alternatively, we might say that 
the ionic dissociation energy is 0.135 hartree in this model and everything 
at least seems to be internally consistent.  

\subsection{LiH}
\label{sec:LiH}

The example of LiH is treated in the SI.

\subsection{O$_2$}
\label{sec:O2}

Our third and last example of applying MSM-DFT is to O$_2$.
Our specific interest is in reactions of singlet O$_2$
($^1$O$_2$) with molecules such as the classic photochemical synthesis of 
the anthelmintic drug ascaridole from $\alpha$-terpene and $^1$O$_2$ 
\cite{WI81,O10,GG16,O16,AOZ+21}.
The presence of $\alpha$-terpene will necessarily lower the symmetry of the
system making ordinary MSM-DFT difficult, hence our desire to use diagrammatic
MSM-DFT to seek alternative approximations which might ultimately free us
from the symmetry-dependence of classic MSM-DFT. 

\begin{figure}
\begin{center}
\includegraphics[width=0.8\textwidth]{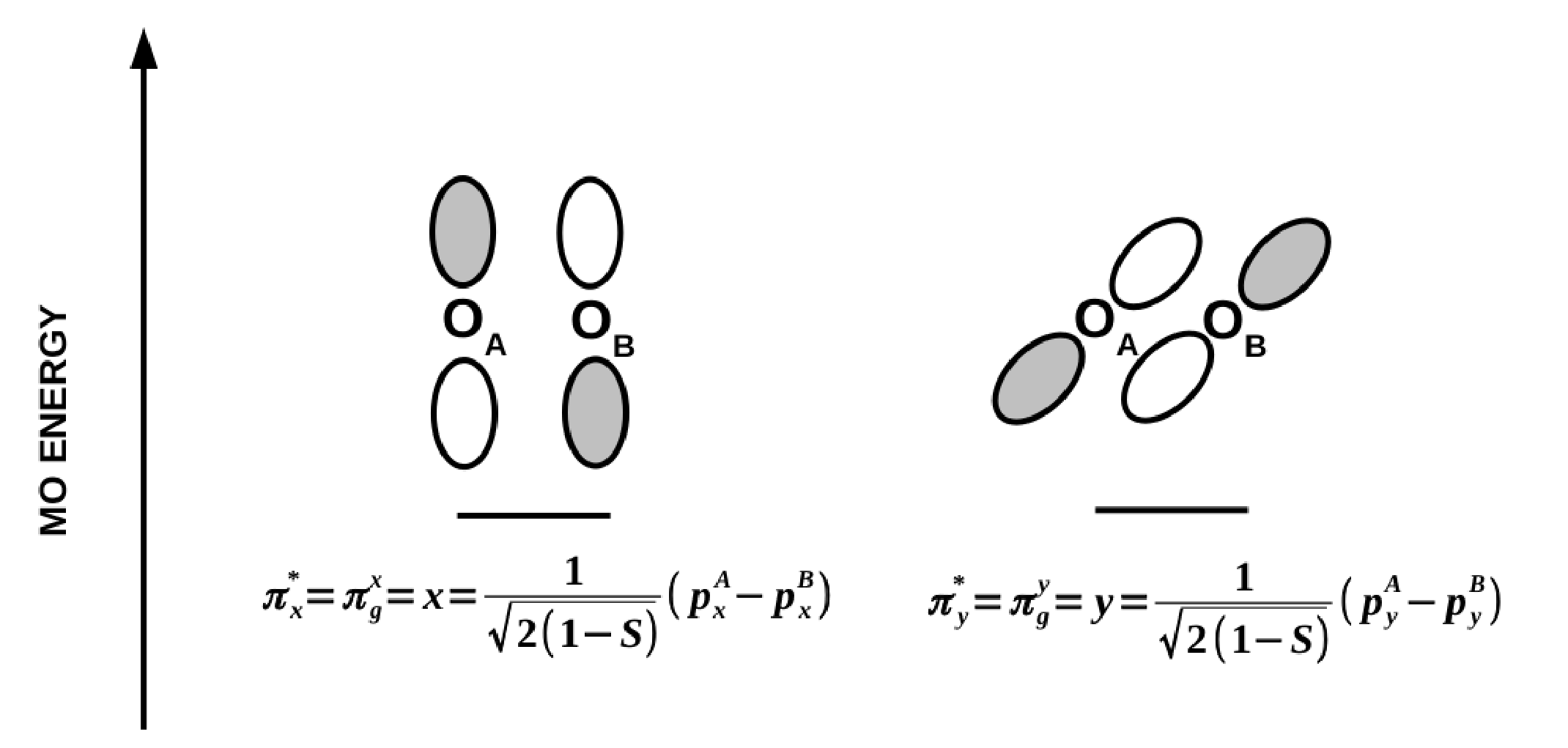}
\end{center}
\caption{
Textbook-like MO diagram for O$_2$.  This molecule is different from
either H$_2$ or LiH because the MOs are energetically degenerate.
The orbitals have been chosen to be real valued and
$S=\langle p_x^A \vert p_x^B \rangle = =\langle p_y^A \vert p_y^B \rangle$
is the overlap integral.  Note the drastic abbreviations $\pi_x^* \rightarrow x$
and $\pi_y^* \rightarrow y$.  [{\em Gerade} (g) and {\em ungerade} (u)
refer to inversion through the center of symmetry.]
\label{fig:O2MOdiagram}
}
\end{figure}
\begin{figure}
\begin{center}
\includegraphics[width=0.8\textwidth]{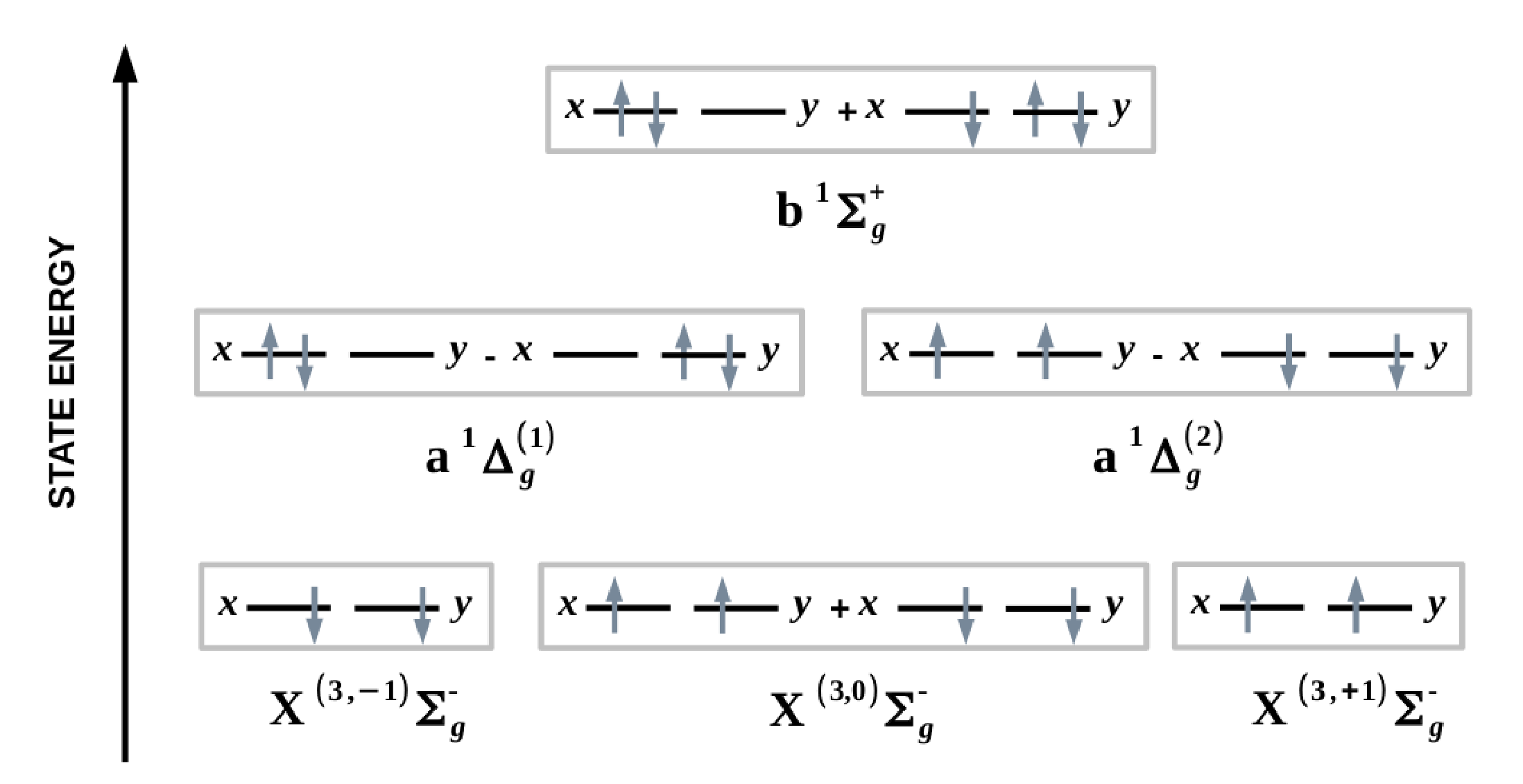}
\end{center}
\caption{
Diagram showing the six states which may be constructed by placing 
two electrons in the two MOs of Fig.~\ref{fig:O2MOdiagram}.
\label{fig:O2statediagram}
}
\end{figure}
A research spin-off \cite{PEMC21} of a relatively new pedagogical 
workbook \cite{OPEC23} presents one of the few, and certainly the 
most thorough, applications of MSM-DFT to the spin- and spacial-symmetry 
multiplet splitting problem in O$_2$.  We will only summarize the highlights 
here and then move on to a diagrammatic analysis.  

A textbook-like representation of the key frontier MOs is shown in 
{\bf Fig.~\ref{fig:O2MOdiagram}}.  Note that we are going to be lazy 
and just abbreviate the $\pi_x^*=\pi_g^x$ and $\pi_y^* = \pi_g^y$ 
MOs as $x$ and $y$ respectively.  As shown in Ref.~\cite{PEMC21},
a TOTEM may be constructed with the states shown in 
{\bf Fig.~\ref{fig:O2statediagram}}.  The appearance of this diagram
depends upon whether real or complex MOs are used \cite{PEMC21}.  We
are using real-valued MOs. It is evident that this state diagram looks
very different from those shown previously (Figs.~\ref{fig:genericstatediagram},
\ref{fig:H2statediagram}, and \ref{fig:LiHstatediagram}). This can
\begin{figure}
\begin{center}
\includegraphics[width=0.8\textwidth]{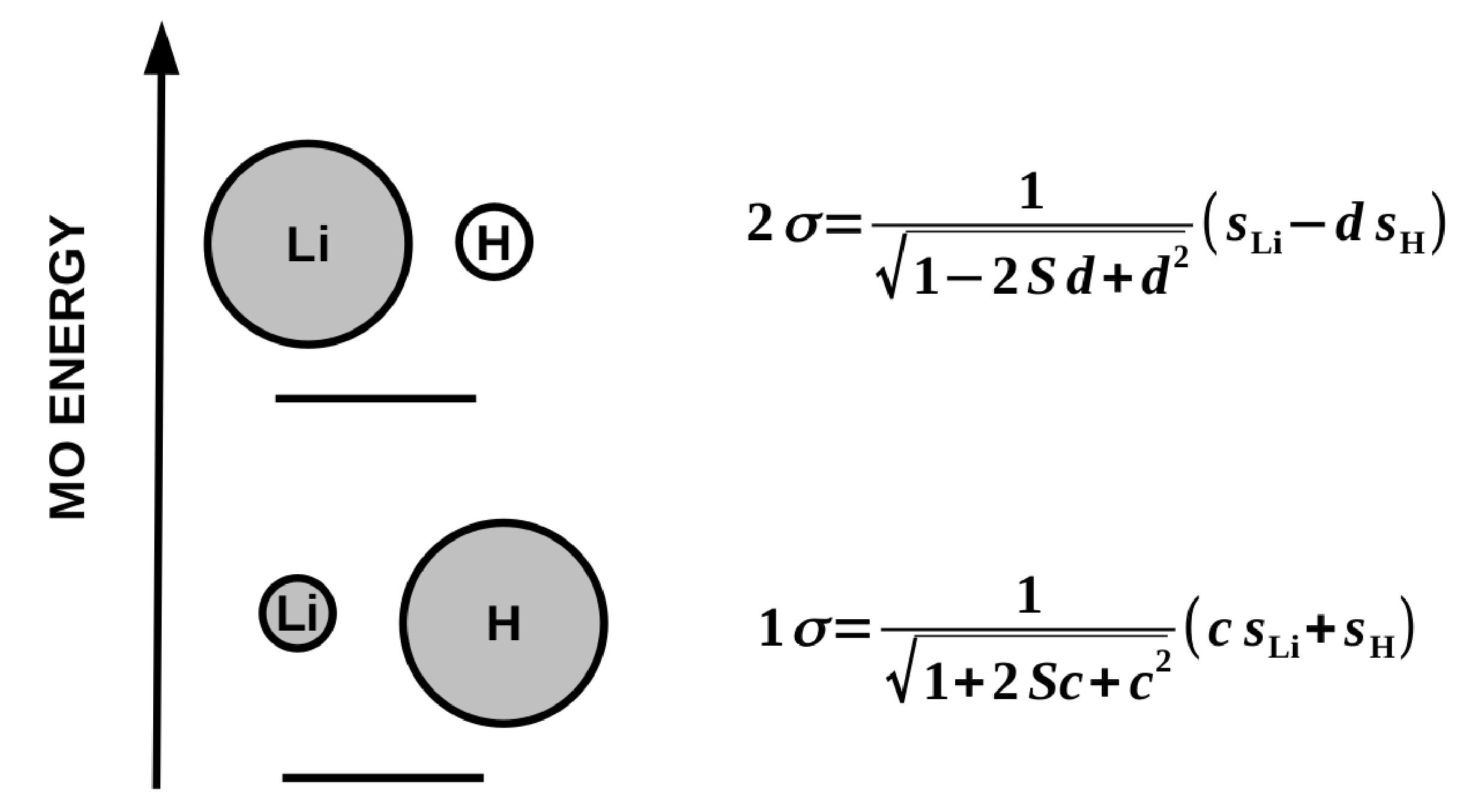}
\end{center}
\caption{
Textbook-like MO diagram for LiH.  There is only axial symmetry in
this molecule, so the valence MOs are labeled as $1\sigma$ and $2\sigma$.
Orthonormality fixes $d=(S+c)/(1+Sc)$ where $S=\langle s_{\mbox{Li}} \vert
s_{\mbox{H}} \rangle$ is the overlap integral.
\label{fig:LiHMOdiagram}
}
\end{figure}
\begin{figure}
\begin{center}
\includegraphics[width=0.8\textwidth]{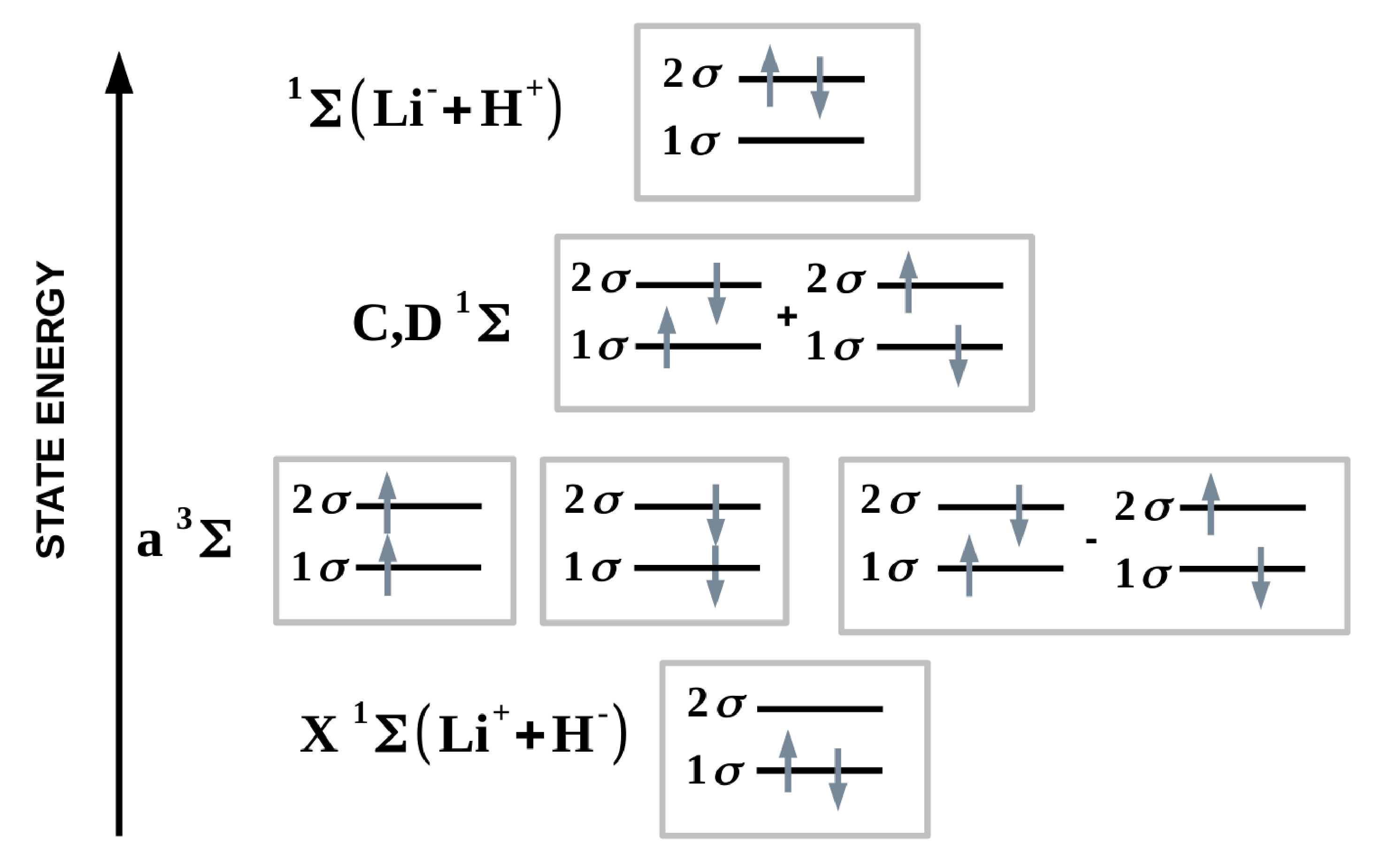}
\end{center}
\caption{
Diagram showing the six states which may be constructed by placing
two electrons in the two MOs of Fig.~\ref{fig:LiHMOdiagram}.
\label{fig:LiHstatediagram}
}
\end{figure}
cause problems when using diagrammatic theory.  On the other hand, 
the $X$ $^{(3,0)}\Sigma_g^-$ and $a$ $^1\Delta_g^{(2)}$ states form 
a natural $\pm$ pair, perfect for MSM, as do the $a$ $^1\Delta_g^{(1)}$ and
$b$ $^1\Sigma_g^+$ states.  We may then easily work out that,
\begin{eqnarray}
  E[X \,^3\Sigma_g^-] & = & E[\vert x,y \vert] \nonumber \\
  E[a \,^1\Delta_g] & = & E[\vert x,y \vert] + 2 F \nonumber \\
  E[b \,^1\Sigma_g^+] & = & E[\vert x,y \vert] + 2 P \, ,
  \label{eq:O2.1}
\end{eqnarray}
where the spin-flip energy,
\begin{equation}
  F = E[\vert x, {\bar y} \vert ] - E[\vert x, y \vert ] 
  \label{eq:O2.2}
\end{equation}
and the spin-pairing energy,
\begin{equation}
  P = E[\vert x, {\bar x} \vert ] - E[\vert x, {\bar y} \vert ] \, .
  \label{eq:O2.3}
\end{equation}
The $M_S=0$ part of the MSM CI matrix is,
\begin{equation}
  {\bf H} = \left[
  \begin{array}{cccc}
  E[\vert x, {\bar y} \vert] & A & 0 & 0 \\
  A & E[\vert y, {\bar x} \vert] & 0 & 0 \\
  0 & 0 & E[\vert x, {\bar x} \vert] & B \\
  0 & 0 & B & E[\vert y, {\bar y} \vert] 
  \end{array}
  \right] \, ,
  \label{eq:O2.4}
\end{equation}
where Brillouin's theorem has been used to zero out some of the matrix 
elements. Clearly $E[\vert x, {\bar y} \vert]=E[\vert y, {\bar x} \vert]$
and $E[\vert x, {\bar x} \vert]=E[\vert y, {\bar y} \vert]$ for the isolated
molecule.  Comparing Eqs.~(\ref{eq:O2.1})-(\ref{eq:O2.5}) shows that,
\begin{eqnarray}
  A & = & F = E[\vert x, {\bar y} \vert] - E[\vert x , y \vert]  \nonumber \\
  B & = & P-F = E[\vert x, {\bar x}\vert ] + E[\vert x, y \vert] 
       - 2 E[\vert x, {\bar y} \vert] \, . 
  \label{eq:O2.5}
\end{eqnarray}
While the expression for the $A$ matrix element is familiar, we have never
before had an MSM expression for the $B$ matrix element, though $B=A$
seemed to be an excellent educated guess in H$_2$.  Indeed the DFT-MSM results
given in Ref.~\cite{PEMC21} suggest that $B=A$ is probably a pretty reasonable
guess! This is the point that we wish now to explore.

\begin{figure}
\begin{center}
\includegraphics[width=0.8\textwidth]{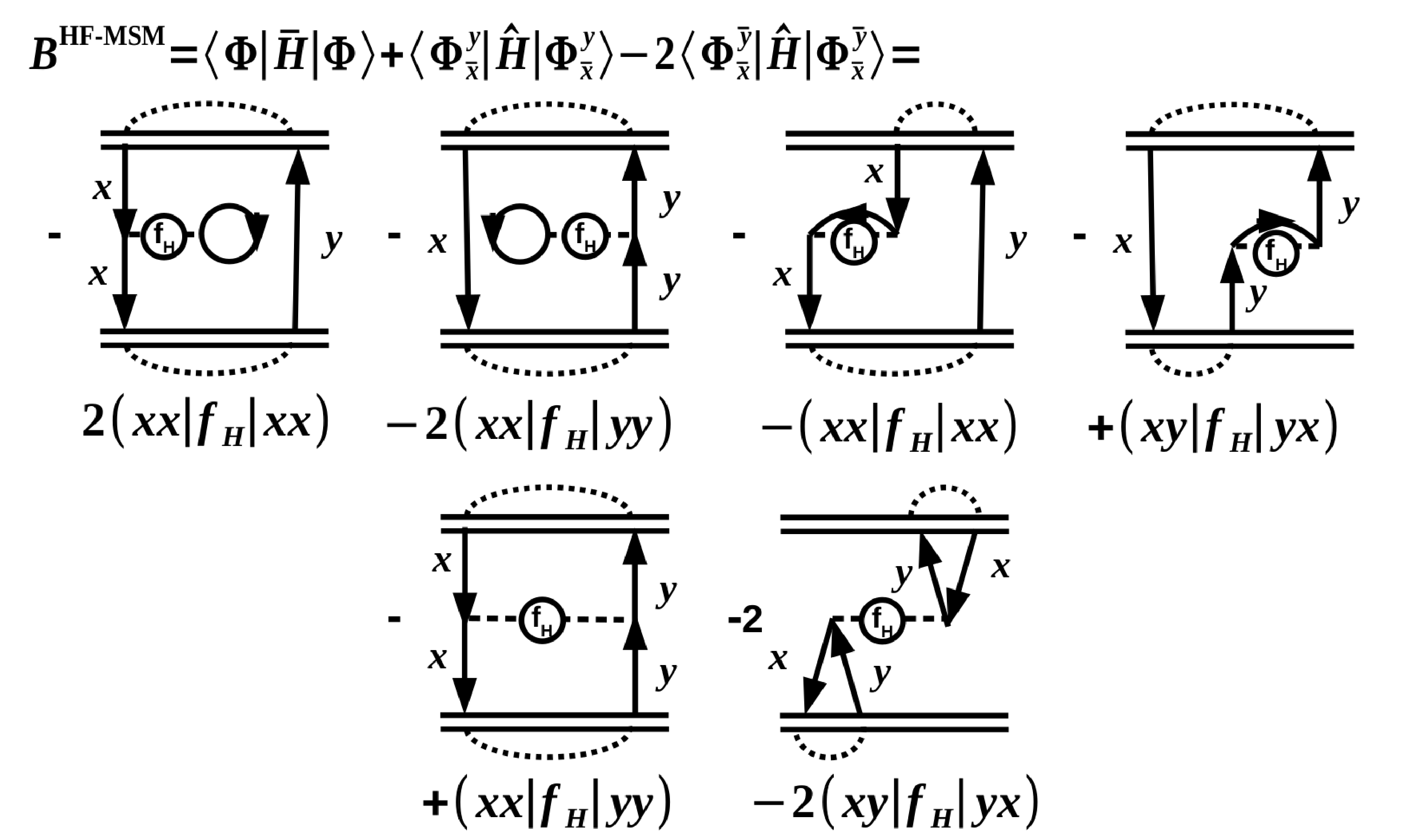}
\end{center}
\caption{
Diagrammatic development of the MSM-HF expression for $B$ in O$_2$.
Note that many terms have been canceled including terms with different
spins but the same algebraic expressions.
\label{fig:O2MSMB}
}
\end{figure}
\begin{figure}
\begin{center}
\includegraphics[width=0.4\textwidth]{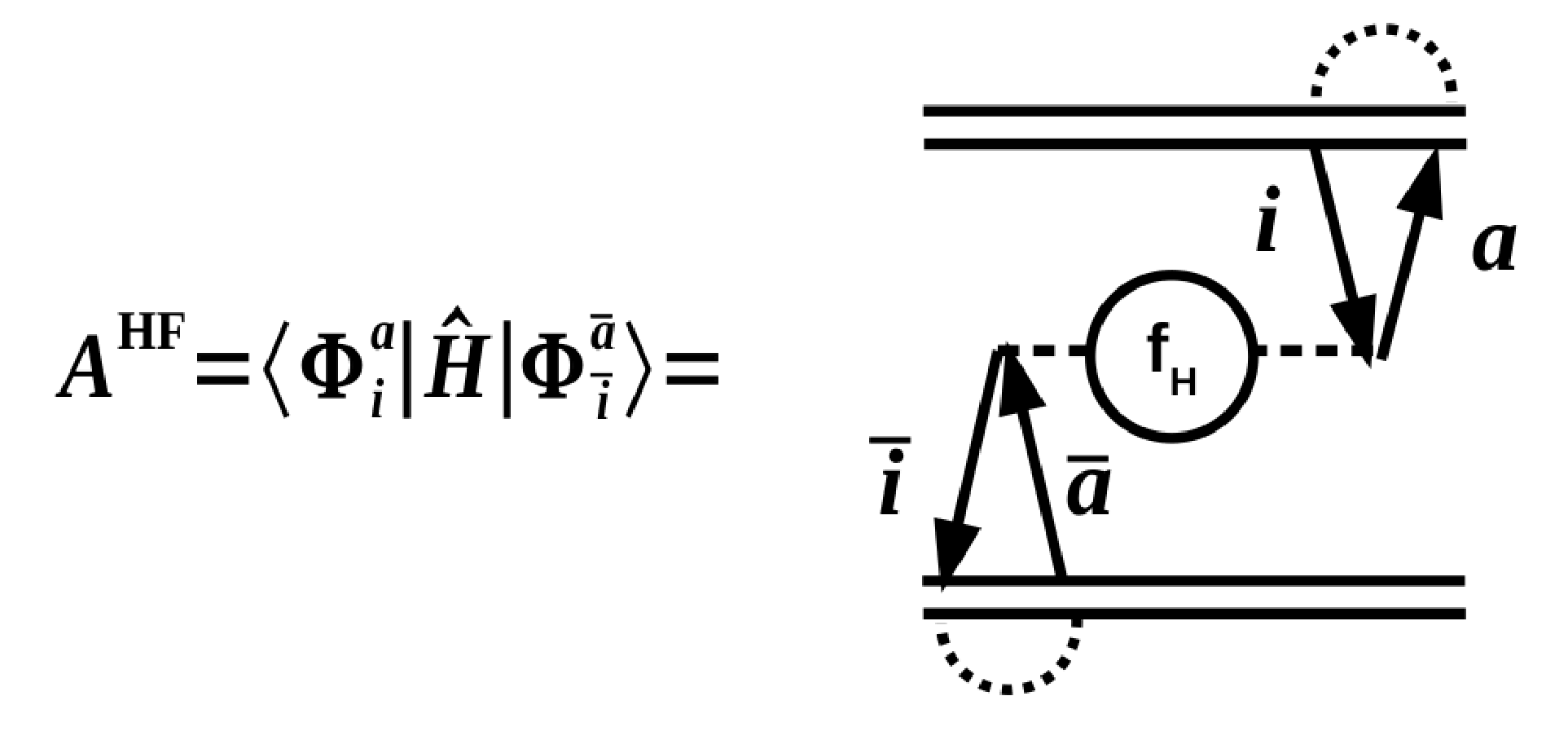}
\end{center}
\caption{
Diagrammatic evaluation of $A^{\mbox{HF}} = \langle \Phi_i^a \vert {\hat H}
\vert \Phi_{\bar i}^{\bar a} \rangle$.
\label{fig:WFTdiags7}
}
\end{figure}

We will use the diagrammatic analysis as a tool to aid us.  The price that
we pay for this is that, although the results will be independent of the
choice of physical vacuum, the appearance of the diagrams (e.g., which are
direct and which are exchange diagrams) depends upon the choice of the
reference.  In order to be as consistent as possible with everything that
has been done up to this point, we will choose $\Phi = \vert x, {\bar x} \vert$
as the physical vacuum. In WFT (either by direct evaluation or by MSM-HF), 
the diagrams of $A$ are shown in Fig.~\ref{fig:WFTdiags7} when $i$ and $a$ 
are replaced respectively by $x$ and $y$.  In either case, the matrix element 
evaluates to 
\begin{equation}
  A^{\mbox{HF}} = (xy \vert f_H \vert yx) \, .
  \label{eq:O2.6}
\end{equation}
However the situation is more complex for the $B$ matrix element.  The
diagram for direct evaluation of $B$ is identical to that of $A$ 
(Fig.~\ref{fig:WFTdiagB} when $i$ and $a$ are replaced respectively 
by $x$ and $y$),  
\begin{equation}
  B^{\mbox{HF}} = (xy \vert f_H \vert yx) \, .
  \label{eq:O2.7}
\end{equation}
On the other hand, the diagrams obtained using the 
MSM formula in Eq.~(\ref{eq:O2.5}) are shown in {\bf Fig.~\ref{fig:O2MSMB}}
and evaluate to 
\begin{eqnarray}
  B^{\mbox{HF}} & = & (xx \vert f_H \vert xx) - (xx \vert f_H \vert yy)
  -(xy \vert f_H \vert yx) \nonumber \\
  & = & (xy \vert f_H \vert yx) + \left[ (xx \vert f_H \vert xx) 
   - (xx \vert f_H \vert yy) - 2(xy \vert f_H \vert yx) \right] \, .
  \label{eq:O2.8}
\end{eqnarray}
Equations~(\ref{eq:O2.7}) and (\ref{eq:O2.8}) are actually identical
but the proof has been relegated to the SI as it is a little involved. 
In particular, it follows that $B=A$ in WFT.  

\begin{figure}
\begin{center}
\begin{tabular}{cc}
(a) & \\
(b) & \includegraphics[width=0.8\textwidth]{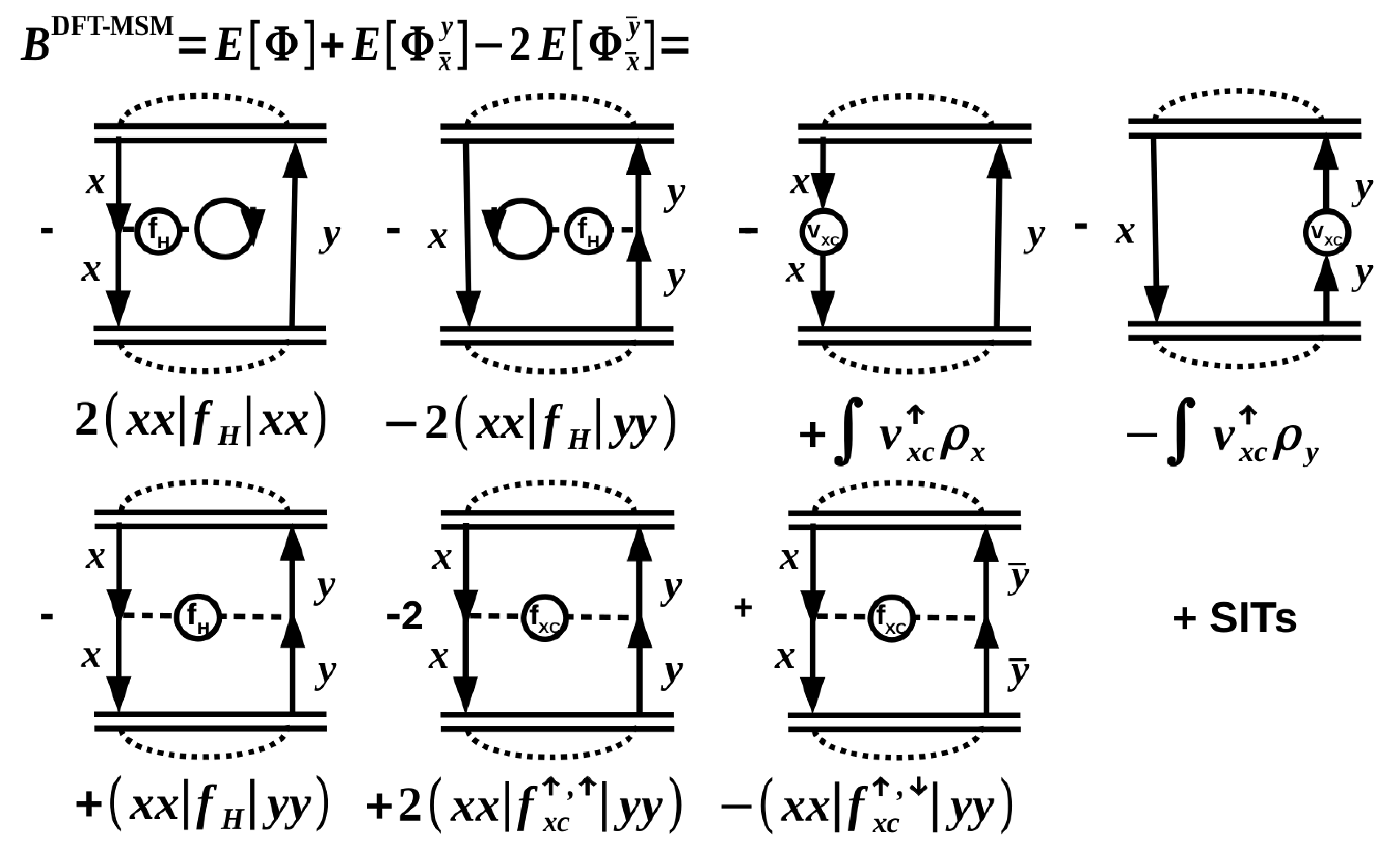} \\
    & \includegraphics[width=0.6\textwidth]{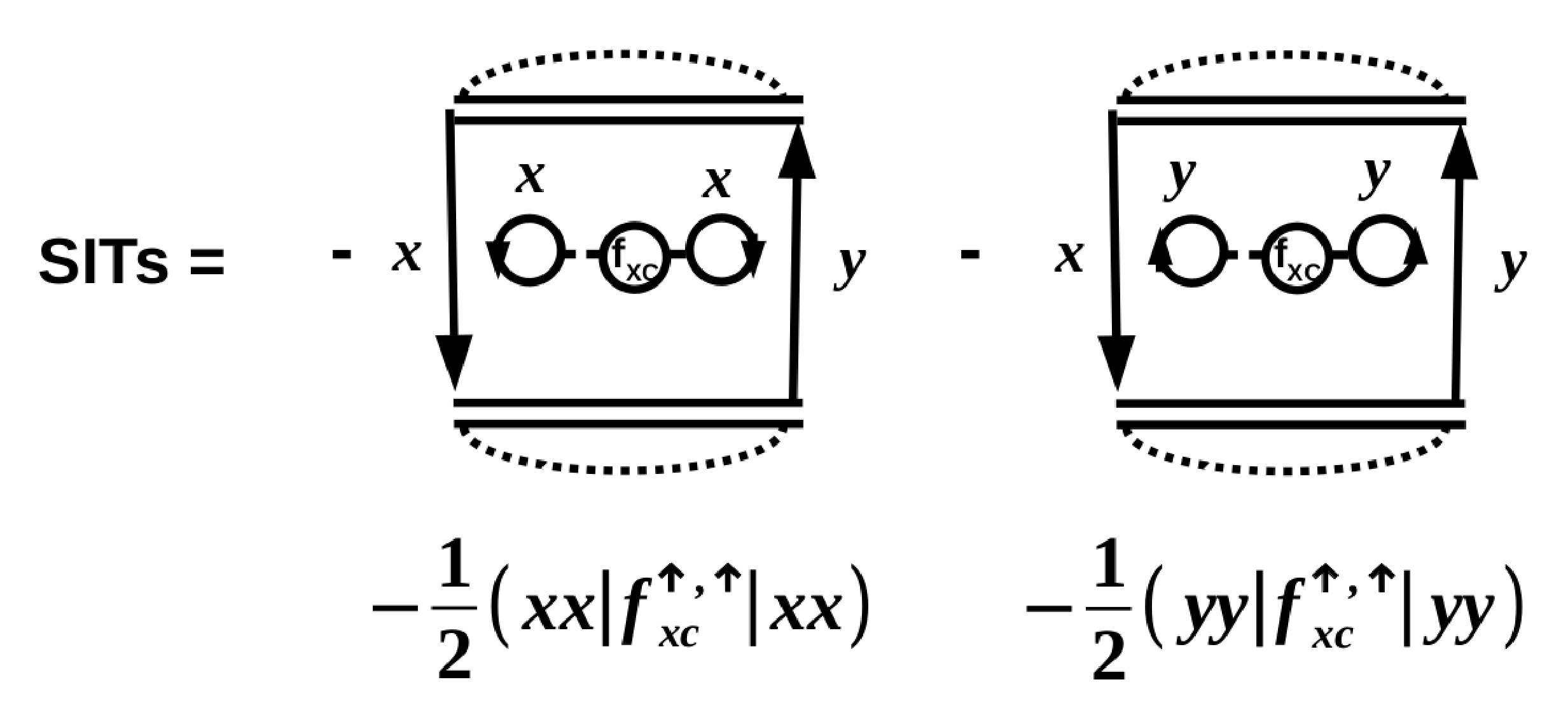} 
\end{tabular}
\end{center}
\caption{
Diagrammatic development of the DFT-MSM expression for $B$ in O$_2$ truncated
to second order.  Note that the xc terms are functionals of the density
corresponding to the reference density with both electrons in $x$ which we
will simply denote as $[\rho_x,\rho_x]$ in the TOTEM.
Also many terms have been canceled including terms with different
spins but the same algebraic expressions.
\label{fig:O2DFTMSMB}
}
\end{figure}
Now let us turn to DFT.  Traditional symmetry-based DFT-MSM arguments
were used in Ref.~\cite{PEMC21} to justify the use of Eq.~(\ref{eq:O2.8}).
Diagrammatic  analysis of the expression gives 
{\bf Fig.~\ref{fig:O2DFTMSMB}} rather than Fig.~\ref{fig:O2MSMB}
for the $B$ matrix element.  Nevertheless the general pattern is consistent
with the correspondance rules set out in Sec.~\ref{sec:theory}. Evaluating
the diagrams gives explicitly,
\begin{eqnarray}
  B^{\mbox{DFT-MSM}} & = & 2(xx\vert f_H \vert xx) -(xx\vert f_H \vert yy)
  \nonumber \\
  & + & \int v_{xc}^\uparrow[\rho_x,\rho_x](1) \rho_x(1) \, d1 
  - \int v_{xc}^\uparrow[\rho_x,\rho_x](1) \rho_y(1) \, d1
  \nonumber \\
  & + & (xx \vert 2f_{xc}^{\uparrow,\uparrow}[\rho_x,\rho_x]
  - 2f_{xc}^{\uparrow,\downarrow}[\rho_x,\rho_x] \vert yy)
  \nonumber \\
  & - & \frac{1}{2} \left\{ (xx \vert f_{xc}^{\uparrow,\uparrow}[\rho_x,\rho_x]
  \vert xx) + (yy \vert f_{xc}^{\uparrow,\uparrow}[\rho_x,\rho_x]
  \vert yy) \right\} \, .
  \label{eq:O2.15}
\end{eqnarray}
This expression is a little problematic should we wish to compare it with
Eqs.~(\ref{eq:O2.7}) and (\ref{eq:O2.8}) because of the presence of terms
involving $v_{xc}^{\uparrow}[\rho_x,\rho_x]$.  We can get around this by
using another expansion,
\begin{eqnarray}
  \int v_{xc}^{\uparrow}[\rho_x,\rho_x](1) \rho_y(1) \, d1 
  & = & \int v_{xc}^{\uparrow}[\rho_y,\rho_y](1) \rho_y(1) \, d1
  + (xx\vert f_{xc}^{\uparrow,\uparrow}[\rho_y,\rho_y] \vert yy)
  - (yy\vert f_{xc}^{\uparrow,\uparrow}[\rho_y,\rho_y] \vert yy)
  \nonumber \\
  & + & (xx \vert f_{xc}^{\uparrow,\downarrow}[\rho_y,\rho_y] \vert yy)
  - (yy \vert f_{xc}^{\uparrow,\downarrow}[\rho_y,\rho_y] \vert yy)
  + \mbox{ HOT} \, .
  \label{eq:02.16}
\end{eqnarray}
As 
\begin{equation}
  f_{xc}^{\sigma,\tau}[\rho_y,\rho_y](1,2) = 
  f_{xc}^{\sigma,\tau}[\rho_x,\rho_x](1,2) + \mbox{ HOT} \, ,
  \label{eq:O2.17}
\end{equation}
then the final second-order expansion for the DFT-MSM $B$ becomes,
\begin{eqnarray}
  B^{\mbox{DFT-MSM}} & = & (xx \vert f_H \vert xx) 
  + \left[ (xx \vert f_H \vert xx) + (xx \vert
  f_{xc}^{\uparrow,\uparrow} \vert xx) \right]  \nonumber \\
  & - & (xx \vert f_H \vert yy) \nonumber \\
  & + & (xx \vert f_{xc}^{\uparrow,\uparrow} \vert yy) \nonumber \\
  & + & \left[ (yy \vert f_{xc}^{\uparrow,\downarrow} \vert yy)
  - 2(xx \vert f_{xc}^{\uparrow,\downarrow} \vert yy) \right] \nonumber \\
  & - & \frac{1}{2} \left[ (xx \vert f_{xc}^{\uparrow,\uparrow} \vert xx)
  + (yy \vert f_{xc}^{\uparrow,\uparrow} \vert yy) \right]
   \, .
  \label{eq:O2.18}
\end{eqnarray}
Note how this fits the general from expected from applying Eq.~(\ref{eq:EXAN})
to Eq.~(\ref{eq:O2.8}).  

\begin{figure}
\begin{center}
\begin{tabular}{cc}
(a) &  \\
(b) & \includegraphics[width=0.6\textwidth]{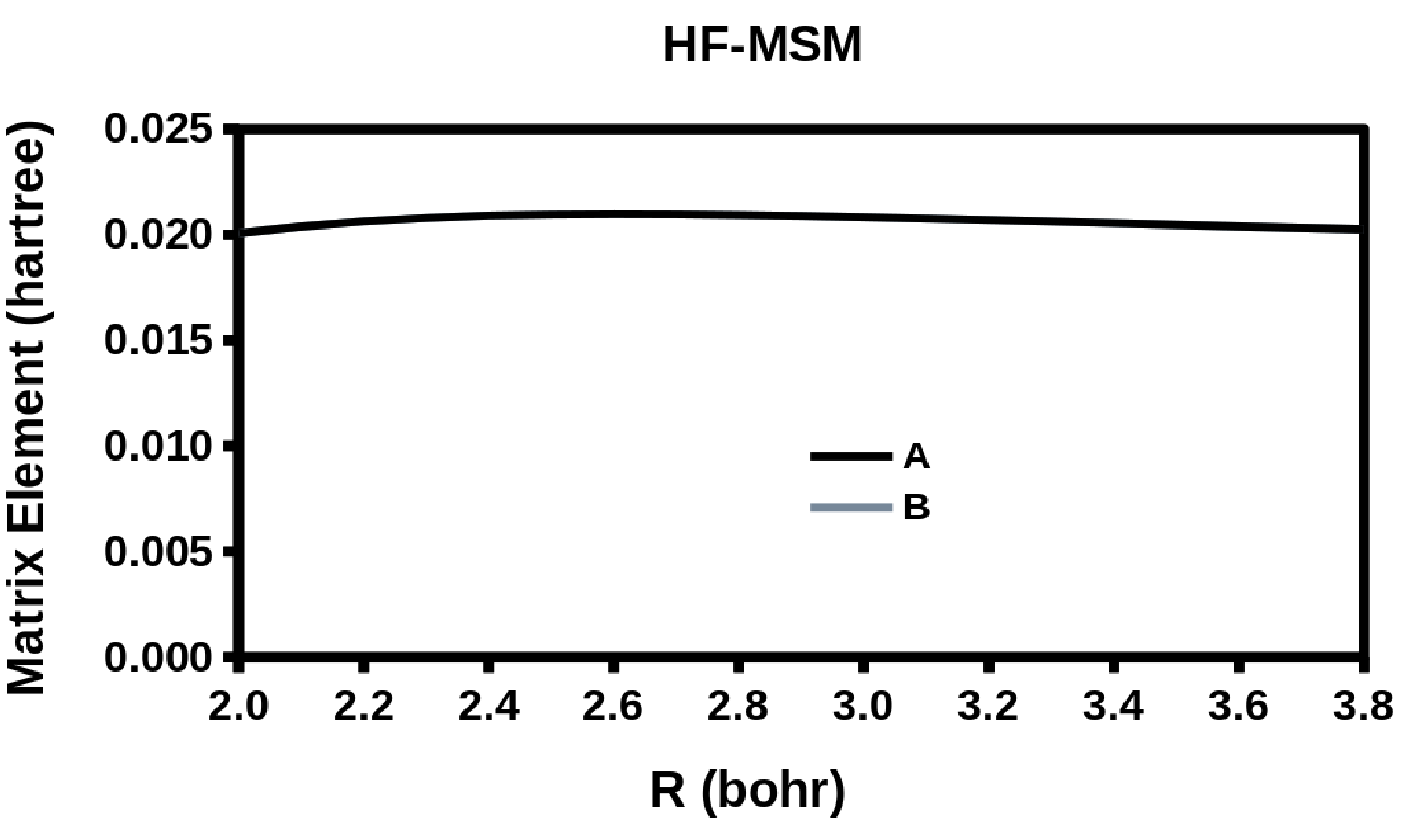} \\
(c) & \includegraphics[width=0.6\textwidth]{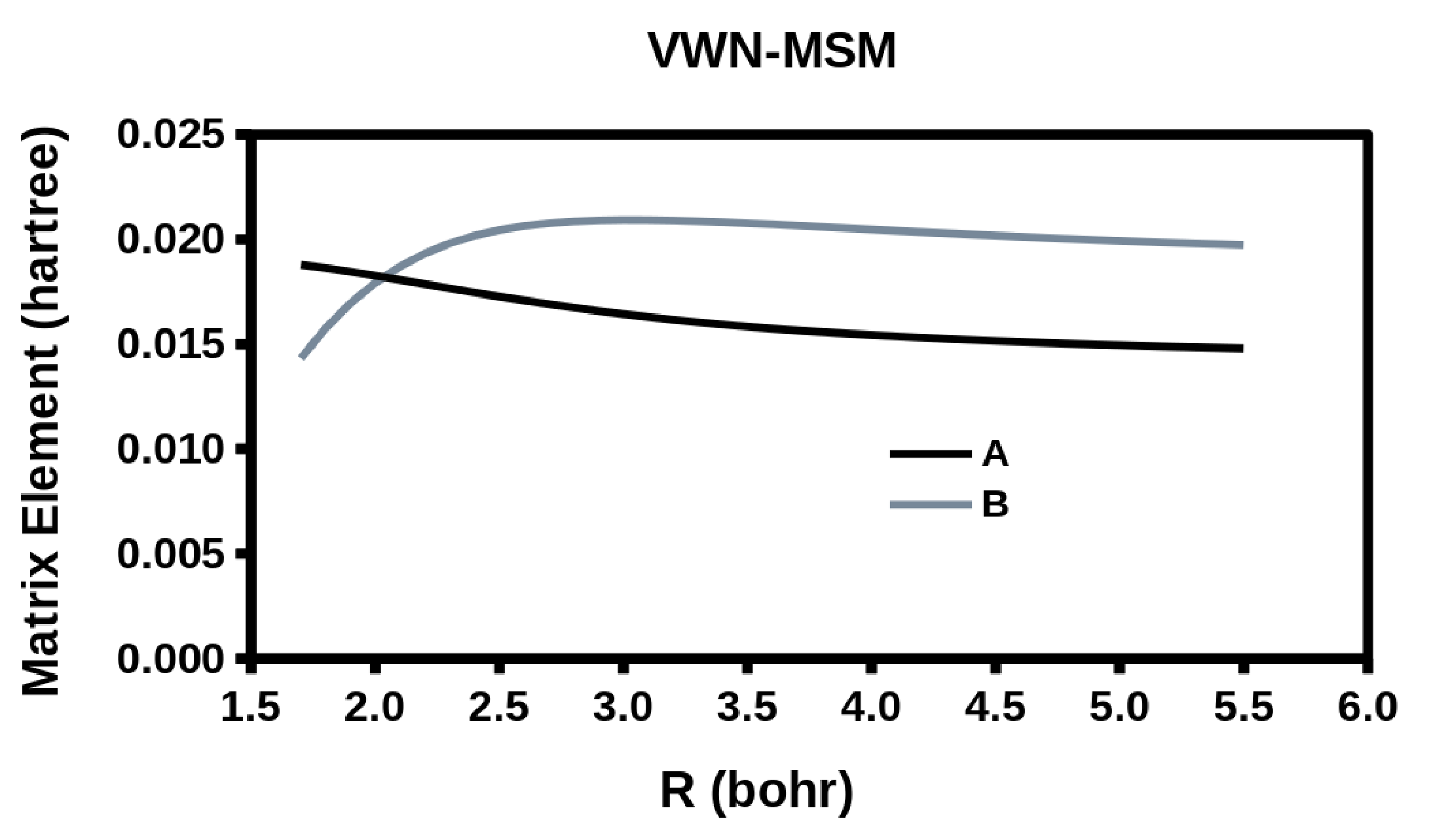} \\
    & \includegraphics[width=0.6\textwidth]{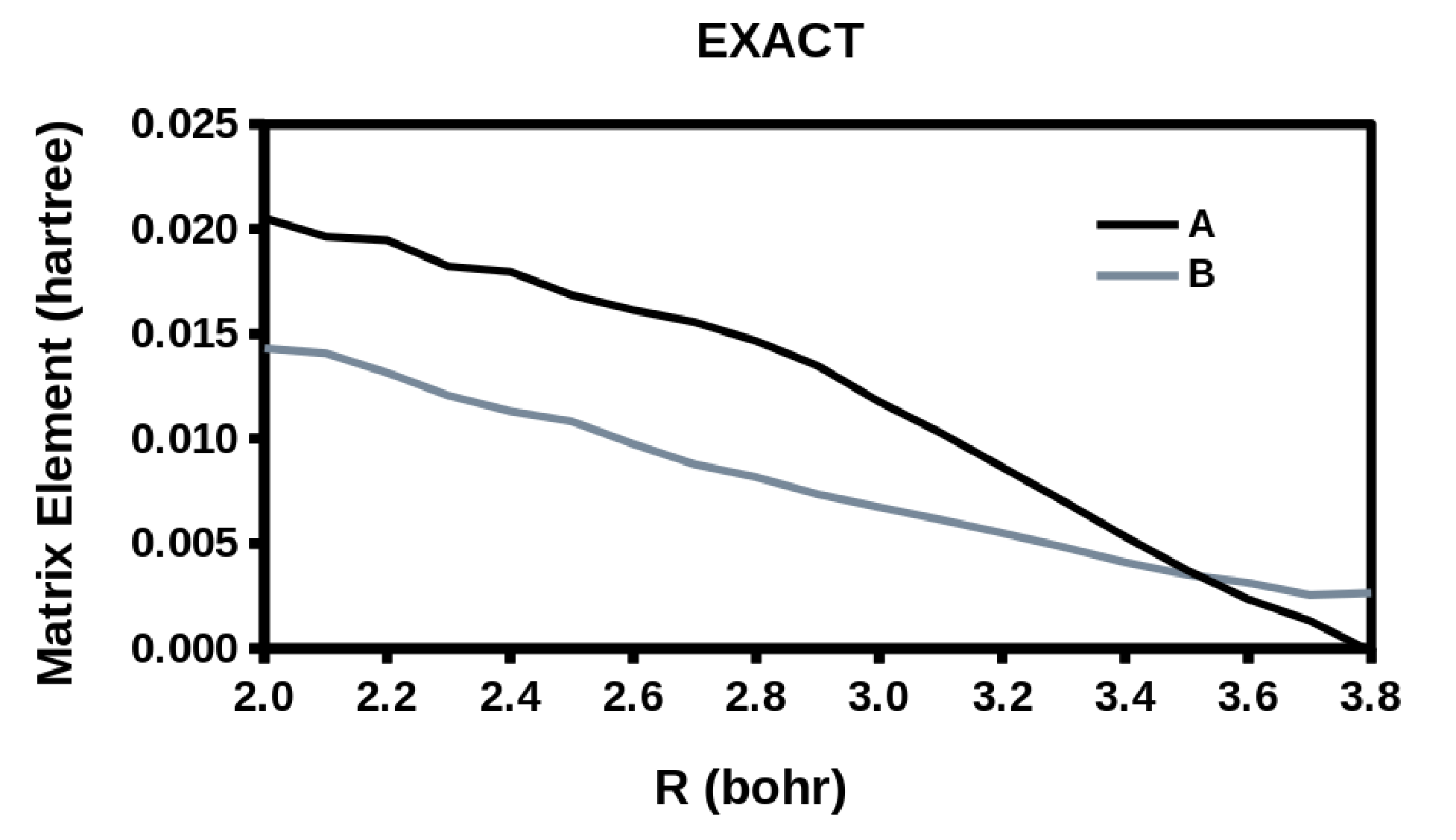}
\end{tabular}
\end{center}
\caption{
Plot of $A$ and $B$ as a function of bond distance $R$: (a) MSM-HF, 
(b) MSM-VWN, (c) EXACT.  Note that the $A$ and $B$ lines are identical 
in part (a).  The EXACT calculations are the same as the one used in
Ref.~\cite{PEMC21}. The equilibrium bond length in the ground state
is 2.3 bohr and it is very close to this value for the two excited states.
\label{fig:O2BvsA}
}
\end{figure}

It is also far from obvious that $B^{\mbox{DFT-MSM}}$ should equal,
\begin{equation}
  A^{\mbox{DFT-MSM}} = (xx \vert f_{xc}^{\uparrow,\uparrow} \vert yy)
  - (xx \vert f_{xc}^{\uparrow,\downarrow} \vert yy) \, ,
  \label{eq:O2.19}
\end{equation}
and they are probably not equal in practice.  Setting them equal to each
other, and hence lifting some of the symmetry-imposed degeneracy, might be
the price that we have to pay to free ourselves from symmetry in the MSM.
If so, we would still like to insure that $B^{\mbox{DFT-MSM}} \approx
A^{\mbox{MSM-DFT}}$.   In order to investigate this further, we carried
out MSM-HF and MSM-VWN calcutions beginning with a reference state with
half an electron of each spin in the $x$ and $y$ orbitals. We followed
the same procedure as for H$_2$ and LiH except for the VWN-MSM calculations
where we encountered excessive problems with symmetry breaking and decided
to use the same reference but with the {\tt FOCK} option so that the 
VWN energies are evaluated using HF MOs.  This is not expected to make
much difference in a small highly symmetric molecule such as O$_2$.
The results are shown in {\bf Fig.~\ref{fig:O2BvsA}}.
These calculations confirm that $B=A$ for the MSM-HF and both are roughly
independent of $R$.  For the MSM-VWN, $B$ is approximately equal to, but
smaller than, $A$ for most $R$.  So MSM-VWN is behaving roughly like MSM-HF
and $B \approx A$ is a reasonable first approximation.  For the EXACT
calculations, both $A$ and $B$ decrease as $R$ increases.  This is not
captured in our simple calculations, but we note that the MSM-DFT is able
to give fairly good results compared to the EXACT results near the 
ground-state equlibrium bond length {\em provided} we are willing to 
climb Jacob's ladder and use more sophisticated functionals \cite{PEMC21}.

%
%

\section{Concluding Discussion}
\label{sec:conclude}

This article began with a discussion of the need for a simple parameter-free
way to incorporate static correlation into DFT as a way to overcome effective
failures of NVR in practical applications of DFT, notably to photochemistry. 
DFT-MSM is a well-established symmetry-based method for introducing some 
multideterminantal character into DFT.  It has some advantages in the simplicity
with which it is able to treat charge transfer and (some) double excitations.
It may be viewed as constructing and diagonalizing a small CI matrix from 
different occupations of frozen orbitals constructed from some 
``state-averaged'' reference configuration.  The diagonal elements
are calculated using DFT with standard approximate xc-functionals.  The 
off-diagonal elements could be calculated from WFT but we have shown in our
examples that such matrix elements are rather inappropriate for use in 
MSM-DFT.

This article has presented a diagrammatic method for analyzing MSM-DFT
with the lofty aim of freeing MSM-DFT from the symmetry-dependence of its
traditional form.  We emphasize that we are neither doing diagrammatic
MBPT on top of DFT as would be done
in $GW$ or Bethe-Salpeter Green's function calculations 
(e.g., Ref.~\cite{ORR02}) nor are we adding MBPT corrections as in 
dressed TD-DFT (e.g., Ref.~\cite{HIRC11} and references therein).
Instead, this is a tool for revealing ``hidden'' similarities between MSM-DFT
and WFT.  ``Hidden'' here means that rather than calculating the same
object in two different ways, the object is playing the same role in the\
two theories.  In particular, we see that certain terms in MSM-DFT
can play a similar role in MSM-DFT as they do in WFT, even though we
have no other justification for this than that the diagrams are similar.

We have given a few examples of MSM calculations for calculating the PECs
of simple molecules within the TOTEM.  A deliberate attempt was made to use 
only the simplest functionals, namely VWN and HF, as we wanted to keep our
examples as elementary as possible.  In general, the HF-MSM calculations do 
better at describing the shape of the open-shell singlet PECs in both H$_2$
and LiH.  Surprisingly, the diagrammatic approach yields a simple but
effective parameter-free way to dissociate H$_2$ without symmetry breaking.
The same approach is not possible for LiH within the TOTEM as it would
most likely require an expanded active space.  On the other hand, the 
VWN-MSM describes charge transfer well which would not be the case in TD-DFT
\cite{FRM11}.  

Diagrammatic MSM-DFT also suggests a way to carry out symmetry-free 
calculations for O$_2$ by equating $B=A$ (Subsec.~\ref{sec:O2}).  This 
necessarily means that degenerate states in O$_2$ will no longer be 
exactly degenerate, but this may be a small price to pay for enlarging
the field of application of MSM-DFT beyond high-symmetry problems.

On the occasion of this {\em Festschrift} in honor of John P.\ Perdew's
80th birthday, we think it appropriate to emphasize the parameter-free
first principles approach taken here, even if not justified within 
formal DFT.  John is one of the strongest advocates of creating
{\em ab initio} functionals that fulfill as many conditions as possible
that should be satisfied by the exact xc functional on the grounds
that such a functional would be most likely to describe a variety
of systems from atoms to molecules equally well.  As discussed in the
introductory section of this article, NVR is known to fail for many
systems and effective failure of NVR is omnipresent in many photochemical
applications of DFT because of the presence of strong static correlation
effects due to quasi-degeneracies.  Hence there is a need for a 
parameter-free multideterminantal DFT which includes static correlation.  The
present article does not yet meet that lofty ideal as we should show,
for example, how to go beyond the TOTEM and how gradients would be calculated
before claiming to have a reasonably complete theory.  However the present
method, as far as we have gone, is indeed parameter-free and is, we think,
not without merit as a step in the right direction.

%
%

\section*{Acknowledgements}
\label{sec:thanks}

The formal part of this project was initiated before the 7th African
School on Electronic Structure Methods and Applications (ASESMA) which
took place in Kigali, Rwanda, in 2023.  However the numerical calculations
began as an ASESMA project with the intent of doing research-level work
that could be continued after ASESMA and submitted.  Calculations were
done on our personal computers running {\sc deMon2k} under {\sc Linux}.
MEC is grateful to Anne Milet for financing his plane tickets to and from
ASESMA.  MEC and AP are grateful to Anne Justine Etindele for many 
helpful discussions.  We are all grateful to Stephen Nyaranga for 
helpful discussions.  He began this project with us but was unfortunately 
unable to continue.  We are all grateful to the many people who have made 
this and previous ASESMA schools possible.  They are too many to be named here.

\section*{Supplementary Information}
\label{sec:suppl}

\begin{itemize}
 \item Multideterminantal Extensions to DFT
 \item Attempts to Unify TD-DFT amd MSM-DFT
 \item LiH Example Calculations
 \item Proof of the Identical Nature of Two Expressions in the MSM-HF
 Treatment of O$_2$
 \item CRediT contributor roles \cite{CRediT}.
\end{itemize}

\bibliographystyle{myaip}
\bibliography{refs}
\end{document}